\documentclass[preprint]{aastex}
\usepackage{epsf}
\font\smcap=cmcsc10

\begin{document}

\title{Metallicity and Kinematics of M31's Outer Stellar Halo from a
Keck\altaffilmark{1} Spectroscopic Survey}

\author{David B.\ Reitzel\altaffilmark{2}}

\affil{Department of Physics \& Astronomy, 8371 Mathematical Sciences
Building, University of California at Los Angeles, Los Angeles, California
90095, USA\\
Email: {\tt reitzel@astro.ucla.edu}}

\and

\author{Puragra Guhathakurta}

\affil{UCO/Lick Observatory, Department of Astronomy \& Astrophysics,\\
University of California, 1156 High Street, Santa Cruz, California 95064,
USA\\
Email: {\tt raja@ucolick.org}}

\altaffiltext{1}{Data presented herein were obtained at the W.\ M.\ Keck
Observatory, which is operated as a scientific partnership among the
California Institute of Technology, the University of California and the
National Aeronautics and Space Administration.  The Observatory was made
possible by the generous financial support of the W.\ M.\ Keck Foundation.}

\altaffiltext{2}{This research was carried out as part of DBR's Ph.D.\
thesis in the Department of Astronomy \& Astrophysics at the University of
California at Santa Cruz.}

\begin{abstract}
We present first results from a spectroscopic survey designed to examine the
metallicity and kinematics of individual red giant branch stars in the outer
halo of the Andromeda spiral galaxy (M31).  This study is based on multislit
spectroscopy with the Keck~II 10-m telescope and Low Resolution Imaging
Spectrograph of the Ca\,{\smcap ii} near-infrared triplet in 99~M31 halo
candidates in a field at $R=19$~kpc on the SE minor axis with brightnesses
from $20<I<22$.  The spectra are used to isolate M31 halo red giants from
foreground Milky Way dwarf stars, faint compact background galaxies, and M31
disk giants.  The observed distribution of radial velocities is well fit by
an equal mix of foreground Milky Way dwarf stars, drawn from a standard
Galactic model and with velocities $v\lesssim0$~km~s$^{-1}$, and M31 halo
giants represented by a Gaussian of width $\sigma_v^{\rm
M31}\sim150$~km~s$^{-1}$ centered on its systemic velocity of $v_{\rm
sys}^{\rm M31}\approx-300$~km~s$^{-1}$.  A secure sample of 29~M31 red giant
stars is identified on the basis of radial velocity ($v<-220$~km~s$^{-1}$),
and, in the case of four intermediate-velocity stars
($-160<v<-220$~km~s$^{-1}$), broadband $B-I$ color.  For this sample of
objects, there is rough agreement between the metallicities derived in
independent ways: two different calibrations of the Ca\,{\smcap ii}
absorption line strength and a photometric estimate based on fitting model
stellar isochrones to an object's location in a ($B-I,~I$) color-magnitude
diagram.  The [Fe/H] distribution of M31 halo giants has an rms spread of at
least 0.6~dex and spans the $\gtrsim2$~dex range over which the abundance
measurement methods are calibrated.  The mean/median metallicity of the M31
halo is about $\rm\langle[Fe/H]\rangle=-1.9$ to $-1.1$~dex (depending on the
details of metallicity calibration and sample selection) and possibly higher:
the high-metallicity end of the distribution is poorly constrained by our
data since the selection function for the secure M31 sample excludes $>80\%$
of the giants in solar/super-solar metallicity range.  Possible reasons are
explored for the apparent discrepancy between the mean [Fe/H] found in our
spectroscopic survey (corrected for metallicity selection bias) and the
slightly higher mean values found in earlier photometric studies.  Field halo
red giants in M31 appear to be somewhat more metal-rich on average than their
Milky Way counterparts.  The M31 halo [Fe/H] distribution is comparable to
that of M31 globular clusters, Galactic globular clusters, and Local Group
dwarf satellite galaxies.  The data in this 19~kpc outer halo field are
broadly consistent with a scenario in which the halo is built from the
accretion of small stellar subsystems.  There are four~stars in the secure
M31 sample which have particularly strong Ca\,{\smcap ii} lines indicating
solar metallicity, at a common velocity of $\approx-340$~km~s$^{-1}$ close to
the galaxy's systemic velocity, similar to what might be expected for M31
disk giants on the minor axis.  An extrapolation of the inner disk brightness
profile, however, falls far short of accounting for these four~stars---the
disk would instead have to be very large ($R_{\rm disk}\gtrsim80$~kpc) and/or
warped.  More likely, these four stars represent a metal-rich debris trail
from a past accretion event in the halo.
\end{abstract}


\section{Introduction}\label{specintro}

Studying the metallicity gradient and spread in metallicity of galactic
spheroids is important for understanding their formation and evolutionary
history.  The dissipational collapse model (Eggen, Lynden-Bell,~\& Sandage
1962; Larson 1974) predicts a strong radial metallicity gradient and a small
spread of metallicities for fields at large distances from the center of the
galaxy.  The gradient is due to progressive chemical enrichment during the
collapse of the protogalactic gas cloud while the small spread in metallicity
is due to the paucity of metals present when the outer halo stars formed.
The accretion model (Searle~\& Zinn 1978), on the other hand, predicts no
strong gradient and a larger metallicity spread.  In this picture, star
formation largely precedes assembly of the galactic spheroid and the stellar
subsystems which go into making the galaxy can have a variety of enrichment
and star formation histories which are mixed together more or less randomly
when the galaxy forms.  Computer simulations and semi-analytic modeling of
galaxy formation have reached new levels of sophistication in terms of
dynamic range and inclusion of relevant physical processes such as star
formation and feedback (cf.~Johnston, Hernquist,~\& Bolte 1996; Johnston
1998; Helmi \& White 1999; Helmi \& de~Zeeuw 2000; Bullock, Kravtsov,~\&
Weinberg 2001).  While the simulations are not yet detailed enough to make
reliable predictions about the metallicity distribution within galaxies, they
generally display a Searle-Zinn flavor of accretion during the formation of
galactic spheroids.

The halo of the Galaxy has been studied in both its globular cluster system
and field stars.  Zinn (1993) finds evidence for two subpopulations of
globular clusters.  An ``old halo'' population of clusters displays a radial
[Fe/H] gradient, a small age spread, and significant rotation, while the
(relatively) ``young halo'' population displays no radial [Fe/H] gradient, a
large age spread, and very little rotation.  Zinn associates the former
population with the initial collapse of the galaxy, and the latter with
accretion of stellar systems throughout the lifetime of the Galaxy.  The
Milky Way halo field star population, however, does not show evidence for a
radial metallicity gradient and has a lower mean metallicity than the
clusters \cite{Carney}.

The Andromeda spiral galaxy (M31) is a good object to test these competing
galaxy formation models as it provides an external perspective of a large
spiral similar to our own and yet is close enough for individual stars to be
studied in detail.  Globular clusters and field red giant branch (RGB) stars
are two visible tracers of M31's halo.  Huchra, Brodie,~\& Kent (1991) showed
that there is evidence for a weak metallicity gradient in a sample of 150~M31
globular clusters, with a mean metallicity of $\rm [Fe/H]=-1.2$~dex, which is
slightly higher than the mean value of $\rm [Fe/H]=-1.4$~dex for Galactic
globular clusters \cite{ZW84}. The M31 clusters also have a range of
metallicities comparable to that of the Galactic globular clusters.  While
globular clusters are relatively prominent by virtue of their luminosity and
unique appearance, M31 field halo stars are more elusive.

Over the last few decades, several groups have studied the metallicity
distribution of field RGB stars in M31's halo.  Fields with projected
distances from the center ranging from $7\lesssim{R}\lesssim34$~kpc have been
targeted in ground-based imaging surveys.\footnote{Throughout this paper,
$D=783$~kpc is adopted as the distance to M31 (Stanek \& Garnavich 1998;
Holland 1998).}  There is rough agreement in the mean metallicity measured in
these studies, with values ranging from $\rm-0.6\leq[Fe/H]\leq-1.0$~dex.  In
particular, the recent $R=20$~kpc minor axis wide-field study by Durrell,
Harris,~\& Pritchet (2001) finds a median [Fe/H] value (0.3~dex lower than
their derived [m/H] value) near the low-metallicity end of this range, but
even this is slightly higher than the mean of the M31 globular cluster
system.  Another striking result is the discovery of an apparent streamer in
the projected distribution of M31 halo red giants, whose metallicity appears
to be similarly high judging from broadband colors of the constituent red
giants \cite{Ibata}.  The reader is referred to Reitzel, Guhathakurta,~\&
Gould (1998, hereafter referred to as RGG) and references therein for a
discussion of earlier ground-based studies.

RGG have studied a $15'\times15'$ field in M31's outer spheroid at $R=19$~kpc
on the SE minor axis.  In that study, deep $UBRI$ images obtained using the
Kitt Peak National Observatory\footnote{Kitt Peak National Observatory of the
National Optical Astronomy Observatories is operated by the Association of
Universities for Research in Astronomy, Inc., under cooperative agreement
with the National Science Foundation.} (KPNO) 4-m telescope, and $I$-band
images obtained with the Keck 10-m telescope and Low Resolution Imaging
Spectrograph (LRIS, Oke et~al.\ 1995), have been used to isolate a sample of
M31 halo candidate RGB stars.  The M31 RGB stars are distinguished from the
more numerous distant field galaxies on the basis of broadband $U-B$, $B-R$,
and $R-I$ colors and image morphology.  The M31 halo field contains a clear
excess of faint red objects ($I\sim20\>$--$\>$23, $B-I\sim2\>$--$\>$3.5)
relative to a well-matched photometric control sample in a comparison field.
The location of this population of faint red objects in the color-magnitude
diagram (CMD) is as would be expected for red giant stars at the distance of
M31.  The overall color and distribution in the CMD of these excess objects
suggests a relatively metal-rich population
$\rm\langle[Fe/H]\rangle\ga-1$~dex, with a spread of $\sim2$~dex.

A couple of studies of the M31 spheroid have used the {\it Hubble Space
Telescope\/} ({\it HST\/}) Wide Field/Planetary Camera~2 (WFPC2) images to
distinguish between stars and distant field galaxies.  Rich et~al.\ (1996a,b,
2002) have analyzed archival images of a field around the M31 globular
cluster G1 ($R=34$~kpc, major axis) along with fields surrounding a few other
clusters distributed across the galaxy's halo; Holland, Fahlman,~\& Richer
(1996) have analyzed fields around two~clusters, G302 ($R=7$~kpc) and G312
($R=11$~kpc) near M31's minor axis.\footnote{We hereafter refer to these as
the ``G1 field'', ``G302 field'', etc., because of their proximity to the
corresponding M31 globular clusters, but should point out that the samples
under discussion contain only field stars and no cluster members.}  The
results of these {\it HST\/} studies are in good agreement with ground-based
results despite the different contamination issues (see below).  These
studies find the mean metallicity of M31's halo to be comparable to that of
47~Tuc ($\rm[Fe/H]=-0.7$~dex), with a spread of nearly 2~dex, suggesting that
there was a greater degree of pre-enrichment during the assembly of M31's
spheroid than in the case of the Galaxy's spheroid.

While {\it HST\/}'s excellent angular resolution can distinguish stars from
most of the background field galaxies, M31 disk red giants, foreground
Galactic dwarf stars, and faint compact galaxies remain as possible
contaminants in studies of the M31 halo.  Many of the {\it HST\/} studies
have targeted the relatively high surface brightness inner portions of the
M31 halo (in order to include a reasonable number of stars within the limited
field of view of the camera) but some of these regions are projected against
M31's bright inner disk so that the danger of contamination by M31 disk
giants can be particularly high.  Moreover, the lack of comparison field data
in most of the {\it HST\/} studies makes it impossible to carry out
statistical subtraction.  For ground-based surveys on the other hand, even
the most careful attempts at statistical subtraction of contaminants are
hampered by field-to-field variations in photometric error, seeing, and the
surface density of foreground stars and background galaxies, along with
asymmetries in M31's disk.  Whether one uses {\it HST\/} for high angular
resolution, or ground-based telescopes for their relatively large fields of
view, sample contamination remains the main obstacle in determining the true
metallicity and metallicity spread in M31's halo.

Spectroscopy of individual RGB stars in M31 is a powerful tool for measuring
the metallicity of M31's halo.  With the advent of 8--10-m class telescopes
such as the Keck telescope and efficient multi-object spectrographs such as
LRIS, it is feasible to observe a sizeable sample of individual stars in M31.
This paper describes spectroscopy of individual RGB star candidates in M31's
halo identified in the RGG study.  Preliminary spectroscopic results have
been presented in Reitzel~\& Guhathakurta (1998, 2000), Guhathakurta \&
Reitzel (1998), and Guhathakurta, Reitzel,~\& Grebel (2000).  The selection
of the sample, observations, and data reduction are described in
\S\,\ref{data_sec}, the kinematics of the sample in \S\,\ref{v_sec}, and the
measurement of the metallicity of M31's halo in \S\,\ref{feh_sec}.  We
discuss the results in the context of other studies of the halo and model
calculations in \S\,\ref{discuss_sec}, and summarize the conclusions in
\S\,\ref{concl_sec}.

\section{Data}\label{data_sec}
\subsection{Spectroscopic Targets: M31 Red Giant Candidates and Control
Sample}\label{targets_sec}

A sample of 99~objects has been targeted for spectroscopy from the RGG
photometric study of the $R=19$~kpc SE minor axis field.  Of the 2078~objects
in that study which have $UBRI$ colors and morphology consistent with a star,
the subset of 284~objects in the apparent magnitude range $I=20\>$--$\>$22
are likely to be bright red giants in M31 (the tip of the RGB is at $I_{\rm
TRGB}\approx20.5$ at the distance of M31), and 99~of these are targeted for
spectroscopy.  The full color range is used in choosing the spectroscopic
targets, so as to avoid biasing the metallicity determination: the shape of
the $B-I$ color distribution of the 99~spectroscopic targets is identical to
that of the parent sample of 284~objects.  The distribution of apparent
$I$-band magnitudes for the spectroscopic targets is slightly skewed in favor
of bright stars relative to that of the parent sample within the range
$I=20\>$--$\>$22, reflecting the slightly higher priority given to brighter
stars during the design of the multislit masks (the significance of this
selection bias is discussed in \S\,\ref{selbias}).  In the RGG work, an
object is assigned a classification of ``star-like'' only if its color excess
significance criterion $\delta<1.5$ and angular size $\theta_{\rm
FWHM}<1\farcs4$ for the KPNO image and $\theta_{\rm FWHM}<0\farcs7$,
$0\farcs8$, $0\farcs9$, and $0\farcs9$ for the four~Keck/LRIS images (see
also Gould et~al.\ 1992).  Astrometric and photometric data for the
99~spectroscopic targets are presented in Table~\ref{phot_tbl}: positions
(good to $\sim0\farcs1$) are derived from several bright US Naval Observatory
catalog astrometric reference stars present in this low-latitude field;
$UBRI$ photometry is from RGG, while $V$ magnitudes are synthesized from
these data and model stellar isochrones (\S\,\ref{feh_spec_ccf_sec}).

\begin{table}
\dummytable\label{phot_tbl}
\end{table}

There are two objects in the list of spectroscopic targets that do not pass
the above $UBRI$ and morphology selection criteria.  This is due to the fact
that these objects were selected before the final refinements to the
photometric and morphological selection were made in the RGG work.  One of
them, t1.01, barely fails the final color selection with $\delta=1.55$, while
the second, t2.22, fails the final morphological selection with $\theta_{\rm
FWHM}=1\farcs5$ on the KPNO image.  It turns out that both objects are likely
foreground Galactic dwarf stars based on the fact that their radial
velocities are close to zero (see \S\,\ref{secure} and Table~\ref{spec_tbl}).

Spectra have also been obtained for 13~objects with $I<20$, which places them
well above the expected tip of M31's RGB ($I_{\rm TRGB}\approx20.5$).  These
objects form a control sample of Galactic stars whose kinematical properties
are compared to the Institute for Advanced Study Galaxy (IASG) star count
model \cite{Ratnatunga} to aid in the rejection of foreground contaminants
from the main sample of M31 halo giant candidates.

\subsection{Observations}\label{obs_sec}

The spectra were obtained using the Keck~II 10-m telescope and LRIS during
two~2-night observing runs in 1996 October and 1997 October.  These observing
runs predate the installation of the blue side of LRIS, so the instrumental
details given below refer to what is now the ``red side'' of LRIS.  The
observations were carried out in multi-slit mode with a typical slit mask
having 20--30~slitlets on M31 halo RGB candidates and control sample stars.
Each slitlet has a width of $1\farcs0$ and the typical slitlet length is
$10''$.  The 1200~lines~mm$^{-1}$ grating provides a dispersion of
0.62\,\AA~pixel$^{-1}$ and was used to cover the spectral range
$\lambda\lambda7550\,$--$\,$8850\,\AA\ containing the near-infrared
Ca\,{\smcap ii} triplet: $\lambda\lambda8498$, 8542, and 8662\,\AA.  The
instrumental spectral/velocity resolution is 1.94\,\AA/68~km~s$^{-1}$ (FWHM)
or a Gaussian $\sigma$ of 0.83\,\AA/29~km~s$^{-1}$.  This is based on a
dispersion of 0.62\,\AA~pixel$^{-1}$, a 1\farcs0 slit width, a pixel scale of
0\farcs215~pixel$^{-1}$ along the spatial direction, an anamorphic
magnification factor of 0.565, and the best spectrograph focus of $\rm
FWHM\sim1.7$~pixel \cite{Phillips}.  The seeing FWHM was typically better
than 1\farcs0, so the actual velocity resolution is slightly better than the
above estimate---e.g.,~$\sigma\sim25$~km~s$^{-1}$ for 0\farcs8 seeing.  A
slight dither of about $1''$ along the slit was done between exposures to
help with fringe removal and to minimize the effect of bad pixels.  Five
multislit masks were used to obtain spectra with total exposure times
$t_{\rm exp}$ of 3.8, 3.7, 4.5, 4.5, and 1.5~hr.  Individual exposure times
were typically 50~min long, although some exposures had to be truncated due
to telescope and instrument problems.  Detailed exposure information is
presented in Table~\ref{obs_tbl}.

\begin{table}
\dummytable\label{obs_tbl}
\end{table}

For the purpose of checking the metallicity calibration, a handful of red
giants were observed in M79 and NGC~6791, two reference Milky Way star
clusters.  Short Keck/LRIS exposures in long-slit spectroscopic mode and
broad-band imaging mode were obtained during twilight/dawn on the nights of
the main M31 observations.  The instrumental setup for these calibration star
long-slit spectra is practically identical to the above setup used for the
M31 multi-slit spectra, and the reduction procedure is essentially the same
as for the main M31 data set (\S\,\ref{datared_sec}).  The calibration star
cluster images are analyzed with DAOPHOT/ALLSTAR (Stetson 1987, 1992) to
derive $V$ magnitudes for the red giants for which we have spectra.

\subsection{Data Reduction}\label{datared_sec}

The data are reduced using standard IRAF\footnote{IRAF is distributed by the
National Optical Astronomy Observatories, which are operated by the
Association of Universities for Research in Astronomy, Inc., under
cooperative agreement with the National Science Foundation.} tasks and the
spectral reduction program Expector \cite{Kelson}.  The main steps are
summarized below, annotated by {\bf [I]} for IRAF or {\bf [E]} for Expector:

\begin{itemize}
\item[{\bf [I]}]{{\bf Overscan subtraction} of the mean bias level is done
for each raw CCD frame and the overscan region is then {\bf trimmed} off.
Two-dimensional {\bf bias subtraction} is carried out using the median of
several 1-s dark exposures.}

\item[{\bf [I]}]{{\bf Cosmic ray} events (identified by object sharpness/peak
pixel brightness) are masked, including a surrounding 1-pixel buffer zone to
capture low level wings.}

\item[{\bf [I]}]{{\bf Geometric distortion} in the images is mapped by
tracing the boundaries between slitlets using night sky atmospheric emission
lines in the M31 multislit spectral exposures (`data frames') and emission
lines in short arc lamp calibration exposures.  The distortion mapping and
correction are done using the tasks {\smcap IDENT}, {\smcap REIDENT}, {\smcap
FITCOORDS}, and {\smcap TRANSFORM}.}

\item[{\bf [E]}]{{\bf Flat field} and {\bf slitlet illumination corrections}
are performed using a spectral dome flat that is well-matched to each data
frame in terms of LRIS flexure effects.  Mechanical flexure in LRIS can cause
the location of a slitlet to shift by up to 2--3~pixels from image to image.
Thus ``flexure matching'' requires that the dome flat observations be
obtained at the same telescope elevation angle and instrument rotator angle
as the data frame to ensure the same orientation of LRIS with respect to the
gravity vector.  The matched spectral dome flat is corrected for geometric
distortion as described above.  The `slit function' for each slitlet is
obtained by collapsing along the dispersion direction (via median-filtering)
the corresponding band in the rectified spectral dome flat.  The `slit
function image', a composite of the slit illumination functions of all the
slitlets, is then applied to the spectral dome flat.  This produces a map of
the Fabry-Perot fringe pattern and fixed-pattern quantum efficiency
variations which is divided into the data frame to achieve pixel-to-pixel
flat fielding.  The slit function image is shifted to empirically match each
data frame (to account for any difference in mechanical flexure) before it is
used to apply the slitlet illumination correction.}

\item[{\bf [E]}]{{\bf Wavelength calibration} is based on numerous night sky
emission lines (of known wavelength) in the Ca\,{\smcap ii} portion of the
spectrum.  An independent wavelength solution is derived for each slitlet on
a given multislit spectral frame, and all slitlets are transformed to a
common wavelength grid.  The exact wavelength coverage for a slitlet varies
with position along the dispersion axis within the multislit mask---i.e.,~its
CCD column number in a direct image of the mask---but all slitlets cover all
three~lines of the Ca\,{\smcap ii} triplet, even after allowing for Doppler
shifts of $\pm5000$~km~s$^{-1}$.}

\item[{\bf [E]}]{{\bf Sky subtraction} is a critical step because of the
large number of strong emission lines in this spectral region.  Each slitlet
is visually inspected and `sky windows' (ranges of row numbers) are defined
on either side of the intended target for that slitlet, taking care to avoid
the point spread function wings of the target and contamination from bright
neighboring objects.  The mean sky spectrum at the location of the target is
linearly interpolated by collapsing the rows within the sky windows, after
clipping out $\sim10\%$ of the highest and lowest pixel values.  Flexure in
LRIS can cause slight differences in the fringe pattern between data frame
and dome flat which are not accounted for in the above procedure; this
occasionally manifests itself in the form of large, systematic residuals in
the vicinity of the brightest sky lines.}

\item[{\bf [I]}]{{\bf Extraction} of one-dimensional spectra (one target per
slitlet) is carried out on each sky-subtracted multislit exposure.  A
16-pixel-wide `target window' is defined centered on each target, and the
rows are collapsed into a weighted average using the optimal scheme described
by Horne (1986).  The 1-D extracted spectra are normalized by dividing by
their mean continuum level.  The individual normalized 1-D spectra of a given
target (one per exposure) are {\bf coadded} by weighted averaging, with the
weight being the inverse of the square of the typical photon noise in a clean
part of the spectrum.}
\end{itemize}

Figure~\ref{sample_spec} shows the spectra and corresponding uncertainty
levels for two representative targets in our sample, along with a typical
night sky spectrum.  The night sky spectrum is dominated by bright emission
lines all across the region of the near-infrared Ca\,{\smcap ii} triplet.
The trio of absorption lines is clearly visible in the spectrum of the bright
object ($I=20.7$) and the two strongest lines ($\lambda\lambda8542$ and
8662\,\AA) are
visible in the spectrum of the faint object ($I=21.7$).  The rest wavelengths
of the Ca\,{\smcap ii} absorption lines are indicated by the bold lines below
the spectra.  The Ca\,{\smcap ii} lines for both objects are shifted to the
blue, as might be expected for RGB stars in M31: $v_{\rm
obs}=-173$~km~s$^{-1}$ (bright object) and $v_{\rm obs}=-500$~km~s$^{-1}$
(faint object).  The $1\sigma$ error in the mean spectrum of a given target
is empirically estimated to be rms/$\sqrt{N}$, where `rms' is the weighted
rms scatter about the weighted mean of the individual spectra extracted from
the $N$ independent exposures of the same object.  The rms is not computed
for objects on multislit mask~\#5 as the number of independent exposures $N$
is only 2 (Table~\ref{obs_tbl}).  The night sky spectrum is useful for
identifying regions where sky subtraction is problematic; the rms for each
object increases at the locations of the bright emission lines because of
increased Poisson noise and residual fringing.  Away from these bright night
sky emission lines, the S/N ratio per pixel is $\gtrsim20$ for the bright
object and in the range 10--20 for the faint one.

A variety of factors play a role in determining how well the strength and
position (i.e.,~radial velocity) of the Ca\,{\smcap ii} absorption line
triplet are measured for a given target.  In addition to the usual dependence
of the Poisson error on the target's apparent $I$-band magnitude, absorption
line strength, and effective exposure time, there are several possible
sources of systematic error: large residuals at the locations of bright night
sky emission lines because of imperfect fringe removal (typically due to
differential LRIS mechanical flexure and/or slight wavelength mismatch
between data frame and corresponding dome flat); scattered light/spectral
ghosts; poor seeing; focus variations across the field of view of the mask;
miscentering of a target within the slitlet due to errors in astrometry
and/or slitmask misalignment; and the occasional placement of a target near
the end of the slitmask which causes its spectrum to land near the edges of
the CCD making sky subtraction and extraction of the spectrum problematic.

\subsection{Cross-Correlation Analysis}\label{ccf_sec}

In order to determine the radial velocity and Ca\,{\smcap ii} line strength
of each object, its final coadded spectrum is cross-correlated against a
template spectrum.  The cross-correlation function (CCF) is computed from
$-1000$ to +1000~km~s$^{-1}$, covering a plausible range of radial velocities
for stars associated with M31.  The cross-correlation technique yields an
unambiguous peak and a reliable radial velocity for 80 of the 99~objects
comprising the main sample of M31 targets and all 13~control objects.  The
CCF for each object is shown in Fig.~\ref{ccf}.

Tests show that the strength, significance, and position of the CCF peak are
insensitive to the exact choice of template (e.g.,~red giant vs.\ dwarf
star), which is understandable given that the CCF signal is driven almost
entirely by the three~lines of the Ca\,{\smcap ii} triplet whose relative
strengths and line shapes are the same across these templates at the
relatively low resolution of our spectra.  The template used in the rest of
this paper is the average spectrum of three bright control sample stars, each
of which is shifted to the rest frame (zero velocity) before coaddition.  The
template is masked so that only a window around each of the three Ca\,{\smcap
ii} lines is used in the cross-correlation analysis, where the window size is
optimized to maximize the S/N ratio in the resulting CCF.

Of the 19~objects for which the CCF fails to yield a reliable velocity, 8
have apparent magnitudes $I<21.4$ and have relatively high S/N spectra, while
11 have $I>21.4$ and are generally characterized by lower S/N data.  A brief
description of the objects is given here; they are excluded from the rest of
the analysis in this paper.

The slitlet for one of the brighter objects, t1.27, happens to be at one end
of mask~1 and its spectrum falls at the edge of the CCD frame causing
spectral extraction to fail (bold solid horizontal line in bottom right panel
of Fig.~\ref{ccf}).  The remaining 7~higher S/N objects display a wide range
of colors, $1.5\leq{B-I}\leq4.5$, and each appears to have a featureless
continuum.  Most of these objects are likely to be distant compact background
field galaxies for which the Ca\,{\smcap ii} lines and any other prominent
spectral features happen to have been redshifted out of the observed
wavelength range.  This would correspond to residual background galaxy
contamination at the level of $\approx5\%$--10\% in the RGG $UBRI$ color- and
morphology-selected sample from which the spectroscopic targets are drawn.  A
couple of the reddest objects might possibly be foreground Galactic late-M
dwarfs; their spectra tend to be a poor match to the cross-correlation
template, lacking the three Ca\,{\smcap ii} absorption lines in particular,
and the resulting CCF contains no peak at the true radial velocity of the
object (or at any other velocity for that matter).  None of these 7~higher
S/N objects is likely to be an M31 RGB star or else the Ca\,{\smcap ii}
triplet lines would have been obvious (as they are for the brighter object in
Fig.~\ref{sample_spec}).

The 11~lower S/N objects, like their higher S/N counterparts, show a large
spread in color, $1.6\leq{B-I}\leq3.9$.  Nine of the 11 contain no
identifiable spectral emission/absorption features that are significantly
above/below the continuum with respect to the noise.  One object with
$I=21.81$ and $B-I=2.13$ (object ID t5.24 in Tables~\ref{phot_tbl} and
\ref{spec_tbl}) shows very faint absorption features that may well be the
Ca\,{\smcap ii} triplet at a radial velocity of $v=-330$~km~s$^{-1}$, which
would make this object an RGB star in M31 (\S\,\ref{secure}).  The last of
these 11~lower S/N objects for which there is no reliable velocity
measurement is object t4.14 with $I=21.78$ and $B-I=2.47$.  It appears to
have {\it two\/} sets of Ca\,{\smcap ii} triplet absorption lines, with
radial velocities of $v=-435$ and $-147$~km~s$^{-1}$; this object is probably
a chance alignment of two stars, an M31 RGB star and a foreground Galactic
dwarf star, respectively (see \S\,\ref{secure}).  Inspection of the Keck/LRIS
$I$-band image reveals a slightly elongated object; the object is not
resolved into two separate peaks, and its ellipticity is not large enough for
it to fail the stellar morphological selection criterion as it is within the
range of values produced by focus variations across the image.

The standard Tonry-Davis (1979) parameter, $r_{\rm TD}=h/(\sqrt{2}\sigma_{\rm
CCF})$, is computed for each CCF, where $h$ is the height of the CCF peak and
$\sigma_{\rm CCF}$ is the rms noise in the CCF.  For 80 of the 99~objects
comprising the main sample of M31 targets and all 13~control objects, there
is an unambiguous peak in the CCF from which $h$ is calculated; for the
19~targets for which there is no clear peak in the CCF, the highest point in
the CCF is used to determine $h$.  A linear baseline is fit to the CCF after
excluding a window of width $\Delta{v}=450$~km~s$^{-1}$ centered on the CCF
peak.  The peak height $h$ and rms (and the area under the CCF peak, $A_{\rm
CCF}$---see \S\,\ref{feh_spec_ccf_sec}) are calculated relative to this
baseline, with the window around the peak excluded from the rms calculation.
The CCF baseline is typically flat and close to zero (Fig.~\ref{ccf}) so the
results are independent of the details of the baseline fit.  The $r_{\rm TD}$
parameter for each object is listed in Table~\ref{spec_tbl}.  The 80~objects
in the main M31 target sample for which the radial velocity determination is
successful have a mean significance level of $\langle{r_{\rm
TD}}\rangle=5.75$, while $\langle{r_{\rm TD}}\rangle=11.01$ for the
13~objects in the control sample.  By contrast, the significance level is far
lower on average for 18 of the 19~targets (excluding t1.27) for which the
radial velocity measurement fails: $\langle{r_{\rm TD}}\rangle=2.07$.

\section{Kinematics}\label{v_sec}

The radial velocity determined from the location of the CCF peak, $v_{\rm
obs}$, is corrected to the heliocentric frame using the task {\smcap RVCOR}
in IRAF.  The $v_{\rm hel}$ values for 80 of the 99~objects comprising the
main sample of M31 targets and all 13~objects in the control sample are
listed in Table~\ref{spec_tbl}.  This section describes: empirical
determination of the velocity measurement error; isolation of a secure sample
of M31 RGB stars using radial velocity and (in a few cases) color
information; and the dynamics of M31's stellar halo.

\begin{table}
\dummytable\label{spec_tbl}
\end{table}

\subsection{Velocity Measurement Error}\label{v_mes}

The measurement error in radial velocity is investigated by splitting the
data for each target in masks~1-4.  A total of 80~objects is used in this
analysis, all of them M31 spectroscopic targets; the control sample objects
are excluded from this analysis.  For each of these objects, the four~longest
available exposures are combined into two independent coadds of two exposures
each.  Each of these coadds is hereafter referred to as a ``coadded pair''.
Reliable radial velocities are measured for both coadded pairs for
58~objects, while the velocity measurement fails for at least one of the two
coadded pairs for the remaining 32~objects.  The failure rate for the coadded
pairs ($\approx30\%$) is higher than the failure rate for the full coadds
($\approx20\%$).  As discussed in \S\,\ref{ccf_sec}, faint galaxies and
Galactic M~dwarfs account for about 10\% of the failed measurements in the
case of the full coadds, and the measurement would of course fail again for
these objects in the case of the coadded pairs.  The rate of failure due to
insufficient S/N in the spectrum is expected to increase for the coadded
pairs (compared to the full coadds) given the reduction in effective exposure
time by over a factor of two.

The two independent velocity measurements for a given object, derived from
its two coadded pairs, are differenced.  The cumulative distribution of
velocity differences for various subsamples of objects is shown in
Fig.~\ref{verr} along with the best-fit Gaussian.  The velocity measurement
accuracy is slightly greater for the brighter half of the sample (29~objects
in the range $20<I<21.1$) than for the fainter half (29~objects in the range
$21.1<I<22$): the rms velocity difference of the coadded pairs is
$\sigma_v^{\rm pair}\rm(bright)=22.8~km~s^{-1}$ and $\sigma_v^{\rm
pair}\rm(faint)=28.0~km~s^{-1}$, respectively.  The strength of the
Ca\,{\smcap ii} absorption line triplet (characterized here by $A_{\rm
CCF}$---see \S\,\ref{feh_spec_ccf_sec}) does not have much of a bearing on
the velocity measurement accuracy: $\sigma_v^{\rm
pair}($strong-lined$)=26.8$~km~s$^{-1}$ versus $\sigma_v^{\rm
pair}($weak-lined$)=24.1$~km~s$^{-1}$ for stars with line strengths
greater/less than the median value, respectively.  As discussed in
\S\,\ref{datared_sec} above, a large number of factors can influence the data
quality so it is no surprise that the velocity measurement error does not
correlate with line strength.

Despite these complicating factors, we make the simplifying assumption that
the velocity measurement error for any given target scales as $t_{\rm
exp}^{-1/2}$.  The exposure time for the coadded pairs ranges from
1.42$\>$--$\>$1.75~hr.  A typical coadded pair exposure time of $t_{\rm
exp}^{\rm pair}=1.7$~hr is adopted for the velocity measurement error
calculation; this is at the high end of the actual range implying that the
velocity error is slightly overestimated in our calculation.  The rms of
velocity differences for the entire sample of coadded pairs is $\sigma_v^{\rm
pair}\rm(all)=25.4~km~s^{-1}$.  The average velocity measurement uncertainty
for the targets, based on the full coadd of the total available exposure
time, is therefore expected to be:
\begin{equation}
\langle{\sigma_v^{\rm tot}}\rangle=\bigl[\sigma_v^{\rm pair}({\rm
all})/\sqrt{2}\bigr]\sqrt{t_{\rm exp}^{\rm pair}/t_{\rm exp}^{\rm tot}}
\label{verr_eqn}
\end{equation}
\noindent
where $t_{\rm exp}^{\rm tot}$ is the total exposure time for the mask in
question (see Table~\ref{obs_tbl}).  The mean $1\sigma$ error in velocity is
approximately 12~km~s$^{-1}$ for targets in masks~1--4 and 19~km~s$^{-1}$ for
those in mask~5.

The velocity measurement error for an individual target is expected to scale
as: $\sigma_v^{\rm ind}=\sigma_v^{\rm TD}(1+r_{\rm TD})^{-1}$ \cite{Tonry}.
The velocity differences of the coadded pairs for the 58~targets are scaled
following Eq.~\ref{verr_eqn} to reflect the reduction of velocity error that
is expected to result from the coaddition of all the exposures:
\begin{equation}
\Delta{v}_{\rm scaled}=\bigl[\Delta{v}_{\rm pair}/\sqrt{2}\bigr]\sqrt{t_{\rm
exp}^{\rm pair}/t_{\rm exp}^{\rm tot}}
\label{deltav_eqn}
\end{equation}
\noindent
These scaled velocity differences are plotted versus $(1+r_{\rm TD})^{-1}$ in
Fig.~\ref{verr_vs_rtd}.  The Tonry-Davis velocity error parameter
$\sigma_v^{\rm TD}$ is equal to the rms of the quantity $\Delta{v}_{\rm
scaled}(1+r_{\rm TD})$; its value is empirically determined to be
77~km~s$^{-1}$ (solid lines), comparable to the instrumental velocity
resolution (\S\,\ref{obs_sec}).

\subsection{Defining a Secure Sample of M31 Red Giants}\label{secure}

Kinematical information derived from spectroscopy is useful for isolating M31
halo red giants and rejecting contaminants.  The $UBRI$ color- and
morphology-based selection in the RGG study removes most (though not all) of
the background field galaxies, but stellar contaminants in the form of
foreground Milky Way dwarf stars and M31 disk red giants remain in the
spectroscopic target sample.  Halo RGB stars in M31 are expected to span a
large range of radial velocities (due to their random motion within the
gravitational potential) centered on the galaxy's systemic velocity, $v_{\rm
sys}^{\rm M31}=-297$~km~s$^{-1}$ \cite{RC3}.  On the other hand, if M31 disk
RGB stars are present in this field, they are expected to be tightly
clustered around a single velocity given the small field of view of our
study.  This velocity is expected to be close $v_{\rm sys}^{\rm M31}$ as our
field of study is on the minor axis, so that the disk rotation velocity
vector is in the plane of the sky with a zero line-of-sight component.

The typical contaminating foreground Galactic star in the sample is a K dwarf
with colors of $B-V\approx{V-I}\approx1.5$ (Fig.~\ref{bi_cmd}), corresponding
to an absolute magnitude of $M_V\approx7$--8.  At apparent magnitudes of
$I\sim21$ or $V\sim22.5$, in the middle of the range from which the
spectroscopic targets are selected, this implies a distance modulus of
$(m-M)_V^0\approx15$ or $D\approx10$~kpc.  Such a star would be a few kpc out
of the plane---$\sin(b)=0.37$ for the Galactic latitude of M31---presumably
in the thick disk of the Milky Way, with a small but negative radial velocity
(the projection of the thick disk rotation curve along the line of sight to
M31 minus the projection of the Solar rotational speed).  Redder foreground
stars (late-K to M dwarfs) in a similar apparent magnitude range are somewhat
less luminous than the star in the above example and are thus only a few kpc
away; bluer stars (near the main-sequence turnoff for an old population) are
about an order of magnitude more luminous ($M_V\approx4$--5), which places
them a few tens of kpc away in the Galaxy's halo, with more negative radial
velocities due to the reflex of the projected Solar motion.

The distribution of radial velocities measured from the Keck/LRIS spectra is
shown in Fig.~\ref{vhist}.  There is a clear peak at $\lesssim0$~km~s$^{-1}$
superimposed on a broader distribution plausibly centered on M31's systemic
velocity.  The IASG star-count model of the Galaxy \cite{Ratnatunga} is used
to estimate the radial velocity distribution expected for foreground Milky
Way dwarf stars along this particular line of sight and for the given
apparent $I$ magnitude range.  A linear combination of the IASG model $v$
distribution and a Gaussian centered on M31's systemic velocity of
$-297$~km~s$^{-1}$ provides an adequate fit to the data (solid curve).  The
fraction of M31 halo stars in our sample is estimated to be
$\approx43\pm6\%$.

The spectroscopic control sample consists of stars in the apparent magnitude
range $I<20$, significantly brighter than $I_{\rm TRGB}$ at the distance of
M31 (Fig.~\ref{bi_cmd}).  {\it A~priori\/}, these stars are likely to be
Milky Way dwarf stars and form a good sample of objects with which to test
the IASG model predictions.  The observed radial velocity distribution of the
spectroscopic control sample is more or less consistent with the IASG model
prediction, although there appears to be a 20$\>$--$\>$50~km~s$^{-1}$ offset
between the two distributions with the former being more negative (solid
histogram vs.\ dotted curve in the lower panel of Fig.~\ref{vhist}).  A
similar offset between model and observed $v$ distributions also appears to
be present in the main M31 target sample (upper panel).

The Gaussian plus IASG model fit to the velocity distribution of the main
sample of M31 targets is used to estimate the fractional foreground
contamination within various radial velocity ranges.  For example, the
Galactic dwarf fraction in the sample is expected to be: 8\% for
$v<-220$~km~s$^{-1}$, 76\% for $v>-220$~km~s$^{-1}$, 12\% for
$v<-160$~km~s$^{-1}$, 84\% for $v>-160$~km~s$^{-1}$, and nearly 100\% for
stars with $v>-50$~km~s$^{-1}$.  The lack of any substantial concentration of
objects at $v_{\rm sys}^{\rm M31}$ in Fig.~\ref{vhist} indicates that there
are few (if any) M31 disk giants in the sample, but this issue will be
revisited later (\S\,\ref{poss_disk_sec}).  For a start, selecting the
25~stars with $v<-220$~km~s$^{-1}$ is likely to yield mostly M31 halo RGB
stars with minimal contamination from Milky Way dwarf stars (at the level of
only $\approx8$\% or 2~stars out of 25; see \S\,\ref{feh_comp_sec}).

Figure~\ref{bi_cmd} shows an $I$ vs.\ $B-I$ CMD for all color- and
morphology-selected stellar candidates in the $R=19$~kpc minor axis M31 field
based on the earlier photometric study by RGG.  The main sample of M31
spectroscopic targets and control sample (representative subset of the
stellar candidates in the range $20<I<22$ and $I<20$, respectively) are shown
as special symbols indicating each object's radial velocity.  It is
reassuring to find that the radial velocity information and the distribution
of objects on the CMD tell a consistent story.  As discussed above, the
control sample is a clear example of this: the photometric selection
criterion indicates these objects should be foreground Galactic dwarfs stars
and they all turn out to have $v>-160$~km~s$^{-1}$ in keeping with the IASG
model prediction for such stars.  The sample of 25~stars with
$v<-220$~km~s$^{-1}$ is another exmaple: these stars have large enough
negative velocities for them to be likely M31 red giants and indeed they
occupy a portion of the CMD that is bracketed by model RGB tracks at the
distance of M31 spanning a plausible range of metallicities.

A similarly consistent picture emerges for the foreground Galactic dwarf
stars present within the main spectroscopic target sample.  Over the apparent
magnitude range of interest ($I\sim18\>$--$\>22$), this dwarf population
contains stars with a variety of masses, absolute magnitudes, and effective
temperatures, with a broad and nearly uniform color distribution from
$1\lesssim{B-I}\lesssim4.5$: see the control sample with $I<20$ in
Fig.~\ref{bi_cmd} and the comparison field data and IASG model predictions in
RGG [Figs.~8(b--d) of that paper].  The IASG model predicts that most of the
objects with $v>-160$~km~s$^{-1}$ in the main spectroscopic sample are
Galactic dwarfs.  As expected, the stars in this radial velocity range are
found to span the full range of $B-I$ colors in the CMD (Fig.~\ref{bi_cmd})
and do not appear to be particularly strongly concentrated in the subsection
of the CMD bracketed by the M31 model RGB tracks.  Moreover, the $20<I<20.5$
portion of the CMD should be largely free of M31 giants since it is above the
nominal tip of the RGB; as expected, 10 out of the 12~stars with reliable
velocity measurements within this $I$ magnitude range of the main sample have
$v>-160$~km~s$^{-1}$ (Fig.~\ref{bi_cmd}).

The $B-I$ and $R-I$ color distributions are compared in Figs.~\ref{bi_cmd}
and \ref{ri_cmd}.  The foreground Galactic dwarf population spans the full
width of both CMDs with a near-uniform distribution.  The theoretical M31 RGB
tracks have somewhat different shapes and locations relative to this
foreground population in the two CMDs.  As expected, secure M31 giants with
$v<-220$~km~s$^{-1}$ are shifted further to the blue in ($R-I,~I$) than in
($B-I,~I$) relative to the rest of the data points.  It should be noted that
$R-I$ is not a good choice of color for metallicity discrimination: color
measurement errors are more prominent in Fig.~\ref{ri_cmd} than in
Fig.~\ref{bi_cmd}, so it is not surprising that $\sim20$\% of the secure M31
halo giants with $v<-220$~km~s$^{-1}$ scatter blueward of the models in the
former CMD.

The situation is somewhat more complicated for stars in the intermediate
radial velocity range, $-220<v<-160$~km~s$^{-1}$, and bears closer
examination.  Five of the 9~intermediate velocity stars have $B-I<2$ which
implies that they are too blue to be M31 red giants (Fig.~\ref{bi_cmd}).
Following our earlier discussion, foreground stars in this color range are
expected to be distant turnoff stars in the halo of the Milky Way, so it is
only natural for them to have a substantial negative velocity due to the
reflex of the Solar rotation vector projected in the direction of M31
($\approx-175$~km~s$^{-1}$).  Figure~\ref{vhist_mod} shows the IASG model
prediction for the radial velocity distribution of dwarfs with $B-I<2$ versus
those with $B-I>2$ (the relative scaling of the two curves is arbitrary).
The blue subset of Galactic dwarfs is expected to have the following velocity
distribution: 1~star with $v<-220$~km~s$^{-1}$, 2 with
$-220<v<-160$~km~s$^{-1}$, and 7 with $v>-160$~km~s$^{-1}$.  The observed
numbers of stars with $B-I<2$ in these velocity ranges are: 0, 5, and
4~stars, respectively (see Fig.~\ref{bi_cmd}), and are entirely consistent
with model predictions given the (large) Poisson uncertainties and
measurement errors in radial velocity and color.  Galactic dwarfs stars with
$B-I>2$ on the other hand are expected to lie almost exclusively in the
velocity range $v>-160$~km~s$^{-1}$ (solid curve in Fig.~\ref{vhist_mod}).

Thus, the 4~stars in the sample with $B-I>2$ and with measured radial
velocities in the range $-220<v<-160$~km~s$^{-1}$ (Fig.~\ref{bi_cmd}) are
almost certainly M31 giants; these stars are hereafter referred to as the
``red intermediate-velocity stars''.  These 4~stars are added to the 25~stars
with $v<-220$~km~s$^{-1}$ (selected on the basis of radial velocity alone) to
construct a {\it secure sample of 29~M31 red giants\/}.  Unless otherwise
mentioned, the rest of the analysis in this paper is restricted to this
secure M31 sample.  We recognize that even this `secure' M31 sample might
contain a couple of non-M31 stars; one such star is discussed in
\S\,\ref{feh_comp_sec}.  The secure sample probably contains about 80\% of
all the M31 red giants in the spectroscopic sample: of the 80~stars with
reliable radial velocity measurements, 35 or so are expected to be M31 RGB
stars (based on the IASG model fit to the velocity distribution), and of
these about 27 or 28 are included in the secure sample.  The remaining 7 or
8 M31 giants outside the secure sample are discussed in
\S\,\ref{feh_comp_sec}.

\subsection{Halo Dynamics}\label{halo_dyn}

Individual field stars in M31's halo serve as excellent dynamical tracers of
the mass distribution in the galaxy.  The IASG model plus Gaussian fit to the
observed radial velocity distribution of the main M31 spectroscopic sample
yields a best-fit Gaussian width of $\sigma_{v}^{\rm
M31}=150_{-30}^{+50}$~km~s$^{-1}$ (90\% confidence limits).  The velocity
dispersion of M31's halo can be determined more directly from the data, as
long as one takes care to avoid the foreground Galactic dwarf contaminants.
The 16~stars with $v<v_{\rm sys}^{\rm M31}$ are all likely to be M31 giants
according to the IASG model.  The rms dispersion of these stars with respect
to $v_{\rm sys}^{\rm M31}$ is 125~km~s$^{-1}$, in good agreement with the
best-fit Gaussian width.

The centroid of the Gaussian is held fixed during the fit at $v_{\rm
sys}^{\rm M31}=-297$~km~s$^{-1}$, the expected mean velocity of the halo in
this minor axis field.  As such, the data show no evidence for an offset
between the mean velocity of M31 stars and $v_{\rm sys}^{\rm
M31}$---i.e.,~there is no sign of rotation along the minor axis.  It is
difficult, however, to pin down an exact upper limit for the minor axis
rotation of M31's halo from these data because of the high degree of Galactic
dwarf contamination all of which appears on one side of $v_{\rm sys}^{\rm
M31}$.

Our measurement of the line-of-sight velocity dispersion of M31 field halo
giants in this $R=19$~kpc field is in excellent agreement with other
dynamical tracers located at a comparable distance from the galaxy's center
(cf.~Evans \& Wilkinson 2000, and references therein).  The published radial
velocities of 17~M31 globular clusters with $R>20$~kpc yield a line-of-sight
velocity dispersion of 158~km~s$^{-1}$ at an average projected distance from
M31's center of 23.8~kpc.  The published velocities of 9~planetary nebulae
with $R>18$~kpc in M31's halo yield a dispersion of 145~km~s$^{-1}$ at an
average/median projected distance of $\langle{R}\rangle\sim20$~kpc.

The above estimates for the velocity dispersion of M31's halo are somewhat
low compared to the dispersion observed for various Milky Way halo
populations, and this may appear surprising at first.  The reader is referred
to some recent work on the dynamical modeling of the Milky Way and M31 by
Wilkinson \& Evans (1999), Evans \& Wilkinson (2000), and Evans et~al.\
(2000) that used the latest and most complete information available on the
H\,{\smcap i} rotation curve of the disk and radial velocities of dwarf
satellite galaxies, globular clusters, and planetary nebulae.  They conclude
that, contrary to popular belief/intuition, M31 is probably {\it less\/}
massive than the Milky Way.

\section{Stellar Metallicity Measurements}\label{feh_sec}

This section describes the measurement of the metallicities of individual M31
RGB stars using two different methods, one based on photometric data and the
other based primarily on spectroscopic data.  The photometric technique
compares the position of each star in the ($B-I,~I$) CMD to a set of
theoretical RGB tracks covering a wide range of metallicities.  The
spectroscopic technique is based on the weighted sum of the equivalent widths
of the Ca\,{\smcap ii} triplet of absorption lines, which is converted to
[Fe/H] via two separate empirical calibration relations derived from Galactic
globular cluster red giants.  Estimates are made of the measurement
uncertainties (both random and systematic) in the photometric and
spectroscopic methods.

Both methods implicitly {\it assume\/} that the object whose metallicity is
being measured is a red giant at the distance of M31.  It has already been
noted though that more than half of the stars in the full sample of M31
targets are probably foreground Galactic dwarf stars (\S\,\ref{secure}), so
the techniques, especially the photometric one, will yield nonsensical
results for these stars.  While the 29~stars comprising the secure M31 sample
are almost certainly red giants in M31, we expect a handful of additional M31
red giants to be hidden in the rest of the sample.  Even with the photometric
and radial velocity information we have in hand, it is generally difficult,
if not impossible, to tell whether a {\it particular\/} star is an M31 giant
or a Galactic dwarf.  We have therefore decided to blindly apply the
photometric technique to all 99~stars in the main M31 spectroscopic target
sample and the spectroscopic technique to the subset (80 out of~99) for which
there is a reliable measurement of the Ca\,{\smcap ii} lines.  Metallicity
measurements are not made for the 13~stars comprising the control sample as
they are definitely Galactic dwarfs.

\subsection{Photometric Estimates}\label{feh_phot_sec}
\subsubsection{Method}\label{feh_phot_det_sec}

The apparent/observed $I$ magnitude and $B-I$ color of each star are
corrected for extinction and reddening, respectively, using the Schlegel,
Finkbeiner,~\& Davis (1998) $E(B-V)$ map and the relation:
$E(B-I)=2.4\>E(B-V)$, which is based on the standard Galactic dust extinction
law with $R_V=3.1$ (Cardelli, Clayton, \& Mathis 1989).  The
extinction-corrected apparent $I$ magnitude is converted to an absolute
magnitude, $M_I$, based on our adopted M31 distance of 783~kpc (Stanek \&
Garnavich 1998; Holland 1998) or a true distance modulus of $(m-M)_0=24.47$.
The absolute $I$-band magnitude and $B-I$ color are then compared to a set of
theoretical RGB tracks in the CMD: 17~model isochrones from Bergbusch \&
VandenBerg (2001) with metallicities spanning the range
$\rm-2.3\leq[Fe/H]\leq-0.3$~dex, $\rm[\alpha/Fe]=+0.3$~dex, and age
$t=14$~Gyr (four of which are shown in Figs.~\ref{bi_cmd} and \ref{ri_cmd}),
along with two model isochrones from the Padova group \cite{padova} with
$\rm[Fe/H]=-0.02$ and +0.18~dex and $t=14.3$~Gyr constraining the
high-metallicity end.  A fourth-order spline is used to interpolate between
isochrones and this yields a photometric estimate of the star's metallicity,
$\rm[Fe/H]_{phot}$.  These estimates are listed in the last column of
Table~\ref{spec_tbl}.  For stars located to the left of (bluer than) the most
metal-poor isochrone in the ($B-I,~I$) CMD shown in Fig.~\ref{bi_cmd}, the
$\rm[Fe/H]_{phot}$ estimate should be regarded as highly uncertain as it is
derived by extrapolation; only one of the stars with $v<-220$~km~s$^{-1}$ in
the secure sample of M31 giants, object ID t3.12, is in this category, and we
conclude in \S\,\ref{feh_comp_sec} that this object is possibly a foreground
Galactic dwarf.

We have chosen to apply reddening/extinction corrections on a star-by-star
basis: the angular resolution of the Schlegel et~al.\ maps is high enough
($\sim2'$) to account for variations in the dust optical depth across the
$15'\times15'$ region of the sky from which the stars are drawn:
$0.12<E(B-I)<0.14$.  For the sake of illustration only, the four~model RGB
tracks shown in Figs.~\ref{bi_cmd} and \ref{ri_cmd} have been reddened by a
typical amount of $\langle{E(B-I)}\rangle=0.13$ while the data points
represent the raw, uncorrected measurements.  The $\rm[Fe/H]_{phot}$
measurement procedure is exactly the opposite of this though: it involves
applying a reddening/extinction correction to the photometry of each star and
comparing it to the original, unreddened model isochrones.

While the $\rm[Fe/H]_{phot}$ measurements in this paper are based largely on
the Bergbusch \& VandenBerg (2001) isochrones, it should be noted that the
older Padova isochrones \cite{Bert} were used for the definition of the
$UBRI$ color-based selection regions in the RGG work.  The former set of
isochrones are favored here as they are computed over a finer grid of
metallicities.  The two sets of isochrones are in very good agreement in
color-color space, so there is no inconsistency between photometric selection
of stellar candidates and photometric measurement of their metallicity.

\subsubsection{Measurement Errors}\label{feh_phot_uncert_sec}

The photometric errors (RGG) and uncertainty in the distance modulus are used
to compute an error ellipse for each star in the ($B-I,~I$) CMD.  The major
axis of the error ellipse is tilted in general because the photometric errors
in $B-I$ and $I$ are correlated.  The distance modulus error is arbitrarily
assumed to be 0.1~mag and is added in quadrature to the $y$-axis alone.  The
resulting photometric-cum-distance error ellipse for a star represents its
probability distribution in ($B-I,~I$) space; this is used as a weighting
function to calculate the weighted average and rms in $\rm[Fe/H]_{phot}$ over
the area of the error ellipse within $5\sigma$ of a star's location in the
CMD.  The difference between the weighted average $\rm[Fe/H]_{phot}$ and the
value derived from the nominal CMD location (most probable $I$ and $B-I$) of
a star is typically only $\approx0.03$~dex; the weighted average is what is
used throughout the rest of this paper.

The weighted rms represents the error in $\rm[Fe/H]_{phot}$ due to errors in
photometry and distance modulus.  It does {\it not\/} take into account
possible errors in the theoretical RGB tracks---e.g.,~any mismatch between
the $UBRI$ photometric system adopted in the model computation versus that
used in the RGG stellar photometry.  The average error in $\rm[Fe/H]_{phot}$
for 28 of the 29~secure M31 RGB stars within the range of the model
isochrones is 0.19~dex.  Extrapolation of the spline fit to the isochrones
suggests that the error is nearly twice as large, 0.36~dex, for objects that
lie just blueward of the calibrated range due to the fact that the
lowest-metallicity isochrones tend to crowd together in the CMD so that a
given color error corresponds to a larger error in metallicity.  The error in
$\rm[Fe/H]_{phot}$ is not easily quantifiable for objects that lie well
beyond the range of the isochrones.

\subsection{Spectroscopic Estimates}\label{feh_spec_sec}
\subsubsection{Outline of the Method}\label{feh_spec_intro_sec}

The Ca\,{\smcap ii} triplet of absorption lines, $\lambda\lambda8498$, 8542,
and 8662\,\AA, are among the strongest features in the near-infrared portion
of the spectrum for most late-type stars and have been widely utilized.
Early studies took advantage of the dependence of the equivalent width of the
Ca\,{\smcap ii} lines on surface gravity, log~$g$, to discriminate between
giant and dwarf stars (Spinrad \& Taylor 1969, 1971; Anderson 1974).
Subsequently, Diaz, Terlevich,~\& Terlevich (1989) used the Ca\,{\smcap ii}
lines to measure metallicity, as most of the variation in line strength is
due to a linear combination of log~$g$ and [Fe/H], with little dependence on
$T_{\rm eff}$.  Since then many groups have used this dependence and found
good correlations with other metallicity indicators (Armandroff \& Zinn 1988;
Da~Costa \& Seitzer 1989; Olszewski et~al.\ 1991; Suntzeff et~al.\ 1992).
Armandroff \& Da~Costa (1991) developed a method to rank red giant stars by
[Fe/H] which is independent of distance and reddening.  We have chosen to
adopt this method as have many others (Armandroff, Da~Costa,~\& Zinn 1992;
Da~Costa, Armandroff,~\& Norris 1992; Suntzeff et~al.\ 1993; Geisler et~al.\
1995; Da~Costa \& Armandroff 1995; Suntzeff \& Kraft 1996; Rutledge et~al.\
1997a; Rutledge, Hesser,~\& Stetson 1997b).

While the details of the method are described in Armandroff~\& Da~Costa
(1991), there have been subsequent improvements to the calibration relations
\cite{Rutledge97a}.  The basis of the method lies in the fact that the
combined equivalent width (strength) of the Ca\,{\smcap ii} lines depends
primarily on metallicity, but also has some dependence on the luminosity of
the red giant (relative to the horizontal branch say: $V_{\rm HB}-V$).  This
is due to the fact that both log~$g$ and $T_{\rm eff}$ decrease as one moves
up the RGB at a fixed metallicity.  The combined equivalent width of the
Ca\,{\smcap ii} triplet is defined to be the weighted sum of the equivalent
widths of the three individual lines \cite{Rutledge97a}:
\begin{equation}
\rm\Sigma{Ca}\,\equiv\,0.5\>EW(\lambda8498\AA)\,+\,1.0\>EW(\lambda8542\>\AA)
\,+\,0.6\>EW(\lambda8662\>\AA)
\label{sigmaca_eqn}
\end{equation}
\noindent
The observed equivalent width, $\rm\Sigma{Ca}$, is translated to a `reduced'
equivalent width, $W^\prime$, by correcting it to a common value of log~$g$
\cite{Rutledge97b}:
\begin{equation}
W^\prime\,=\,\Sigma{\rm Ca}\,-\,0.64(V_{\rm HB}-V)~~~ \rm\AA
\label{wprime_eqn}
\end{equation}
\noindent
The reduced equivalent width is converted to a metallicity estimate using two
separate calibration relations \cite{Rutledge97b}.  The first is a cubic
relation:
\begin{equation}
{\rm[Fe/H]_{ZW}}\,=\,-3.005\,+\,0.941\>W^\prime\,-\,0.312\>(W^\prime)^2\,+\,
0.0478\>(W^\prime)^3~~~ {\rm dex}
\label{fehzw_eqn}
\end{equation}
\noindent
for the Zinn~\& West (1984, hereafter referred to as Zinn-West or simply ZW)
metallicity scale and is calibrated over the range
$\rm-2.3\leq[Fe/H]_{ZW}\leq-0.3$~dex.  The second is a linear relation:
\begin{equation}
{\rm[Fe/H]_{CG97}}\,=\,-2.66\,+\,0.42\>W^\prime~~~ {\rm dex}
\label{fehcg_eqn}
\end{equation}
\noindent
for the metallicity scale based on the high-dispersion spectroscopic studies
of Carretta~\& Gratton (1997, hereafter referred to as Carretta-Gratton or
simply CG97), and is calibrated over the range
$\rm-2.2\leq[Fe/H]_{CG97}\leq-0.6$~dex.

\subsubsection{From Cross Correlation to Metallicity}\label{feh_spec_ccf_sec}

The strength of the CCF peak is an excellent diagnostic of the strength of
the Ca\,{\smcap ii} absorption line triplet.  Tests using simulated, noisy
spectra show that: (1)~the area under the CCF peak, $A_{\rm CCF}$, is a more
robust measure than the peak height $h$ since the width of the CCF peak can
vary from exposure to exposure due to variations in seeing/guiding errors and
even from star to star on a given exposure due to focus variations; and
(2)~$A_{\rm CCF}$ is less affected by noise than direct measurements of the
equivalent width since the former is essentially based on a matched-filter
technique which optimizes the S/N ratio of the measurement.  We have
therefore opted to estimate the line strength $\rm\Sigma{Ca}$ by appropriate
scaling of $A_{\rm CCF}$.

The CCF area is measured within a 450~km~s$^{-1}$-wide (13\,\AA-wide) window
centered on the peak after subtracting off a linear baseline that is fit to
regions just outside this window on either side of it.  The equivalent widths
of the three Ca\,{\smcap ii} lines are also measured directly from the
spectrum, using three 350~km~s$^{-1}$-wide (10\,\AA-wide) windows centered on
the three lines each flanked by a pair of windows of the same width for
measurement of the local continuum level.  The three measurements are
combined into a weighted sum, $\rm\Sigma{Ca}$, following
Eq.~(\ref{sigmaca_eqn}).  Figure~\ref{ew_vs_ccfarea} shows a plot of $A_{\rm
CCF}$ versus $\rm\Sigma{Ca}$ for all 93~objects for which there is a reliable
peak in the CCF (80 out of 99~objects in the main M31 target sample and all
13 in the control sample).  A straight line through the origin is fit to
these data; the best-fit slope is the scale factor needed to convert $A_{\rm
CCF}$ into the effective combined equivalent width (in \AA\ units):
$\Sigma{\rm Ca}_{\rm CCF}=0.0915{A}_{\rm CCF}$.\footnote{The specific value
of the slope is tied of course to the rather arbitrary units in which $A_{\rm
CCF}$ in measured, which in turn depends on the widths of the object/template
spectral bins ($\rm0.62\AA$) and velocity bins (10~km~s$^{-1}$) used in the
CCF computation.  The $A_{\rm CCF}$ values displayed in Table~3 and
Fig.~\ref{ew_vs_ccfarea} can be converted to units of \AA~km~s$^{-1}$ by
multiplying by the product of the bin widths, 6.2\,\AA~km~s$^{-1}$; naturally
the slope would then have to be reduced by a corresponding factor to
$0.0915/6.2=0.0148$.}

The correction from effective equivalent width to reduced equivalent width
requires knowledge of the $V$-band luminosity of the star relative to the
horizontal branch.  The RGG photometric dataset covers $UBRI$ only, so it is
necessary to synthesize a $V$-band magnitude from these data.  Stars display
a tight relationship between $B-I$ and $V-I$ across a wide range of
metallicities and ages: this is illustrated in Fig.~\ref{bvi_iso} for
Bergbusch \& VandenBerg (2001) model stellar isochrones in the range
$\rm-2.3\leq[Fe/H]\leq-0.3$~dex and $8\leq{t}\leq18$~Gyr.  This relation and
the extinction-corrected $B$ and $I$ magnitudes of a star are used to derive
$V_{\rm synth}$.  A horizontal branch magnitude of $V_{\rm HB}=25.17$ is
adopted from the Holland et~al\. (1996) {\it HST\/} study of M31 halo stars
in the G312 field; the KPNO data used in the RGG work are not deep enough to
detect, let alone measure, the horizontal branch.

Figure~\ref{w_vs_vhb} shows the combined equivalent width derived from the
CCF, $\rm\Sigma{Ca}_{CCF}$, as a function of the $V$ luminosity relative to
the horizontal branch for the 25~stars in the secure M31 sample with
$v<-220$~km~s$^{-1}$.  The tilted dashed lines are loci of constant
metallicity for the different calibration relations.  The quantities
$\rm\Sigma{Ca}_{CCF}$ and $V_{\rm synth}$ are used to compute the reduced
equivalent width, $W^\prime$, via Eq.~(\ref{wprime_eqn}) for the 80~stars in
the main M31 spectroscopic target sample for which there is a reliable
measurement of the Ca\,{\smcap ii} lines.  The quantity $W^\prime$ is then
used in Eqs.~(\ref{fehzw_eqn} and \ref{fehcg_eqn}) to calculate the
metallicities on the Zinn-West and Carretta-Gratton scales, $\rm[Fe/H]_{ZW}$
and $\rm[Fe/H]_{CG97}$, respectively.  These metallicity estimates, along
with the different measures of the Ca\,{\smcap ii} line strength, $A_{\rm
CCF}$, $\rm\Sigma{Ca}_{CCF}$, and $W^\prime$, are listed in
Table~\ref{spec_tbl}.

\subsubsection{Measurement Errors}\label{feh_spec_uncert_sec}

As explained in \S\,\ref{feh_spec_ccf_sec}, the directly measured equivalent
width $\rm\Sigma{Ca}$ is likely to be a noisier statistic than the CCF area
$A_{\rm CCF}$, so that the scatter in Fig.~\ref{ew_vs_ccfarea} is probably
dominated by the former.  Nevertheless, we make the conservative assumption
that {\it all\/} of the scatter in the plot is caused by measurement errors
in $A_{\rm CCF}$.  On the other hand, the errors in $A_{\rm CCF}$ and
$\rm\Sigma{Ca}$ are undoubtedly correlated---a noise feature near a
Ca\,{\smcap ii} absorption line will cause excursions of the same sign (even
if not exactly the same magnitude) in both quantities---and this causes the
scatter in Fig.~\ref{ew_vs_ccfarea} to be smaller than the true noise.  The
rms scatter in $A_{\rm CCF}$ is~5.57, which corresponds to an rms error of
$\rm5.57\times0.0915=0.51\,\AA$ in the effective combined equivalent width
$\rm\Sigma{Ca}_{CCF}$.  The $1\sigma$ error in $V_{\rm synth}$ is
conservatively assumed to be the quadrature sum of $B-I$ color error (RGG)
and 0.1~mag, the latter to account for possible systematic errors in the
model isochrones used for the interpolation.

Propagating the $\rm\Sigma{Ca}_{CCF}$ and $V_{\rm synth}$ errors for a
typical star through the various calibration relations yields a $1\sigma$
error of about 0.05~dex in the metallicity within the calibrated ranges of
the ZW and CG97 scales.  This is roughly consistent with the metallicity
error estimates of Rutledge et~al.\ (1997b), which must be somewhat
fortuitous given the obvious oversimplifications (leading to
over-/under-estimates) in our input parameter error analysis.  The rms error
of 0.05~dex should be treated as the accuracy with which the stars can be
assigned relative ranks in metallicity.  The ability to determine a ``true''
[Fe/H] on either metallicity scale is probably uncertain by at least 0.2~dex
as a result of systematic errors---e.g.,~departure of the $\rm[\alpha/Fe]$
ratio from the assumed value of $+0.3$~dex.  We therefore add in quadrature
to the random error in [Fe/H] a systematic error component of 0.2~dex.

An external check of the [Fe/H] calibration is carried out using Keck/LRIS
observations of RGB stars in two~Galactic star clusters of known metallicity:
M79 ($\rm[Fe/H]_{ZW}=-1.68$ and $\rm[Fe/H]_{CG97}=-1.37$~dex---Carretta~\&
Gratton 1997) and NGC~6791 ($\rm[Fe/H]_{ZW}=+0.19$~dex---Friel~\& Janes
1993).  Figure~\ref{w_vs_vhb_calib} is a plot of $\Sigma{\rm Ca}$ versus
$V-V_{\rm HB}$ for these stars.  The stars in both clusters appear to follow
the slope of the constant metallicity loci thereby confirming the validity of
the 0.64 factor in the $W^\prime$ formula [Eq.~(\ref{wprime_eqn})].  On
average, the data for M79 giants agree with both ZW and CG97 calibration
relations to within 0.05~dex or better.  The metallicity of NGC~6791 is
beyond the upper end of the range over which the ZW calibration relation is
defined and the cubic formula diverges here: the value of $\rm[Fe/H]_{ZW}$
derived from our NGC~6791 data is nearly 2~dex higher than its standard
published value.  Our empirical estimate of $\rm[Fe/H]_{CG97}$ for NGC~6791
is 0.3~dex lower than its standard ZW value, although this comparison ignores
any systematic offset between the CG97 and ZW scales.  In summary, the
spectroscopic metallicity estimates appear to be generally uncertain in the
vicinity of solar values, particularly for the ZW and empirical ``phot+spec''
calibrations (\S\,\ref{feh_comp_sec}).

There is an independent indication that the uncertainty in the spectroscopic
[Fe/H] estimates might be greatest at the high metallicity end.  The scatter
in the $A_{\rm CCF}$ versus $\rm\Sigma{Ca}$ plot appears to be highest at the
upper end (Fig.~\ref{ew_vs_ccfarea}).  This could be due to the fact that the
highest-metallicity M31 giants are, on average, fainter and consequently have
lower S/N spectra than their lower-metallicity counterparts (because of the
drop in the tip of the RGB magnitude with increasing [Fe/H]; see
\S\,\ref{selbias}).  It could also be due to anomalies (say in the absorption
line shapes) of the four~M31 giants with the strongest Ca\,{\smcap ii} lines.

\subsection{Comparing Photometric and Spectroscopic
Estimates}\label{feh_comp_sec}

In this section, we compare the above spectroscopic metallicity estimates to
the photometric estimate on a star-by-star basis.  The spectroscopic and
photometric metallicity determination methods and their associated systematic
errors are different from/independent of each other, so that a comparison
like this should help set rough upper limits on the ``external'' errors in
our metallicity estimates.  Moreover, this comparison helps shed some more
light on the question of which stars in the sample are likely to be members
of M31 and which ones are not.

Figure~\ref{fehcomp_mem} shows $\rm[Fe/H]_{CG97}$ versus $\rm[Fe/H]_{phot}$
and $\rm[Fe/H]_{ZW}$ versus $\rm[Fe/H]_{phot}$ (upper and lower panels,
respectively) for the 25~stars in the secure M31 sample with radial
velocities $v<-220$~km~s$^{-1}$.  In broad terms, there is a reasonably
good correlation between spectroscopic and photometric [Fe/H] estimates
within the calibrated ranges of the measurement methods.  A more detailed
comparison shows that, within the calibrated range, the $\rm[Fe/H]_{CG97}$
and $\rm[Fe/H]_{ZW}$ estimates are about 0.2 and 0.5~dex (respectively) more
metal-poor on average than the photometric estimate.  If M31 halo giants
happen to have solar abundance ratios, instead of the assumed
$\rm[\alpha/Fe]=+0.3$~dex enhancement level, this would bring the
spectroscopic and photometric metallicity estimates into closer agreement
(\S\,\ref{halo_comp_sec}).

The rms difference between $\rm[Fe/H]_{CG97}$ and $\rm[Fe/H]_{phot}$ is
0.38~dex within the calibrated range and 0.76~dex outside the range, while
the corresponding $\rm[Fe/H]_{ZW}$ versus $\rm[Fe/H]_{phot}$ rms differences
are 0.55 and 3.52~dex, respectively.  The divergence of $\rm[Fe/H]_{ZW}$ for
objects with strong Ca\,{\smcap ii} lines is due to the extrapolation of the
cubic formula; the relative metallicity ranking for objects beyond the
calibrated range should be correct, but the actual values of $\rm[Fe/H]_{ZW}$
are probably far from the true metallicities.  There are other possible
reasons for the rms being high outside the calibrated range: the four~stars
with the strongest Ca\,{\smcap ii} lines may have atypical properties
(\S\,\ref{poss_disk_sec}), while the bluest star (lowest $\rm[Fe/H]_{phot}$
value) may be a foreground Galactic dwarf star (see below).

It is tempting to think of the above rms differences as the quadrature sum of
the errors in $\rm[Fe/H]_{phot}$ and $\rm[Fe/H]_{CG97}$ or $\rm[Fe/H]_{ZW}$,
but there are a couple of complicating factors.  Firstly, any residual
foreground contamination in the sample would cause the errors in [Fe/H] to be
overestimated.  Secondly, the errors in the photometric and spectroscopic
[Fe/H] estimates are correlated in principle: the same $B$ and $I$ magnitudes
from which $\rm[Fe/H]_{phot}$ is determined are also used to compute $V_{\rm
synth}$, which is used to make a surface gravity correction to the
Ca\,{\smcap ii} line strength in the spectroscopic [Fe/H] measurement scheme.
If the correlation between photometric and spectroscopic [Fe/H] errors is
positive, the rms difference would tend to underestimate the errors, and vice
versa if the correlation is negative.  The observed lack of correlation
between photometric and spectroscopic [Fe/H] estimates for objects with
$v>-220$~km~s$^{-1}$ (see below) indicates that any correlation between the
errors, positive or negative, is probably quite small.

In contrast to the $v<-220$~km~s$^{-1}$ sample, there is no obvious overall
correlation between $\rm[Fe/H]_{CG97}$ or $\rm[Fe/H]_{ZW}$ and
$\rm[Fe/H]_{phot}$ for objects with $v>-220$~km~s$^{-1}$
(Fig.~\ref{fehcomp_nmem}).  Instead, most lie in a band of near-constant line
strength (which gets mapped to $\rm[Fe/H]_{CG97}\sim[Fe/H]_{ZW}\sim-2$~dex)
despite their wide range of $B-I$ colors.  The lack of correlation is only to
be expected for Milky Way dwarf stars: both the photometric and spectroscopic
metallicity determination methods are based on the premise that the object
being measured is a red giant at the distance of M31, so [Fe/H] estimates for
foreground objects are totally incorrect and in different ways for the two
methods.  There could of course be a small fraction of M31 red giant stars
hidden in Fig.~\ref{fehcomp_nmem}; in fact, we {\it expect\/} about 11~such
stars to be present.  Firstly, the four bold circles---red subset of stars in
the intermediate velocity range, $-220<v<-160$~km~s$^{-1}$---belong to the
secure M31 red giant sample and they indeed lie relatively close to the $x=y$
line in both panels.  Secondly, the IASG model plus Gaussian fit to the
velocity distribution predicts that about 16\% of the 46~stars in the main
sample with $v>-160$~km~s$^{-1}$, or about 7~stars, are M31 giants
(\S\,\ref{secure}).  The horizontal band formed by the small circles in
Fig.~\ref{fehcomp_nmem} inevitably crosses the $x=y$ diagonal line, but there
are a handful of stars even outside this area of intersection for which the
spectroscopic and photometric [Fe/H] estimates are in rough agreement and
these stars are probably M31 red giants---e.g.,~the five or so stars mixed in
with the bold circles and possibly the star in the lower left corner of the
upper panel with $\rm[Fe/H]_{CG97}\approx[Fe/H]_{phot}\approx-3$~dex.

Conversely, one or two stars in the velocity range $v<-220$~km~s$^{-1}$ are
expected to be Milky Way dwarf stars in the foreground of M31.  Following the
reasoning given in \S\,\ref{secure}, the most likely area of foreground
contamination in this velocity range is at the blue end of the color
distribution.  The star with the bluest color ($B-I=2.2$) and lowest
$\rm[Fe/H]_{phot}$ in Fig.~\ref{fehcomp_mem}, object ID t3.12 in
Tables~\ref{phot_tbl} and \ref{spec_tbl}, is likely to be a star near the
main sequence turnoff in the Galactic halo.  The IASG model prediction for
the velocity distribution of blue stars has a tail toward negative
velocities, extending past the radial velocity of the star in question,
$v=-306$~km~s$^{-1}$ (Fig.~\ref{vhist_mod}).  Moreover, neither
$\rm[Fe/H]_{CG97}$ nor $\rm[Fe/H]_{ZW}$ is close to $\rm[Fe/H]_{phot}$ for
this star, the difference being nearly 1~dex; instead, the spectroscopic
[Fe/H] estimates are within the horizontal band at $\approx-2$~dex occupied
by most Milky Way dwarfs in Fig.~\ref{fehcomp_nmem}.

It is instructive to directly check the degree of correlation between the
reduced Ca\,{\smcap ii} line strength and the photometric metallicity
estimate derived from the ($B-I$,~$I$) CMD.  For the 25~stars in the M31
secure sample with $v<-220$~km~s$^{-1}$, an empirical linear relation of the
form:
\begin{equation}
{\rm[Fe/H]_{phot+spec}}\,=\,-2.13\,+\,0.25\>W^\prime~~~ {\rm dex}
\label{fehphotspec_eqn}
\end{equation}
\noindent
produces the optimal mapping from $W^\prime$ to $\rm[Fe/H]_{phot}$.  The
metallicity derived using this empirical calibration relation, hereafter
referred to as $\rm[Fe/H]_{phot+spec}$, is plotted versus $\rm[Fe/H]_{phot}$
in Fig.~\ref{feh_photspec}.  The rms scatter in this plot is 0.35~dex within
the calibrated range and 0.61~dex outside it.  In fact, a linear mapping
results in a somewhat lower $\chi^2$ value than a quadratic or cubic mapping
of $W^\prime\rightarrow\rm[Fe/H]$.  It should come as no surprise that the
correlation between $\rm[Fe/H]_{phot+spec}$ and $\rm[Fe/H]_{phot}$ is tighter
than the correlation between either $\rm[Fe/H]_{CG97}$ or $\rm[Fe/H]_{ZW}$
and $\rm[Fe/H]_{phot}$.  The empirical ``phot+spec'' calibration relation is
designed to optimize this correlation, while the ZW and CG97 [Fe/H] estimates
are based on calibration formulae that have been derived from {\it
independent\/} data sets.

We emphasize that $\rm[Fe/H]_{phot+spec}$ should {\it not\/} be treated as a
true metallicity scale, or as an alternative to the CG97 or ZW scales; it is
merely designed to highlight the agreement between photometric and
spectroscopic [Fe/H] estimates.  The constant metallicity lines for the
empirical ``phot+spec'' calibration in the right panels of
Figs.~\ref{w_vs_vhb} and \ref{w_vs_vhb_calib} are evenly spaced just as they
are for the linear CG97 relation (middle panels), but the former set of lines
have a much wider spacing reflecting the smaller coefficient of the linear
term in the calibration relation [Eq.~(\ref{fehphotspec_eqn})].  The
``phot+spec'' relation is clearly inaccurate at the high metallicity end: it
underpredicts the metallicity of the slightly super-solar calibration cluster
NGC~6791 by 0.8~dex (right panel of Fig.~\ref{w_vs_vhb_calib}).  This end of
the empirical relation is largely constrained by the four~stars with the
highest spectroscopic metallicities and it is possible that their [Ca/Fe]
abundance ratios are abnormally high (\S\,\ref{poss_disk_sec}).  If these
four~stars were to be excluded from the analysis, the linear coefficient in
Eq.~(\ref{fehphotspec_eqn}) would nearly double and this would bring the
empirical calibration relation more or less in line with the CG97 relation.

Figure~\ref{feh_diff} shows the difference between the photometric and
spectroscopic [Fe/H] estimates (CG97 and ZW) as a function of mean
metallicity for the 29~stars comprising the secure M31 red giant sample.  The
mean metallicity, $\rm\langle[Fe/H]\rangle$, is simply the average of
$\rm[Fe/H]_{CG97}$, $\rm[Fe/H]_{ZW}$, and $\rm[Fe/H]_{phot}$; this is not a
particularly accurate measure of metallicity and is only used here as a
common $x$-axis scale to facilitate comparisons between the two panels.  For
stars with $\rm\langle[Fe/H]\rangle$ in the sub-solar range, the
spectroscopic metallicity estimates are slightly lower on average than the
photometric estimate, by 0.2 and 0.5~dex for the CG97 and ZW scales,
respectively.  For the four~stars with solar or super-solar
$\rm\langle[Fe/H]\rangle$ values (\S\,\ref{poss_disk_sec}),
$\rm[Fe/H]_{CG97}$ is 0.75~dex higher on average than $\rm[Fe/H]_{phot}$,
while the cubic ZW relation diverges at the high Ca\,{\smcap ii} line
strengths that characterize these stars.

\section{Discussion}\label{discuss_sec}
\subsection{Observed Metallicity Distribution of M31's Field Halo
Stars}\label{halo_feh_sec}

The secure sample of 29~M31 red giants is practically free of foreground
contamination so it is used to investigate the metallicity distribution of
M31's stellar halo.  We should emphasize that all of the discussion in this
section is based on the {\it observed\/} [Fe/H] distribution.  Our sample
selection efficiency is {\it not\/} uniform with metallicity; thus, the
mean/median and rms estimates of the [Fe/H] distribution of M31's stellar
halo are revised once the selection efficiency is properly accounted for
(next section).

The observed [Fe/H] distributions for the secure M31 sample are shown in the
three~upper panels of Fig.~\ref{feh_hist} for the ZW and CG97 spectroscopic
calibrations and photometric estimate.  All three~distributions show signs of
bimodality: note the two peaks centered at $\rm[Fe/H]_{phot}=-1.0$ and
$-1.8$~dex.  This is roughly similar to the peaks at roughly $-0.8$ and
$-1.4$~dex found by Durrell et~al.\ (1999, 2001) in their wide-field
photometric study.  Each of the three [Fe/H] distributions displays an
overall spread of $\gtrsim2$~dex.  The sample mean and median for each of the
[Fe/H] calibrations are listed in Table~\ref{table4} and are in the range
$-1.0$ to $-1.4$ and $-1.3$ to $-1.7$~dex, respectively.  Removal of the
four~red intermediate-velocity stars from the sample does not change these
numbers appreciably.

\begin{table}
\dummytable\label{table4}
\end{table}

While the three~estimates of the [Fe/H] distribution of M31 stellar halo are
broadly similar, there are some differences in detail, especially near the
extremes of the distributions.  For example, there are four~stars with
particularly strong Ca\,{\smcap ii} absorption lines that lie beyond the
upper end of the calibrated range of the spectroscopic scales, but there is
no corresponding high-metallicity tail in the $\rm[Fe/H]_{phot}$
distribution.  The $\rm[Fe/H]_{ZW}$ estimates for these four stars are
spuriously high and this causes the mean metallicity of the secure sample to
be slightly higher on the ZW scale than on the CG97 and photometric scales.
The upward extrapolation of cubic ZW relation is not to be trusted and it may
be preferable to instead rely on the photometric and CG97 estimates for solar
or slightly super-solar metallicity giants.  It should be cautioned though
that the agreement between the CG97 and photometric scales is far from
perfect, especially at these high metallicities.  The median [Fe/H] is
insensitive to outliers and therefore robust; it is lowest for the ZW scale
and highest for the photometric scale.

The rms spread of the [Fe/H] distributions is boosted by these outlying
stars, most strongly in the case of the ZW calibration (Table~\ref{table4}).
We argue in \S\,\ref{poss_disk_sec} that the four strong-lined stars are
possibly members of the M31 disk, not its halo.  Thus, these stars are
excluded in order to obtain a conservative lower limit to the rms spread in
[Fe/H] in M31's halo: 0.5$\>$--$\>$0.6~dex.  This is substantially larger
than the (internal and external) measurement uncertainties in [Fe/H]
(\S\S\,\ref{feh_phot_uncert_sec}, \ref{feh_spec_uncert_sec}, and
\ref{feh_comp_sec}) and thus represents the intrinsic spread in metallicity
among M31 halo stars.

\subsection{Selection Efficiency Function: Bias Against Metal-Rich
Giants}\label{selbias}

In this section, we quantify the bias against high-metallicity red giants
that is introduced by the apparent magnitude cut ($I<22$) imposed during the
selection of spectroscopic targets.  Such a bias occurs due to the decrease
in the luminosity of the tip of the red giant branch with increasing
metallicity for $\rm[Fe/H]\gtrsim-1$~dex.  Figure~\ref{i_vs_feh} is a plot of
apparent $I$-band magnitude versus $\rm[Fe/H]_{phot}$ for the 29~stars
comprising the secure M31 sample.  Also shown is the expected locus of the
tip of M31's RGB as a function of metallicity derived from theoretical
isochrones computed by the Padova group (Bertelli et~al.\ 1994; Girardi
et~al.\ 1996) and Bergbusch~\& VandenBerg (1992): solid and dashed lines,
respectively.  The model isochrones are shifted by an apparent distance
modulus of $(I-M_I)=24.57$, based on $D_{\rm M31}=783$~kpc and a typical
reddening amount of $E(B-V)=0.06$ which corresponds to $A_I=0.1$
(\S\,\ref{feh_phot_det_sec}).  As expected, the upper envelope of the
distribution of stars in Fig.~\ref{i_vs_feh} generally appears to follow the
tip of the RGB locus.  There is the occasional star located above one or both
of the loci and this could be due to a variety of factors: photometric
errors, error in the M31 distance modulus/depth effect, uncertainties in the
models, and stellar evolutionary effects (e.g.,~presence of intermediate-age
asymptotic giant branch stars).

While the number of stars per unit $I$ magnitude bin in Fig.~\ref{i_vs_feh}
increases with decreasing brightness over most of the displayed range, it is
noticeable that the increase does {\it not\/} continue monotonically all the
way to the faint end of the range over which spectroscopic targets were
selected, $I=22$.  This is in contrast to what might be expected for an RGB
population that is simply truncated at a specific magnitude limit.  The
phenomenon is more clearly illustrated in Fig.~\ref{lum_func} which shows the
distribution of apparent $I$ magnitudes for the secure sample of 29~M31 red
giants (bold histogram).  The dashed and dotted lines show two model RGB
luminosity functions (LFs) for comparison, derived from the Padova isochrones
(Bertelli et~al.\ 1994; Girardi et~al.\ 1996) and they both continue to rise
to $I=22$ irrespective of the details of [Fe/H] weighting.\footnote{Although
the location of the RGB tip is dependent on metallicity, the shape of the
composite LF is only weakly dependent on the assumed [Fe/H] distribution of
the RGB population.  The two model LFs in Fig.~\ref{lum_func} are based on
Gaussian and top-hat [Fe/H] distributions and yet the resulting LFs are very
similar.}  The decrease in the relative completeness fraction of the secure
sample toward the faint end is a result of two~factors: assignment of lower
priorities in the mask design process and lower rate of spectroscopic success
for faint RGB stars compared to brighter counterparts.  For example, 11 of
the 19~low-S/N cases for which the radial velocity measurement fails
correspond to objects with $I>21.4$ with most clustered near the faint limit
of $I=22$ (see \S\,\ref{ccf_sec} and Fig.~\ref{bi_cmd}).  These two sources
of magnitude bias in the secure sample are difficult (if not impossible) to
quantify from first principles; we instead make an empirical determination of
the combined effect of both factors.

The relative completeness fraction as a function of apparent $I$ magnitude is
parameterized as a pair of exponentials:
$$C(I)\,=\,1\>-\>0.1\>{\rm exp}\>[(I-I_{90})/\Delta_C]~~~~~~~~~~~~~~~(I<I_{90})$$
\begin{equation}
 ~~~~~~~=\,0.25\>+\>0.65\>{\rm exp}\>[-(I-I_{90})/\Delta_C]~~~~~~~(I>I_{90})
\end{equation}
\noindent
where $I_{90}=21.55$ is the 90\% completeness limit, and $\Delta_C=0.05$ is
the $e$-folding range of the completeness function.  These parameters have
been chosen to optimize the match between the observed $I$-band LF of the
secure M31 sample on the one hand and the product of $C(I)$ and the model RGB
LF on the other (histogram vs.\ solid line in Fig.~\ref{lum_func}).  The 50\%
completeness limit, $I_{50}=I_{90}+\Delta_C{\rm ln}(2.6)=21.60$, is indicated
in Fig.~\ref{i_vs_feh}.  The apparent 25\% floor in $C(I)$ at the faint end
could be the result of some masks having substantially better data quality
than the rest (for some of the reasons listed in \S\,\ref{datared_sec}) so
that their spectroscopic success rate is essentially perfect even for objects
as faint as $I=22$.  The detailed form of $C(I)$ should not be taken too
literally, however---the observed secure sample LF from which $C(I)$ is
derived suffers from photometric errors and large Poisson errors, and there
is some uncertainty in the model RGB LFs.  Fortunately, the details of $C(I)$
are not too important for metallicity-bias correction either: in fact, the
corrected mean/median [Fe/H] values in Table~\ref{table4} would, for the most
part, be lowered by $\lesssim0.1$~dex if $C(I)$ were ignored altogether (see
below).

Theoretical stellar isochrone population functions are used to compute the
mass fraction (as a function of [Fe/H]) contained within the apparent
magnitude range $20<I<22$ over which the M31 spectroscopic targets are
selected.  As shown in Fig.~\ref{i_vs_feh}, the tip of the RGB cuts
progressively deeper into the upper part of this range with increasing
metallicity: $I_{\rm TRGB}\lesssim20.6$ for $\rm[Fe/H]<-1$~dex but drops to
$I_{\rm TRGB}\sim21.3$ for $\rm[Fe/H]=0$~dex.  This causes the mass fraction
to decrease with increasing metallicity for $\rm[Fe/H]>-1$~dex.  The
steepness of the decrease is slightly accentuated by the increasing degree of
incompleteness toward the faint end of the range, even though $C(I)$ itself
is metallicity-independent.  Figure~\ref{sel_eff} shows the resulting mass
fraction or metallicity selection efficiency function for the secure M31
sample: the solid line represents a calculation based on the Padova group's
isochrones (Bertelli et~al.\ 1994; Girardi et~al.\ 1996) taking $C(I)$ into
account---this is hereafter adopted for the metallicity-bias correction; the
dashed and dotted lines represent calculations based on the Padova and
Bergbusch~\& VandenBerg (1992) isochrones, respectively, in which $C(I)$ is
ignored (set to unity).  The selection efficiency curves are normalized at
low [Fe/H] as we are only interested in their shapes.  The significant
difference between the selection efficiency curves derived from the two~sets
of isochrones reflects one of the inherent sources of uncertainty in the
metallicity-bias correction.  Both sets of isochrones correspond to an age
$t\approx14$~Gyr with a mass function slope close to the Salpeter (1955)
value of $x=+1.35$; it should be noted though that the metallicity-dependence
of the selection efficiency is the same for all mass function slopes and all
ages older than a few~Gyr, even though the upper RGB mass fraction at any
given metallicity depends strongly on these parameters.

To account for metallicity selection bias, every star is assigned a relative
weight based on its $\rm[Fe/H]_{phot}$ value, the weight being the reciprocal
of the relative [Fe/H] selection efficiency (solid line in
Fig.~\ref{sel_eff}).  The weight is approximately unity for each
low-metallicity star, but increases to 1.4, 2, and 5 for stars with
$\rm[Fe/H]_{phot}=-1$, $-0.5$, and 0~dex, respectively.  One of the
four~stars in the kinematically-coherent metal-rich group (object ID
t1.17---\S\,\ref{poss_disk_sec}) has the highest $\rm[Fe/H]_{phot}$ value of
all stars in the secure M31 sample, $-0.24$~dex.  The weighted
(i.e.,~corrected) [Fe/H] histograms for the secure M31 sample are shown in
the lower panels of Fig.~\ref{feh_hist}.  The histograms are normalized by
the mean weight of the stars in the secure sample; thus the net area under
each [Fe/H] distribution is the same before and after correction and is equal
to 29, the observed number of secure M31 red giants.

A comparison between the uncorrected and corrected $\rm[Fe/H]_{phot}$
distributions shows that the high-metallicity end is systematically boosted
in the latter.  This is also true for the [Fe/H] distributions on the ZW and
CG97 scales but the effect is not nearly as systematic, as the weight is
derived from $\rm[Fe/H]_{phot}$ even for the spectroscopic [Fe/H] histograms
and the correlation between the spectroscopic and photometric metallicity
estimates is not exactly one to one.  The selection bias correction causes
the mean and median values of all the [Fe/H] distributions to increase
slightly (the corrected values are listed in Table~\ref{table4}).  The upward
shift in the mean/median of the $\rm[Fe/H]_{phot}$ distribution is a few
tenths of a dex.  Disregarding the very high (and almost certainly spurious)
$\rm[Fe/H]_{ZW}$ estimates for the four~metal-rich stars in a coherent group,
the shift in the spectroscopic [Fe/H] distributions is somewhat smaller than
for $\rm[Fe/H]_{phot}$, about $+0.1$~dex.

It should be kept in mind that the type of selection bias discussed in this
section is not unique to our study---the drop in $I_{\rm TRGB}$ with
increasing [Fe/H] makes the highest-metallicity giants harder to detect and
photometer than their low-metallicity counterparts in all photometric
studies.  Photometric studies of the M31 halo, especially ones based on {\it
HST\/} images, often reach as deep as 5~mag into the RGB LF so the selection
bias is not as severe as in our $I=20\>$--$\>$22 spectroscopic sample which
contains only the brightest RGB stars.  Other tracers of the halo are
affected too: for example, a high-metallicity globular cluster has a fainter
integrated luminosity than a low-metallicity cluster of the same total mass
and mass function slope as a result of the drop in $I_{\rm TRGB}$, and this
causes a relative underrepresentation of the former in any
apparent-magnitude-limited sample.

\subsection{Possible Disk Stars or Metal-Rich Debris?}\label{poss_disk_sec}

Figure~\ref{v_vs_feh} shows a plot of metallicity (two spectroscopic
estimates and photometric estimate) versus radial velocity for all
80~spectroscopic targets with reliable measurements.  The objects in the
range $v<-220$~km~s$^{-1}$ appear to cluster into three~distinct groups:
(a)~5~objects with the most negative velocities, $v\leq-460$~km~s$^{-1}$,
mostly have low [Fe/H]; (b)~16~objects with $-340<v<-220$~km~s$^{-1}$ exhibit
a somewhat larger spread in [Fe/H] over low to moderate values; and
(c)~4~objects with $v\approx-340$~km~s$^{-1}$ have relatively high [Fe/H]
values.  The lack of objects in the range $-460<v<-360$~km~s$^{-1}$ appears
to be a real feature of our sample; it is not caused by reduced detectability
of the Ca\,{\smcap ii} absorption line triplet in the vicinity of strong
might sky emission lines (top panel of Fig.~\ref{sample_spec}).

The last group of 4~stars---t1.17, t2.01, t1.06, and t1.04---lies close to
M31's systemic velocity in our minor axis field of study: thus, their
kinematical properties are similar to what one might expect for red giants in
M31's disk.  While these stars do not stand out in any particular way in the
velocity histogram (Fig.~\ref{vhist}), they are distinguished from the rest
by their {\it combination\/} of high metallicity and common (close to
systemic) velocity.  In fact the 4~stars really stand out in terms of their
spectroscopic [Fe/H] estimates: they rank 3rd--6th, respectively, out of the
full sample of 93~targets (80 in the main M31 target sample, 13 in the
control sample) in terms of reduced Ca\,{\smcap ii} equivalent width $W'$.
The only 2~stars with stronger Ca\,{\smcap ii} lines, t5.13 and t3.28, have
radial velocities of $v=+239$ and $-141$~km~s$^{-1}$, respectively, and this
makes it likely that both are foreground Milky Way dwarf stars.  This
hypothesis is strengthened by the observation that the difference between
$\rm[Fe/H]_{CG97}$ and $\rm[Fe/H]_{phot}$ for these stars is 6.2 and 1.8~dex,
respectively (upper panel of Fig.~\ref{fehcomp_nmem}).  By contrast, the
4~stars ranked 3rd--6th lie closer to the $\rm[Fe/H]_{CG97}=[Fe/H]_{phot}$
line in the upper panel of Fig.~\ref{fehcomp_mem}, as might be expected for
red giants at the distance of M31 (the possible implications of their
$\sim+0.8$~dex offset are discussed below).  The $\rm[Fe/H]_{ZW}$ estimates
for the 4~stars are significantly greater than $\rm[Fe/H]_{phot}$ but the
difference appears to be a systematic function of metallicity (lower panel of
Fig.~\ref{feh_diff}) which suggests that this is due to the divergence of the
cubic ZW relation beyond the directly-calibrated range.  The spuriously-high
$\rm[Fe/H]_{ZW}$ estimates for these stars are to be disregarded whether or
not the stars turn out to be members of the M31 halo.

In light of the possibility, albeit small (see below), that this
kinematically-coherent group of 4~metal-rich stars belongs to M31's disk
rather than its halo, the mean, median, and rms of the halo [Fe/H]
distribution are recomputed with these stars excluded from the secure M31
sample (Table~\ref{table4}).  The 4~stars are at or near the high end of the
observed [Fe/H] distribution, so their exclusion results in a
0.2$\>$--$\>$0.3~dex drop in the mean/median [Fe/H] values.  The mean
(median) $\rm[Fe/H]_{CG97}$ and $\rm[Fe/H]_{phot}$ values for just the
kinematically-coherent group of 4~stars are also listed in
Table~\ref{table4}: $+0.11$ (0.00) and $-0.64$ ($-0.62$)~dex, respectively.

The External Galaxy Model (Hodder 1995), an adaptation of the Bahcall~\&
Soneira (1984) Galactic star count model to an external spiral galaxy,
predicts that the M31 disk contamination fraction in our $R=19$~kpc minor
axis spectroscopic field should be as low as 0.01\% (RGG).  If the
kinematically-coherent group of 4~metal-rich are indeed members of the M31
disk, they would represent a $>10$\% disk contamination fraction (4~out
of~29), three orders of magnitude higher than expected based on extrapolation
of M31's inner disk brightness profile.  The M31 disk is highly inclined with
an apparent 4:1 aspect ratio ($1/{\rm cos}(i)=\rm1/cos(77^\circ)\gtrsim4$);
thus, the projected distance of our field from the galaxy center, $R=19$~kpc,
corresponds to a radial distance of over 80~kpc in the plane of the disk, and
this would be anomalously large for a spiral galaxy disk.  A favorable warp
in the disk could cause it to intersect the line of sight to our
spectroscopic field without it needing to be this large.  As a matter of
fact, old photographic plates, modern CCD mosaic images, and H\,{\smcap i}
data all point to the presence of substantial (major and minor axis) warps in
M31's stellar and gaseous disks (Newton~\& (Emerson 1977; Innanen, Harris,~\&
Webbink 1983; Brinks~\& Shane 1984; Guhathakurta, Choi,~\& Reitzel 2000).  It
is unclear from these data, however, whether the M31 disk warp favors the
detection of disk stars in our particular SE minor-axis field or works
against it.  Sarajedini~\& Van Duyne (2001) claim to have detected an M31
thick disk population with $\langle[Fe/H]\rangle\sim-0.2$~dex and it is
possible that our group of 4~metal-rich kinematically-coherent stars belongs
to this population.  The properties of M31's thick disk are not well
characterized at the moment, however, so one cannot make a reliable estimate
of the probability of detecting 4~thick disk stars in our field of study.

The Ibata et~al.\ (2001) large-scale study of the M31 halo may be useful for
gaining insight into the nature of this kinematically-coherent group of
4~metal-rich giants found in our spectroscopic study.  In addition to the
giant streamer that Ibata et~al.\ draw attention to, there is a hint of a
much smaller ``feature'' (possibly a minor debris trail of some sort) at the
location of our spectroscopic field: $\sim1.4^\circ$ from the galaxy's center
on the minor axis (see Fig.~1 of their paper).  Thus, these 4~stars may have
nothing to do with M31's disk after all, and may instead represent a coherent
piece of metal-rich debris from a past accretion event in the M31 halo.

Figure~\ref{radec_vel} shows the sky-positions of all spectroscopic targets
grouped by radial velocity.  Figure~\ref{radec_feh} shows the sky-positions
of only the 29~secure M31 giants grouped by spectroscopic [Fe/H] estimate.
There is no obvious correlation in these plots between radial velocity or
[Fe/H] and a star's position on the sky.  The 4~kinematically-coherent
metal-rich stars appear to lie in a straight line, but the small number of
stars and the odd shape of the ``footprint'' of the spectroscopic targets
make this a tantalizing suggestion at best, with little or no statistical
significance.

The average $\rm[Fe/H]_{CG97}$ value for this group of 4~stars is
0.7$\>$--$\>$0.8~dex higher than their average $\rm[Fe/H]_{phot}$ value
(Fig.~\ref{fehcomp_mem} and Table~4).  It is possible that the 4~stars have
distinctly higher [Ca/Fe] ratios that the other secure M31 giants in the
sample, perhaps anomalously high ratios: they truly stand out in terms of
their Ca\,{\smcap ii} line strengths but not in terms of their
$\rm[Fe/H]_{phot}$ values.  Their $\rm[Fe/H]_{CG97}$ versus
$\rm[Fe/H]_{phot}$ discrepancy is unlikely to be a result of extrapolation
error in the CG97 relation beyond the upper end of the calibrated range
[Eq.~(\ref{fehcg_eqn})].  Our empirical check using spectra of red giants in
NGC~6791, a cluster of known metallicity ($\rm[Fe/H]_{ZW}=+0.19$~dex), shows
that if anything the CG97 relation {\it underpredicts\/} the metallicity of
this cluster by 0.3~dex (Fig.~\ref{w_vs_vhb_calib}).  Another manifestation
of this possible [Ca/Fe] anomaly in these 4~stars (relative to NGC~6791
giants) can be found in the third panel of Fig.~\ref{w_vs_vhb_calib}.  The
empirical ``phot+spec'' calibration relation is designed to achieve equality
between spectroscopic and photometric [Fe/H] estimates for the 4~metal-rich
stars, but in doing so it underpredicts the metallicity of NGC~6791 by
0.8~dex (\S\,\ref{feh_comp_sec}).

It may be impossible to tell (certainly from the data at hand, and perhaps in
general) whether this kinematically-coherent group of 4~metal-rich giants has
an ``internal'' origin---i.e.,~M31's warped and/or disrupted disk---or an
``external'' one---i.e.,~one of the satellites currently undergoing tidal
disruption (M32 or NGC~205) or a recently-accreted satellite galaxy that has
been completely disrupted.  In any case, a growing body of evidence has
accumulated over the years showing that galaxy interactions are probably
quite common in the M31 subgroup (Byrd 1979; Innanen et~al.\ 1983; Ibata
et~al.\ 2001; Choi, Guhathakurta,~\& Johnston 2002).

\subsection{Comparison to Photometric Studies: Is M31's Halo
Metal-Rich?}\label{discrep}

Previous photometric studies of the M31 halo---wide-field ground-based
studies by Durrell, Pritchet,~\& Harris (1994), RGG, and Durrell et~al.\
(1999, 2001), complemented by deeper {\it HST\/}/WFPC2 studies of smaller
fields by Rich et~al.\ (1996a,b, 2002) and Holland et~al.\ (1996)---all yield
mean [Fe/H] values for the M31 halo that are systematically (if slightly)
{\it higher\/} than the findings of our present spectroscopic study.  In
fact, the claim is made in some of these photometric studies that field halo
stars in M31 have a significantly higher average metallicity than its
globular cluster system, with a mean [Fe/H] comparable to or in excess of the
Galactic globular cluster 47~Tuc ($-0.7$~dex).  This is very different from
the case of the Milky Way where the field halo population has a slightly
lower mean metallicity than its globular clusters (Fig.~\ref{feh_hist_comp}).

These results must be treated with caution though.  The various M31 halo
samples discussed above are drawn from many different parts of the galaxy so
any intercomparison may be affected by the possible presence of large-scale
gradients or metallicity substructure (\S\,\ref{impl_sec}).  Moreover, even
the photometric metallicity determination methods are not the same across the
various studies (use of different colors, RGB fiducials vs.\ theoretical
isochrones, different metallicity scales, etc.); unlike our present study,
these earlier studies did not have the benefit of independent spectroscopic
determination of [Fe/H].  Finally, sample selection criteria and
contamination levels are quite varied across the samples.  While the inner
halo fields are expected to contain a smaller fraction of foreground Milky
Way dwarf contaminants than the outer halo fields, they tend to contain a
greater fraction of M31 disk star contamination.  The Holland et~al.\ and
Rich et~al.\ {\it HST\/} studies of the G302, G312, and G1 fields made no
correction for sample contamination in their photometric metallicity
estimates.  While statistical subtraction of foreground and background
contaminants was attempted in the ground-based studies, these attempts were
only partly successful at removing contaminants.  The matching of comparison
field data to the main data set is hardly ever perfect: differences in
seeing, depth, crowding, Galactic latitude and longitude, etc.\ invariably
limit the fidelity of the statistically-subtracted sample.  The upshot of
this is that any conclusions about the M31 halo metallicity or structure
drawn from the above photometric studies are bound to be somewhat uncertain.

The spectroscopically-selected secure sample of M31 halo giants presented in
this study is probably the cleanest sample of bright RGB stars in the M31
outer halo to date, but it suffers from severe incompleteness at the
high-metallicity end of the distribution (\S\,\ref{selbias}).   While
attempts have been made to correct for this metallicity bias, our detection
efficiency for near-solar-metallicity giants is simply too low for us to be
able to constrain the high end of the [Fe/H] distribution.  Some metal-rich
stars have in fact been detected in our survey, but not a single one with
$\rm[Fe/H]\gtrsim0$~dex.  The observed raw (i.e.,~uncorrected) [Fe/H]
distribution for the secure M31 sample spans more than 2~dex, and the
bias-corrected distribution appears to cut off quite sharply just shy of
solar [Fe/H] where the selection function of our spectroscopic survey also
drops off [see panel~(f) of Fig.~\ref{feh_hist} for example].  Thus, it is
entirely feasible that our spectroscopic survey is missing giants of
near-solar [Fe/H] and that the M31 halo is just as metal-rich as the
photometric studies indicate.  In other words, the apparent discrepancy
between the mean [Fe/H] found in our spectroscopic study and the generally
higher values found in earlier photometric studies may simply be an artifact
of selection effects in our study.

The discrepancy cannot be attributed to a systematic difference between
spectroscopic and photometric [Fe/H] measurement methods.  As discussed in
\S\,\ref{feh_comp_sec} and illustrated in Fig.~\ref{fehcomp_mem}, a
star-by-star comparison reveals a reasonably good agreement between these
two~[Fe/H] estimates.  Put differently, even the mean {\it photometric\/}
[Fe/H] estimate for our spectroscopically-confirmed sample of secure M31 halo
giants is lower than the mean value found in other photometric surveys.
Thus, the origin of the difference must lie in the red giant samples
themselves.

A few other possible reasons for the discrepancy can be ruled out as well.
Metallicity substructure in the M31 halo cannot be the culprit: the RGG and
Durrell et~al.\ (1999, 2001) studies were carried out in the same $R=19$~kpc
minor axis field as our spectroscopic study and yet these studies find a
slightly higher mean [Fe/H] value.  Sampling effects and pre-selection
criteria for the spectroscopic targets cannot be blamed either: the RGG study
and this spectroscopic study use the same $UBRI$-based photometric criteria
to isolate red giant candidates; in fact, the shape of the $B-I$ color
distribution of our 99~spectroscopic targets is identical to that of the
parent sample of 284~red giant candidates in the RGG study ($UBRI$- and
morphology-selected objects in the $20<I<22$ range).

As part of a large, ongoing spectroscopic survey of the M31 halo, we have
collected similar Keck/LRIS spectra for about 40--75 RGB candidates per field
in each of the G1, G302, and G312 fields, the same fields that were targeted
in earlier {\it HST\/}/WFPC2 photometric studies (Holland et~al.\ 1996; Rich
et~al.\ 1996a,b).  Analysis of these additional LRIS spectra is in progress
(Guhathakurta 2002; Guhathakurta et~al.\ 2002) and this should ultimately
shed some light on the origin of the apparent [Fe/H] discrepancy.  These
spectroscopic data will provide a measure of the foreground Galactic
dwarf/M31 disk giant contamination fraction in the {\it HST\/} photometric
samples.  Moreover, the greater than two-fold increase in the overall
spectroscopic sample size in the M31 halo should improve the statistics on
any rare population of high-[Fe/H] giants, which are of course made even more
scarce by the metallicity bias in our selection function.

\subsection{Comparison to Models and Other Halo Tracers}\label{halo_comp_sec}

Simple chemical enrichment models are a useful guide to help interpret the
observed metallicity distribution in the secure sample of M31 stars.  A
steady-gas-loss model (Da~Costa et~al.\ 2000; Hartwick 1976) is one where the
instantaneous recycling approximation is made and where the gas loss is
assumed to be proportional to the rate of star formation.  In such a model,
the metallicity distribution may be written as:
\begin{equation}
f(z) = (1/y){\rm exp}[-(z-z_0)/y] ~~~~~~~~~~  z \geq z_0
\end{equation}
\noindent
where $z_0$ is the initial abundance and $y$ is the yield.  The yield and the
mean abundance are related via: $y=\langle{z}\rangle-z_0$.  The two~free
parameters for this model are $\langle{z}\rangle$ (or equivalently
$\rm\langle[Fe/H]\rangle$) and $z_0$ (or $\rm[Fe/H]_0$).

The model which best fits the uncorrected [Fe/H] distributions has
$\rm[Fe/H]_0=-10.0$~dex and $\rm\langle[Fe/H]\rangle=-1.5$~dex.  This model
is plotted as a thin solid line over the metallicity distributions in each
panel of Figs.~\ref{feh_hist} and \ref{feh_hist_comp} and as a bold solid
line in panel~(b) of the latter figure.  The dashed and dotted lines in
panel~(b) of Fig.~\ref{feh_hist_comp} are models with
$\rm\langle[Fe/H]\rangle=-2.0$ and $-0.5$~dex, respectively, with
$\rm[Fe/H]_0=-10.0$~dex in both cases.  The best fit model does a reasonable
job of matching the overall metallicity distribution (see Fig.~\ref{feh_hist}
and \ref{feh_hist_comp}).  If the hint of bimodality seen in the observed
[Fe/H] distributions is confirmed with larger samples of stars, a single
model will clearly not fit the data, but a two-component model might.
Moreover, a great deal of pre-enrichment does not appear to be needed to
explain the [Fe/H] distribution of the secure sample of M31 halo stars.  The
details of this ``toy'' chemical evolution model are not to be taken too
literally: there are many simplifying assumptions on which such models are
founded and the real set of physical conditions in place during the formation
of the M31 halo are likely to have been considerably more complex.  Moreover,
the match between the model and observed [Fe/H] distributions in no way
guarantees the uniqueness of this particular theoretical interpretation.  We
merely use the models as a very rough guide to understanding the chemical
enrichment history of the M31 halo, and as a point of reference for
intercomparison of the [Fe/H] distributions observed for the different M31
and Milky Way halo tracer populations.

The metallicity distribution of the secure M31 field halo giant sample is
compared to other halo tracers in the Local Group (Fig.~\ref{feh_hist_comp}).
The corrected distribution of $\rm[Fe/H]_{CG97}$ values is plotted in
panel~(a), and the moments of this distribution appear in Table~\ref{table4}.
It should be noted that different studies compute metallicity differently and
the reader is referred to each reference for the details of each metallicity
calculation.  The data are scaled to match the overall number of objects in
the secure M31 halo sample, 29, in order to facilitate comparison between the
various samples.

The metallicity distribution of M31's globular cluster system is presented
in panel~(c) of Fig.~\ref{feh_hist_comp}.  The distribution is quite similar
to that of the M31 field stars.  Both have a spread of about 2~dex and a mean
value around $-1.2$~dex \cite{Barmby}.  The Milky Way globular clusters have
a slightly lower mean metallicity, $-1.4$~dex \cite{Harris} and a more
strongly bimodal distribution [panel~(d)].  Carney et~al.\ (1990) find a mean
metallicity of $\rm[Fe/H]=-1.72$~dex for field halo stars in the Milky Way
and a metallicity probability distribution function \cite{Laird} which has a
spread of 2~dex and a lower mean than M31's stellar halo [panel~(e)].

Panel~(f) of Fig.~\ref{feh_hist_comp} shows the metallicity distribution of
the Local Group dwarf satellites galaxies from the compilation of Grebel
(1999, 2000).  The histogram represents a direct sum over all satellite
galaxies, with each galaxy getting the same weight.  Alternatively, each
satellite galaxy can be represented by a Gaussian whose mean/width is equal
to the satellite's measured mean/spread in [Fe/H].  Each Gaussian is weighted
by the luminosity of the satellite it represents, and the individual
Gaussians are then combined.  The bold solid curve is the luminosity-weighted
sum over all satellites, while the dashed curve is the sum over all except
the Large Magellanic Cloud and M32.  The direct sum and restricted
luminosity-weighted sum are a good match to the observed [Fe/H] distribution
of M31 field halo giants, while luminosity-weighted sum over the full sample
is dominated by a few luminous systems with relatively high metallicity (the
Magellanic Clouds, M32, and NGC~205) and is a poor match.  It should be noted
that while the halos of M31 and the Milky Way may have been assembled from
dwarf satellites, the properties of the surviving satellites seen today could
be different from the properties of the satellites that were accreted to form
these halos.  Thus, one should not read too literally into the match/mismatch
between the observed M31 halo [Fe/H] distribution and that of the present-day
satellites.

\subsection{Implications}\label{impl_sec}

The field red giant population of the M31 halo is more metal-rich on average
than that of the Milky Way.  This implies that M31 underwent a larger amount
of enrichment before the halo was assembled than did the Milky Way.  In
addition, M31 displays a slightly larger spread of metallicities.  The large
spread in [Fe/H] observed in M31 tends to support the accretion model for
halo formation.  {\it At face value\/}, a comparison of {\it HST\/} studies
of M31's inner halo to our spectroscopic study of its outer halo suggests
that the mean metallicity decreases radially outwards.  If such a radial
gradient is truly present (see \S\,\ref{discrep} caveats below), it would be
consistent with the prediction of the monolithic collapse model.

While these implications appear to be inconsistent, the accretion models are
currently too simplistic.  A radial gradient could result even in the context
of the accretion scenario, as the more massive satellites tend to have higher
metallicities, and these are unlikely to be completely disrupted until they
stray close to the center of the parent galaxy.  By contrast, the less
massive, lower-metallicity satellites will be completely disrupted before
dynamical friction has had a chance to cause them sink too far into the
parent galaxy potential.

One complication in comparing the observed metallicity distribution of M31 to
other halo tracers is the underlying assumption that M31 giants have an
enhancement in their alpha elements similar to that of the oldest Galactic
globular clusters, $\rm[\alpha/Fe]=+0.3$~dex.  If the M31 halo is instead
younger with solar element ratios, $[\alpha/{\rm Fe}]=0$~dex, as is the case
for the young Milky Way globular clusters Pal~12 and Ruprecht~106 (Brown,
Wallerstein,~\& Zucker 1997), then the spectroscopic metallicity estimates
would have to be revised upwards by 0.3~dex.  The photometric estimates would
also be revised, but the change would not be nearly as large as
$+0.3$~dex.\footnote{One way to understand this is to consider a star of a
given $B-I$ color and a theoretical isochrone that passes through the star's
location in the CMD.  If alpha elements were the {\it only\/} ones doing the
line blanketing, changing $\rm[\alpha/Fe]$ from $+0.3$ to 0~dex could be
compensated by an equal and opposite change in [Fe/H] (i.e.,~a 0.3~dex
increase) to keep the color of the isochrone fixed.  This would mean that the
$\rm[Fe/H]_{phot}$ estimate for the star would increase by this amount.  If,
on the other hand, the line blanketing from alpha elements were a negligible
fraction of the overall amount of line blanketing, the change in
$\rm[\alpha/Fe]$ would not affect the color of the isochrone at all, and the
$\rm[Fe/H]_{phot}$ estimate for the star would stay unchanged.  The real
situation is somewhere between these two~extreme examples, perhaps closer to
the latter, which implies that the boost in $\rm[Fe/H]_{phot}$ due to the
change in $\rm[\alpha/Fe]$ would be significantly less than $+0.3$~dex.}

\section{Summary}\label{concl_sec}

The main points of this paper may be summarized as follows:

\begin{itemize}
\item[$\bullet$]{A combination of imaging and spectroscopic data are used to
examine the structure, metallicity, and dynamics of M31's stellar halo.  Deep
Kitt Peak 4-m $UBRI$ CCD images of a 19~kpc minor axis field in M31 are used
to screen a sample of candidate field halo red giants using photometric and
morphological criteria; a subset consisting of 99~such candidates, with
brightnesses in the range $20<I<22$, form the target sample for the
spectroscopic study.}

\item[$\bullet$]{Multi-slit spectroscopy with the Keck~II 10-m telescope and
the Low Resolution Imaging Spectrograph of the Ca\,{\smcap ii} near-infrared
triplet is used to make secure identification of M31 halo red giant branch
stars by discriminating them from foreground Galactic dwarf stars, M31 disk
red giants, and background field galaxies.   Spectra of 13 additional bright
objects ($I<20$), foreground Milky Way stars along the line of sight to M31,
have also been obtained and these serve as a control sample.}

\item[$\bullet$]{The distribution of radial velocities of the main M31
spectroscopic target sample is well fit by an equal mix of foreground dwarfs
(drawn from a standard Galactic model, with a peak close to
$v\approx0$~km~s$^{-1}$) and giants in M31's halo represented by a Gaussian
of width $\sigma_v^{\rm M31}\sim150$~km~s$^{-1}$ centered on its systemic
velocity of $v_{\rm sys}^{\rm M31}\approx-300$~km~s$^{-1}$.  A secure,
largely uncontaminated sample of 29~M31 halo red giants is identified
primarily on the basis of their large negative radial velocities, with
limited use of broadband color information (in the context of a Galactic
model) to reject foreground dwarf star contaminants at intermediate
velocities.}

\item[$\bullet$]{Metallicity measurements are made on a star-by-star basis
using two independent methods: the strength of the Ca\,{\smcap ii} absorption
line triplet (calibrated via the Zinn-West or Carretta-Gratton relations for
Galactic globular clusters) and the location of the star in the $B-I$ vs.\
$I$ color-magnitude diagram relative to model red giant fiducials.}

\item[$\bullet$]{The photometric and spectroscopic [Fe/H] estimates are in
reasonable agreement with each other for the secure M31 giants, especially
those with metallicities within the calibrated range of the [Fe/H]
determination methods; the extrapolation of the cubic Zinn-West relation
beyond the upper end of the calibrated range predicts substantially higher
metallicities than the other [Fe/H] estimates.}

\item[$\bullet$]{The median metallicity of M31 halo red giants is in the
range, $\rm\langle[Fe/H]\rangle=-1.8$ to $-1.1$~dex, for various sample
selection criteria and metallicity measurement methods.  It should be
cautioned, however, that our survey is very inefficient at finding M31 giants
of solar/super-solar metallicity; in fact, none are detected with
$\rm[Fe/H]_{phot}\gtrsim0$~dex.  Thus, even though the quoted median values
are corrected for selection bias against high-metallicity stars, they may be
substantially lower than the true median metallicity of M31's halo.  The mean
$\rm[Fe/H]_{ZW}$ values scatter over a somewhat larger range than the
corresponding median values due to a few stars with very high spectroscopic
metallicities (see below) for which the extrapolation of the cubic Zinn-West
relation diverges.  In all cases, the observed M31 halo [Fe/H] distribution
has an rms spread of at least 0.6~dex, comparable to that of Milky Way field
halo giants, M31 globular clusters, Galactic globular clusters, and Local
Group dwarf satellite galaxies, and spans the full 2~dex range over which the
metallicity measurement methods are calibrated,
$\rm-2\lesssim[Fe/H]\lesssim0$~dex.}

\item[$\bullet$]{Photometrically-determined [Fe/H] values from {\it Hubble
Space Telescope\/} studies of the inner halo of M31, 7 and 11~kpc from the
center on the minor axis, and from ground-based studies of the outer halo,
the same $R=19$~kpc minor axis field targeted in this spectroscopic study,
yield a systematically higher mean metallicity (by up to 1~dex) than our
study and a comparably large spread of about 2~dex.  As noted above, however,
the mean [Fe/H] of M31 halo giants may be severely underestimated in our
spectroscopic survey because it is virtually incapable of detecting stars
with solar or greater metallicity.  Moreover, the previous photometric
studies use different sample selection criteria than this study: in
particular, their lack of spectroscopic data leaves open the possibility that
the samples are contaminated by M31 disk giants and/or foreground Galactic
dwarf stars.  These factors make it difficult to compare the average [Fe/H]
values measured in different parts of M31's halo.}

\item[$\bullet$]{Within the secure sample of M31 red giants, there are
four~stars with red $B-I$ colors and exceptionally strong Ca\,{\smcap ii}
absorption lines indicating solar or super-solar metallicities.  These stars
lie within a relatively narrow velocity range at $\approx-340$~km~s$^{-1}$.
This combination of high metallicity and common velocity close to M31's
systemic velocity is what might be expected for M31 disk giants in this
minor-axis field.  More likely, these stars represent metal-rich debris from
a past accretion event.}
\end{itemize}

\bigskip
{\bf Acknowledgments}\\
\noindent
We would like to acknowledge the support and help of several people who made
this project possible: Linda Pittroff and Johnny Warren for reduction of
cluster red giant calibration data, Drew Phillips for assistance with LRIS
multislit mask design, mask alignment, and calibration procedures, Jeff
Lewis, Bill Mason and Dave Cowley for mask fabrication, Dan Kelson for help
with the Expector spectral data reduction software, Eva Grebel for providing
data on Local Group dwarf satellites, and Tom Bida, Randy Campbell and the
rest of the Keck observatory staff for their expert guidance during the
observing runs.  We would also like to thank Mike Bolte, Phil Choi, Andy
Gould, Bob Kraft, Ruth Peterson and Linda Schroeder for valuable suggestions
and fruitful discussions, and the anonymous referee for a careful reading of
the manuscript and detailed report which has led to key improvements in the
paper.  DBR is grateful to Mike Rich for his support during the final stages
of the project.  DBR and PG were funded in part by a CalSpace grant and by
NASA grants NAG~5-3232 (Long Term Space Astrophysics) and NAG~5-8348
(Astrophysics Data Program).


\centerline{\epsfxsize=7in \epsfysize=9in
\epsfbox{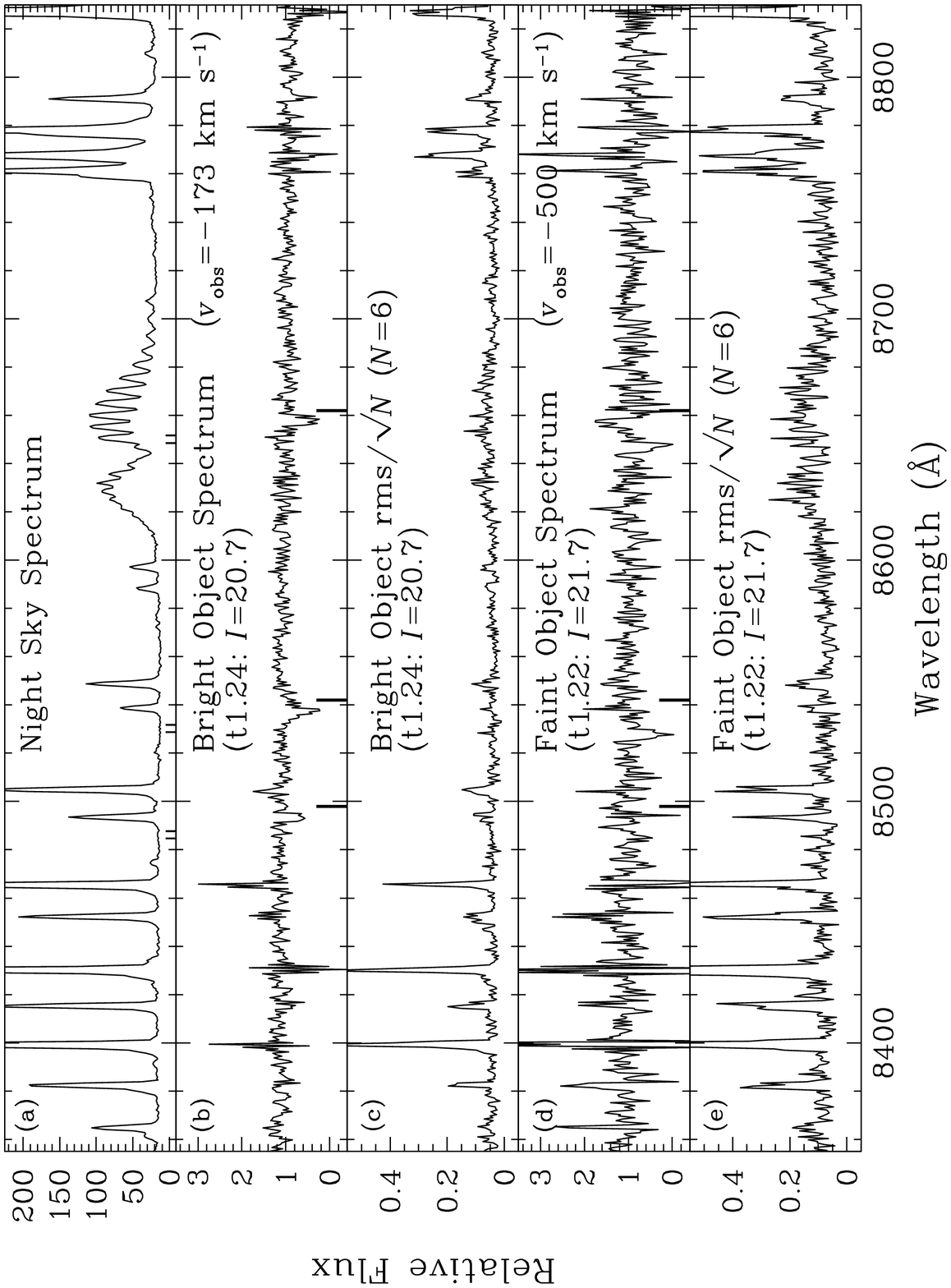}}
\figcaption[Reitzel.fig01.eps]{\label{sample_spec}{
   ~~~({\it a\/})~Typical night sky emission-line spectrum in the region of
the Ca\,{\smcap ii} near-infrared triplet.  The 3~pairs of bold line segments
along the bottom of the panel mark the regions in which the Ca\,{\smcap ii}
lines would appear for objects in the radial velocity range,
$-460<v<-360$~km~s$^{-1}$; these regions are in relatively clean parts of the
spectrum which are not too badly affected by night sky emission lines.  Thus,
the observed lack of M31 giants in this velocity range is {\it not\/} a
result of selection bias (see Fig.~\ref{v_vs_feh} and
\S\,\ref{poss_disk_sec}).
   ~~~({\it b\/})~The spectrum of a typical bright ($I=20.7$) object, t1.24
in Table~\ref{phot_tbl}, with the continuum normalized to unity.  The
rest-frame wavelengths of the Ca\,{\smcap ii} near-infrared triplet are
denoted by 3~bold line segments below the spectrum.  The absorption lines in
the spectrum are blueshifted by an amount corresponding to the observed
radial velocity of the object, $v_{\rm obs}=-173$~km~s$^{-1}$ (prior to
heliocentric correction).
   ~~~({\it c\/})~The rms error per pixel of the mean spectrum, obtained by
dividing the observed rms dispersion among the 6~individual spectral
exposures by the square root of the number of exposures, $N=6$.  The rms
error of the mean is about 0.05 or less for the most part, corresponding to a
signal-to-noise ratio $>20$, except in the vicinity of strong night sky lines
where the rms is boosted by increased photon noise and fringing.
   ~~~({\it d\/})~Same as ({\it b\/}) for a typical faint ($I=21.7$) object,
t1.22, for which $v_{\rm obs}=-500$~km~s$^{-1}$.
   ~~~({\it e\/})~Same as ({\it c\/}) for the faint object whose spectrum is
shown in ({\it d\/}).  The rms level is typically between 0.05 and 0.10
between strong night sky lines, corresponding to a signal-to-noise ratio
between 10 and 20.}}

\centerline{\epsfxsize=7in \epsfysize=9in
\epsfbox{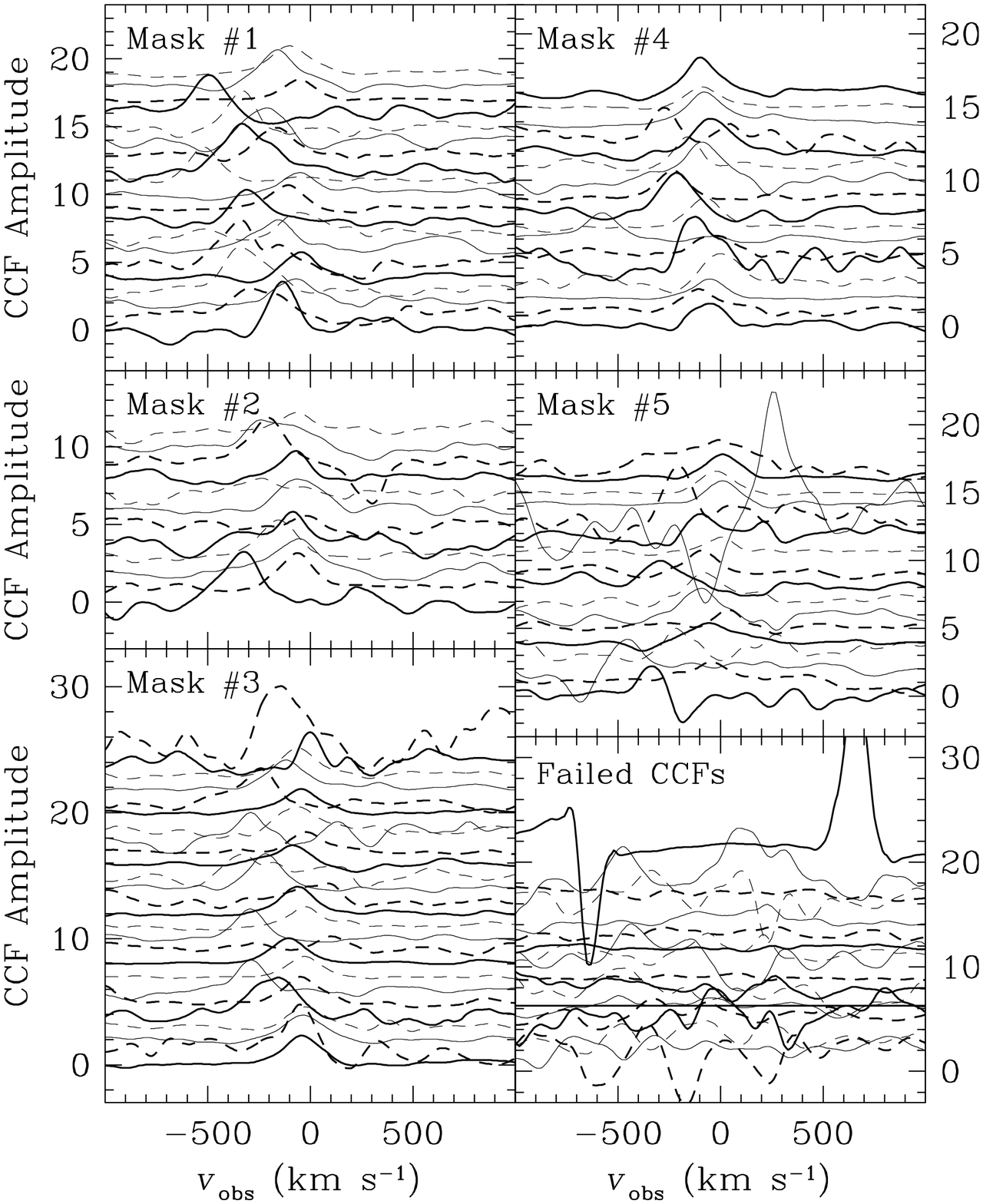}}
\figcaption[Reitzel.fig02.eps]{\label{ccf}{Cross correlation functions (CCFs)
for all 112~objects for which spectroscopy is available: 99~M31 red giant
candidates and 13~control sample stars.  The 93~objects for which reliable
velocity determination is possible are grouped by mask number: 20, 12, 26,
17, and 18~objects for masks 1--5, respectively.  The CCFs in each panel are
sorted in order of increasing object ID number (Tables~\ref{phot_tbl} and
\ref{spec_tbl}) from bottom to top following the sequence of line types: bold
solid, bold dashed, thin solid, thin dashed, bold solid, etc..  The CCF peaks
tend to be concentrated around $v\approx-500$ to 0~km~s$^{-1}$, a velocity
range occupied by M31 red giants and foreground Milky Way dwarf stars.  The
19~objects for which the velocity measurement fails are grouped together in
the bottom right panel, sorted as before by object ID number and in the order
mask~1--5.  These CCFs are noticeably noisier than the rest.}}

\centerline{\epsfbox{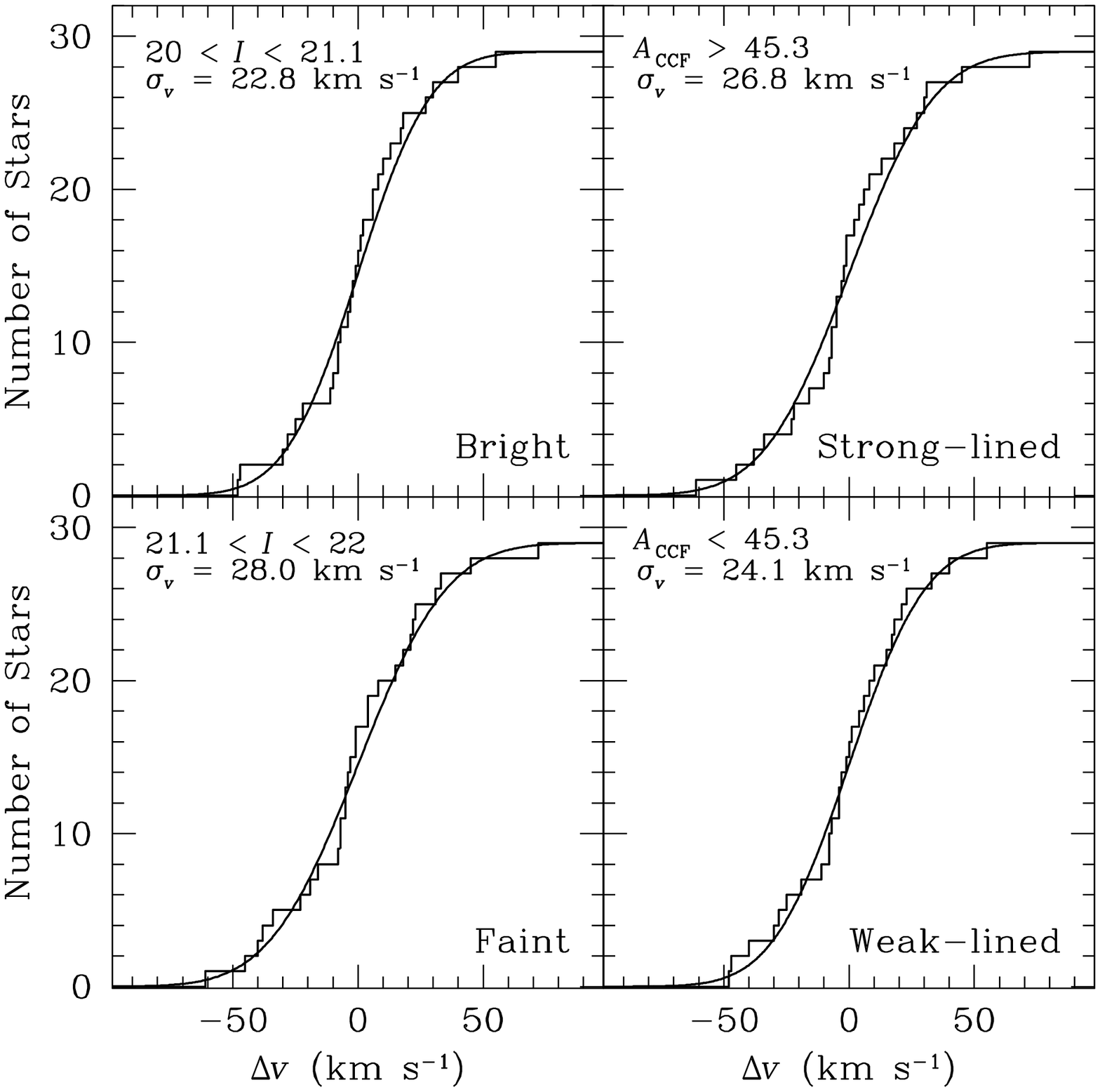}}
\figcaption[Reitzel.fig03.eps]{\label{verr}{Cumulative distribution of radial
velocity differences, derived by splitting the exposure time for each target
in masks~1--4 into two~independent coadds of two~exposures each (see
\S\,\ref{v_mes}).  The bright ($I<21.1$) and faint ($I>21.1$) subsamples are
shown in the upper-left and lower-left panels, respectively; the strong-lined
($A_{\rm CCF}>\langle{A}_{\rm CCF}\rangle$) and weak-lined ($A_{\rm
CCF}<\langle{A}_{\rm CCF}\rangle$) subsamples are shown in the upper-right and
lower-right panels, respectively.  The solid curve is the best-fit Gaussian
with $\sigma_v^{\rm pair}$ given for each distribution.  The four~subsamples
show roughly similar distributions.  The bright subsample has the smallest
velocity measurement error.  The strength of the Ca\,{\smcap ii} absorption
lines does not appear to have a large effect on the velocity measurement
error.}}

\centerline{\epsfbox{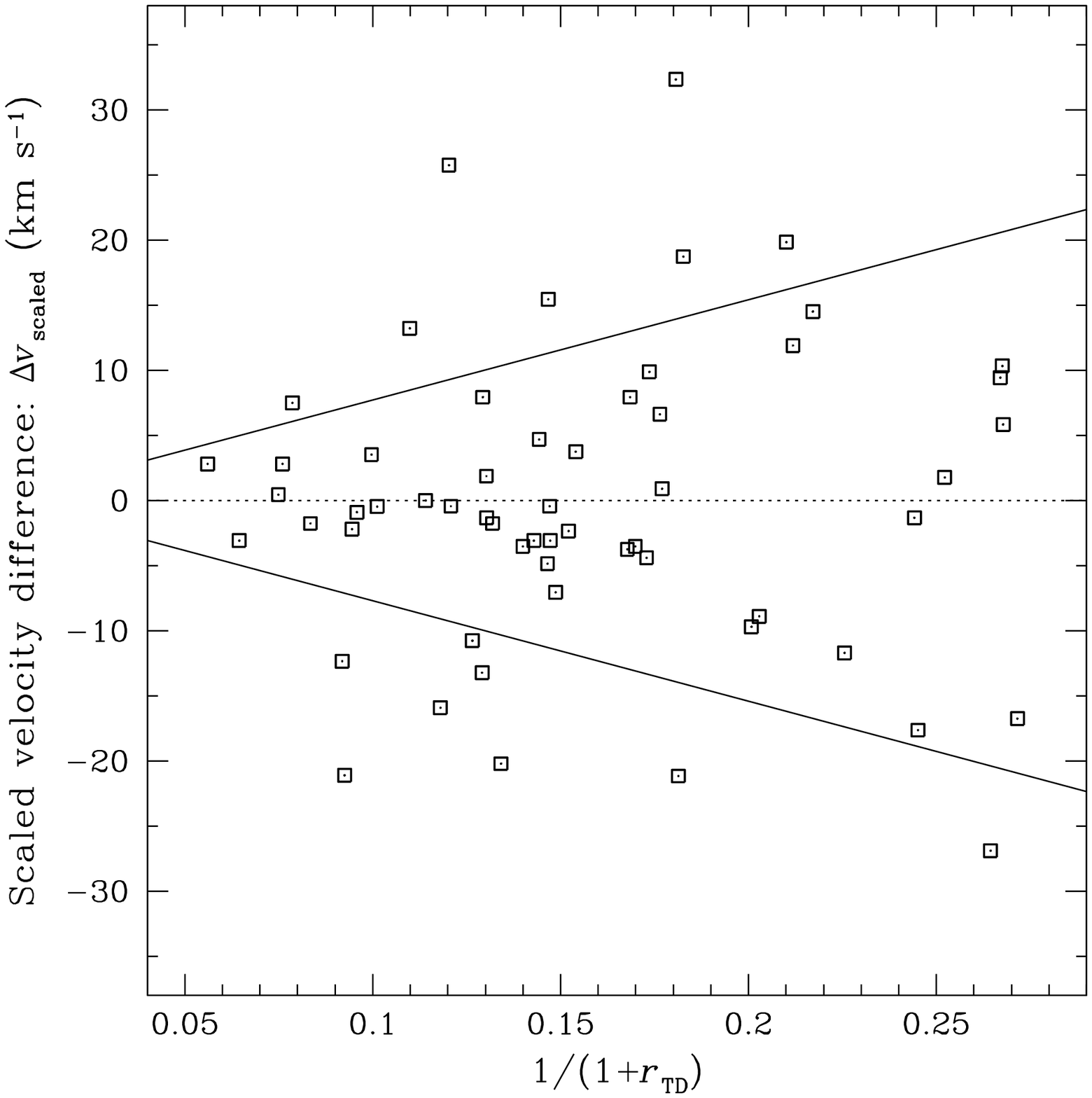}}
\figcaption[Reitzel.fig04.eps]{\label{verr_vs_rtd}{Radial velocity difference
for each pair of split exposure coadds for targets in masks~1--4, scaled
(down) to correspond to the total exposure time, plotted as a function of
$(1+r_{\rm TD})^{-1}$, where $r_{\rm TD}$ is the usual Tonry-Davis (1979)
parameter indicating the significance of the peak in the overall coadded CCF.
The solid lines show an empirical formula for measuring the $\pm1\sigma$
velocity error as described in \S\,\ref{v_mes}: $\rm\pm77~km~s^{-1}/(1+r_{\rm
TD})$.}}

\centerline{\epsfbox{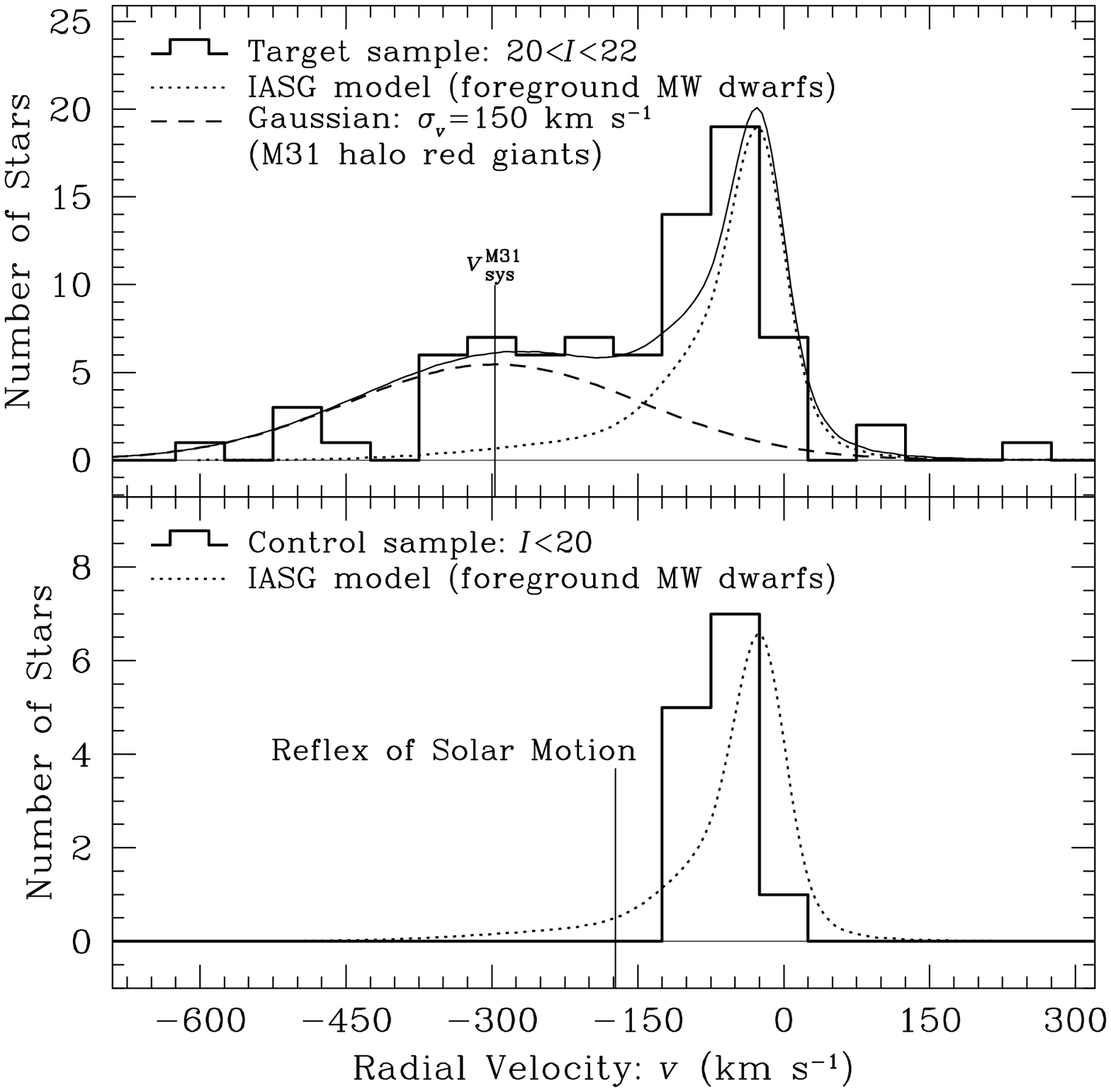}}
\figcaption[Reitzel.fig05.eps]{\label{vhist}{Distribution of heliocentric
radial velocities for the main sample of M31 spectroscopic targets (upper
panel) and control sample (lower panel).  The main sample (histogram in upper
panel) is well fit by the combination of a Gaussian of width
$\sigma=150$~km~s$^{-1}$ centered on M31's systemic velocity of $v_{\rm
sys}^{\rm M31}=-297$~km~s$^{-1}$ (dashed line) and the prediction of the IASG
star count model of the Milky Way (Ratnatunga \& Bahcall 1985), the taller
dotted curve peaked at $v\approx-25$~km~s$^{-1}$.  The solid curve is the sum
of these two models and it is normalized to match the total number of objects
in the histogram, 80.  The IASG model curve in the lower panel is normalized
to the number of objects in the control sample, 13.  The reflex of the solar
motion is indicated, the inverse of the projection of the solar rotation
velocity vector in the direction of M31.  The observed peak of the velocity
distribution of foreground Galactic dwarf stars appears to be about
20$\>$--$\>$50~km~s$^{-1}$ more negative than the prediction of the IASG
model (both panels).}}

\centerline{\epsfbox{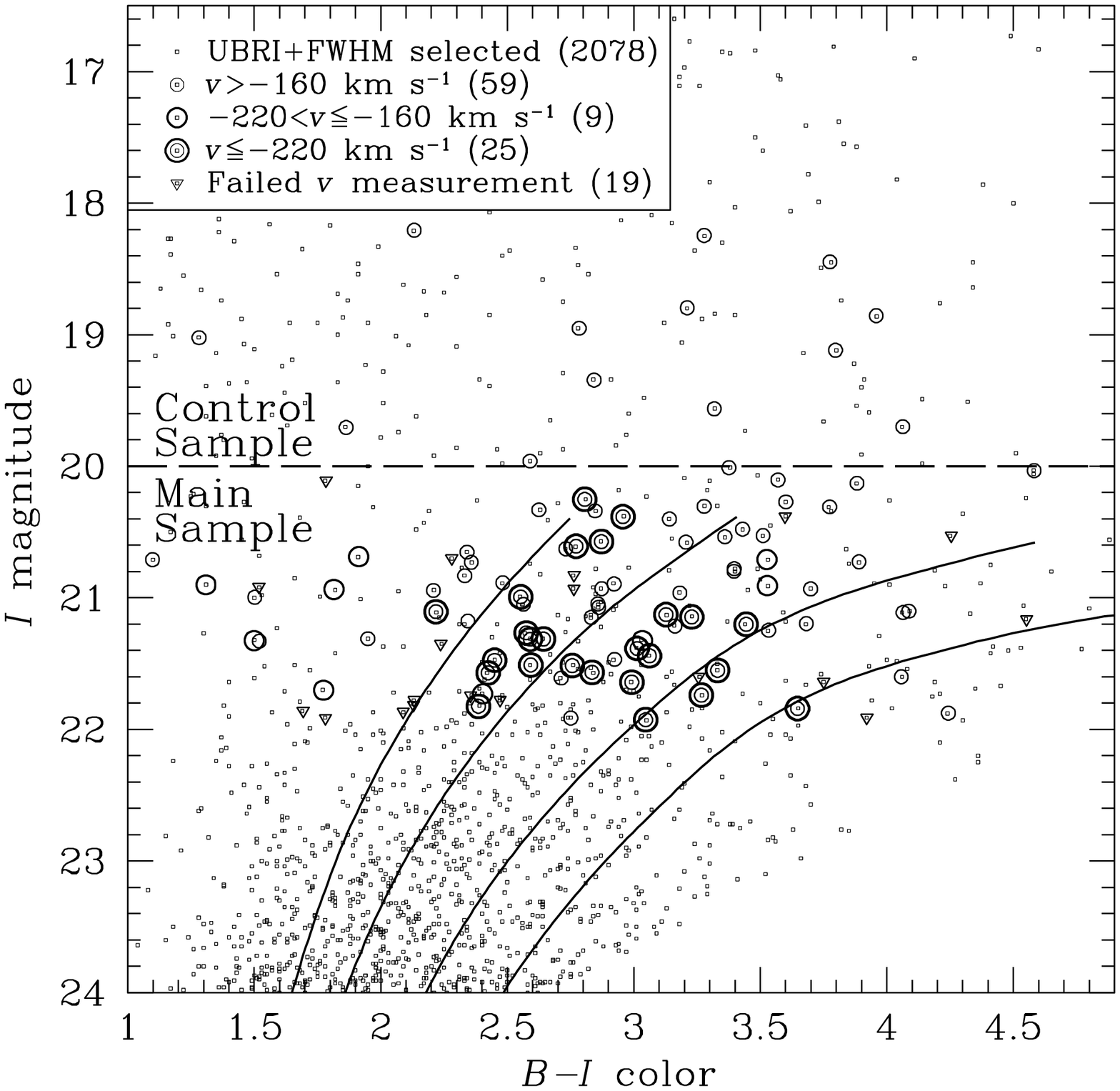}}
\figcaption[Reitzel.fig06.eps]{\label{bi_cmd}{Color-magnitude diagram showing
all $UBRI$- and FWHM-selected stellar candidates from the RGG photometric
study (small squares).  Larger circumscribed symbols indicate spectroscopic
targets with: $v>-160$~km~s$^{-1}$ (thin circles),
$-220<v\leq-160$~km~s$^{-1}$ (bold circles), $v\leq-220$~km~s$^{-1}$
(bold+thin double circles), and those for which there is no reliable velocity
measurement (triangles).  The number of objects in each velocity category is
indicated in parentheses; the number in the $v>-160$~km~s$^{-1}$ category
includes all 13 control sample stars.  Model red giant branch fiducials from
Bergbusch \& VandenBerg (2001) reddened by $\langle{E(B-I)}\rangle=0.13$, the
typical value derived from the Schlegel et~al.\ (1998) dust map over the
field of view of the spectroscopic targets, are plotted as solid lines with
[Fe/H] increasing from left to right: $-2.31$, $-1.41$, $-0.71$, and
$-0.30$~dex.  The $v\leq-220$~km~s$^{-1}$ objects are secure M31 members;
these span the range of colors occupied by the model isochrones indicating a
roughly 2~dex metallicity spread in M31's stellar halo.  The bold and thin
circles occupy increasingly larger $B-I$ color ranges due to increasing
fractions of foreground contaminants in these subsamples.}}

\centerline{\epsfbox{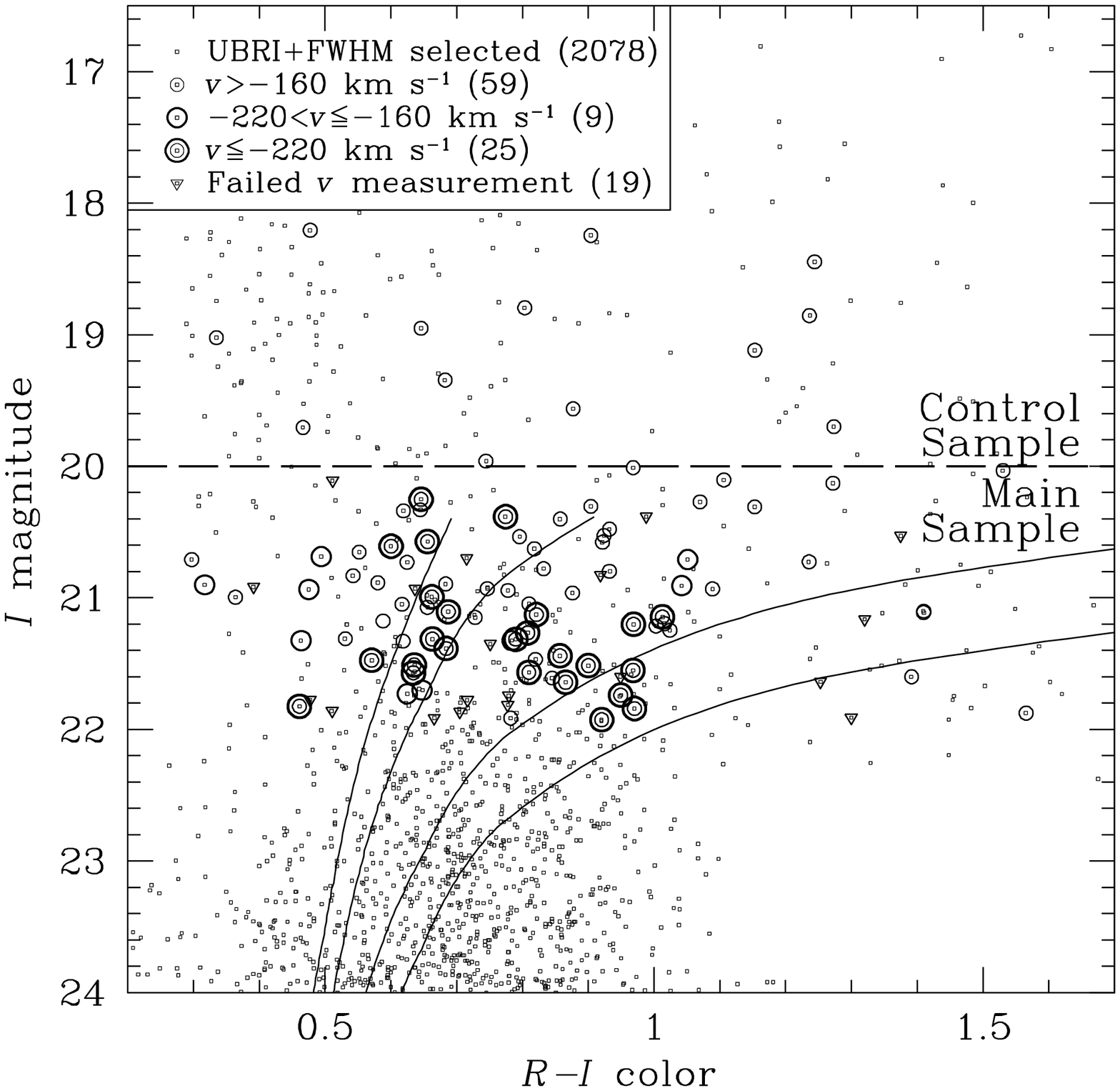}}
\figcaption[Reitzel.fig07.eps]{\label{ri_cmd}{Same as Fig.~\ref{bi_cmd},
except $R-I$ is plotted on the abscissa instead of $B-I$. The location of M31
red giants relative to foreground Milky Way dwarfs and background field
galaxies is expected to be different in ($B-I$, $I$) versus ($R-I$, $I$)
CMDs.  For example, note the difference in the shape and position of the set
of isochrones in the two CMDs in relation to the distribution of foreground
stars with $I<20$.  As expected, the secure sample of M31 giants
($v<-220$~km~s$^{-1}$) is shifted further to the blue (left) in this figure
than in Fig.~\ref{bi_cmd} relative to the rest of the data points.}}

\centerline{\epsfbox{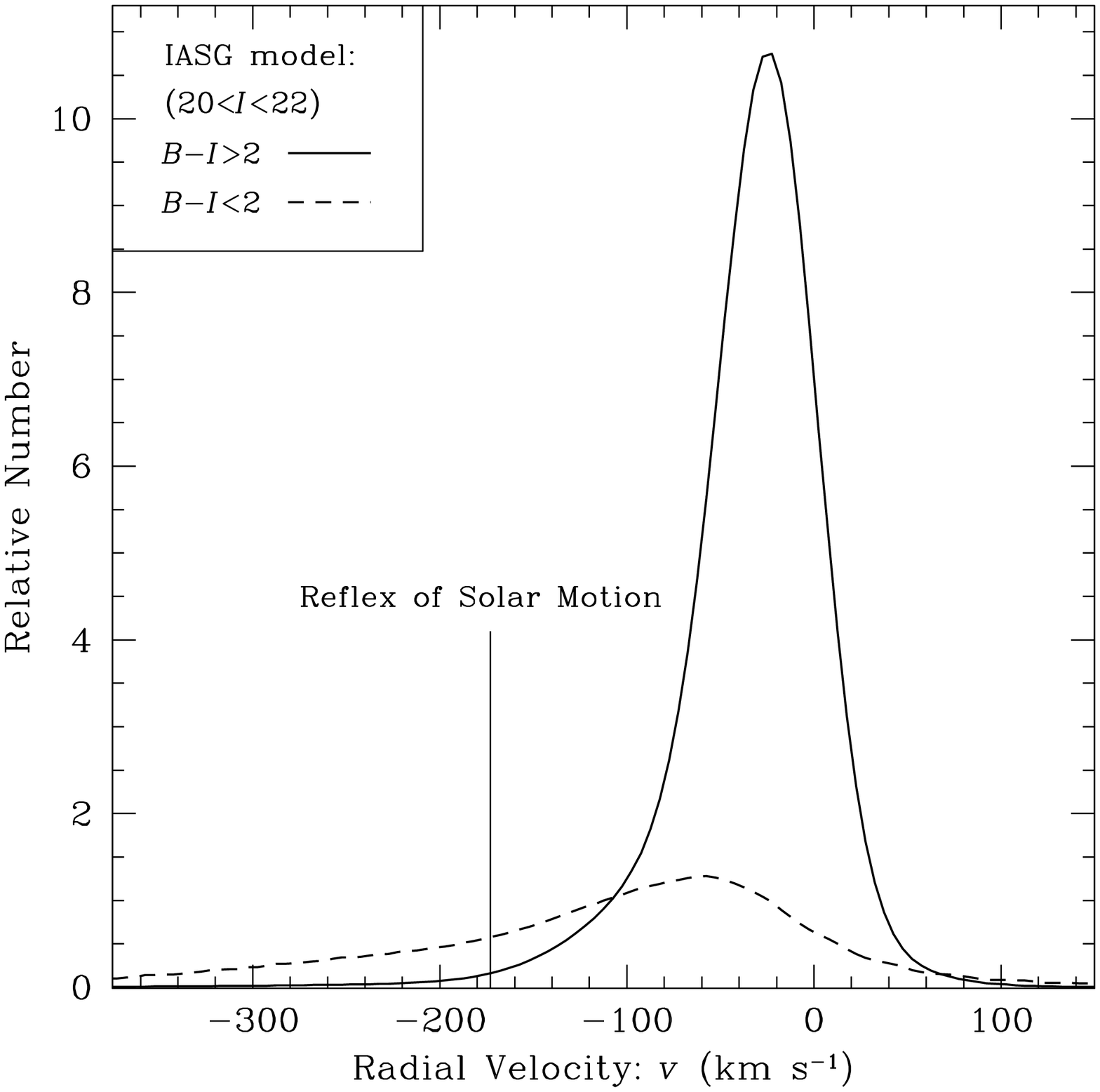}}
\figcaption[Reitzel.fig08.eps]{\label{vhist_mod}{The predicted radial
velocity distribution of foreground Galactic dwarf stars in direction of M31
from the IASG model (Ratnatunga \& Bahcall 1985).  The solid curve is for red
stars ($B-I>2$) and the dashed curve is for blue stars ($B-I<2$), both for
the apparent-magnitude range, $20<I<22$.  The foreground contamination for
the $v<-160$~km~s$^{-1}$ subsamples is almost entirely in the form of turnoff
stars in the Milky Way with $B-I<2$.  The solar reflex motion in the
direction of M31 is indicated (same as in lower panel of Fig.~\ref{vhist}).}}

\centerline{\epsfbox{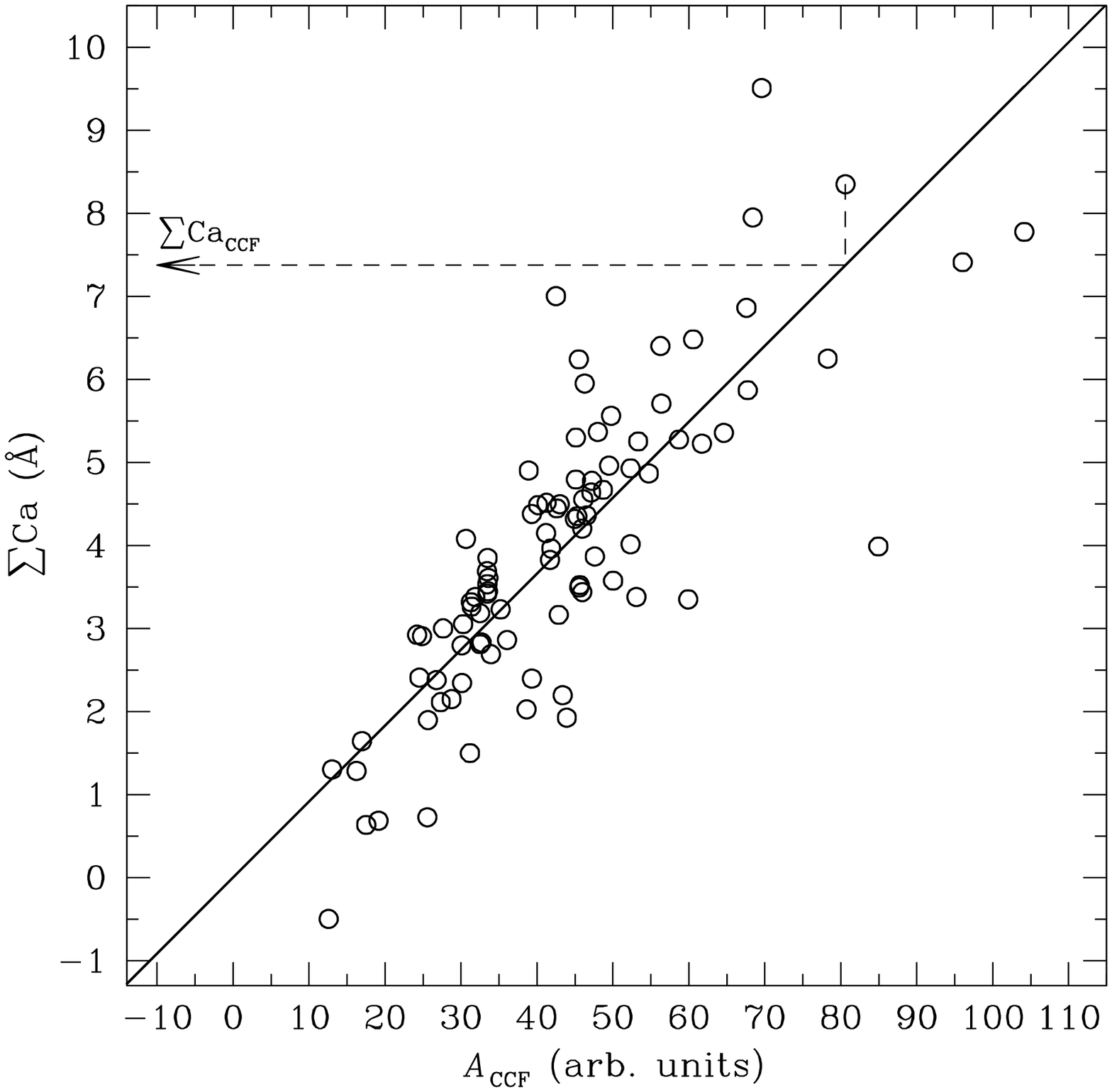}}
\figcaption[Reitzel.fig09.eps]{\label{ew_vs_ccfarea}{The weighted, summed
Ca\,{\smcap ii} line strength, $\rm\Sigma Ca$, plotted against the strength of
(area under) the cross correlation peak, $A_{\rm CCF}$.  Tests show that the
$A_{\rm CCF}$ statistic is less noisy than the directly measured line
strength, $\rm\Sigma{Ca}$, so we prefer to use the former for metallicity
determination.   The dashed lines and arrow demonstrate how the {\it
effective\/} summed line strength, $\rm\Sigma{Ca}_{CCF}$, is calculated from
$A_{\rm CCF}$ using the mean empirical relation shown as a solid line
(\S\,\ref{feh_spec_ccf_sec}).}}

\centerline{\epsfbox{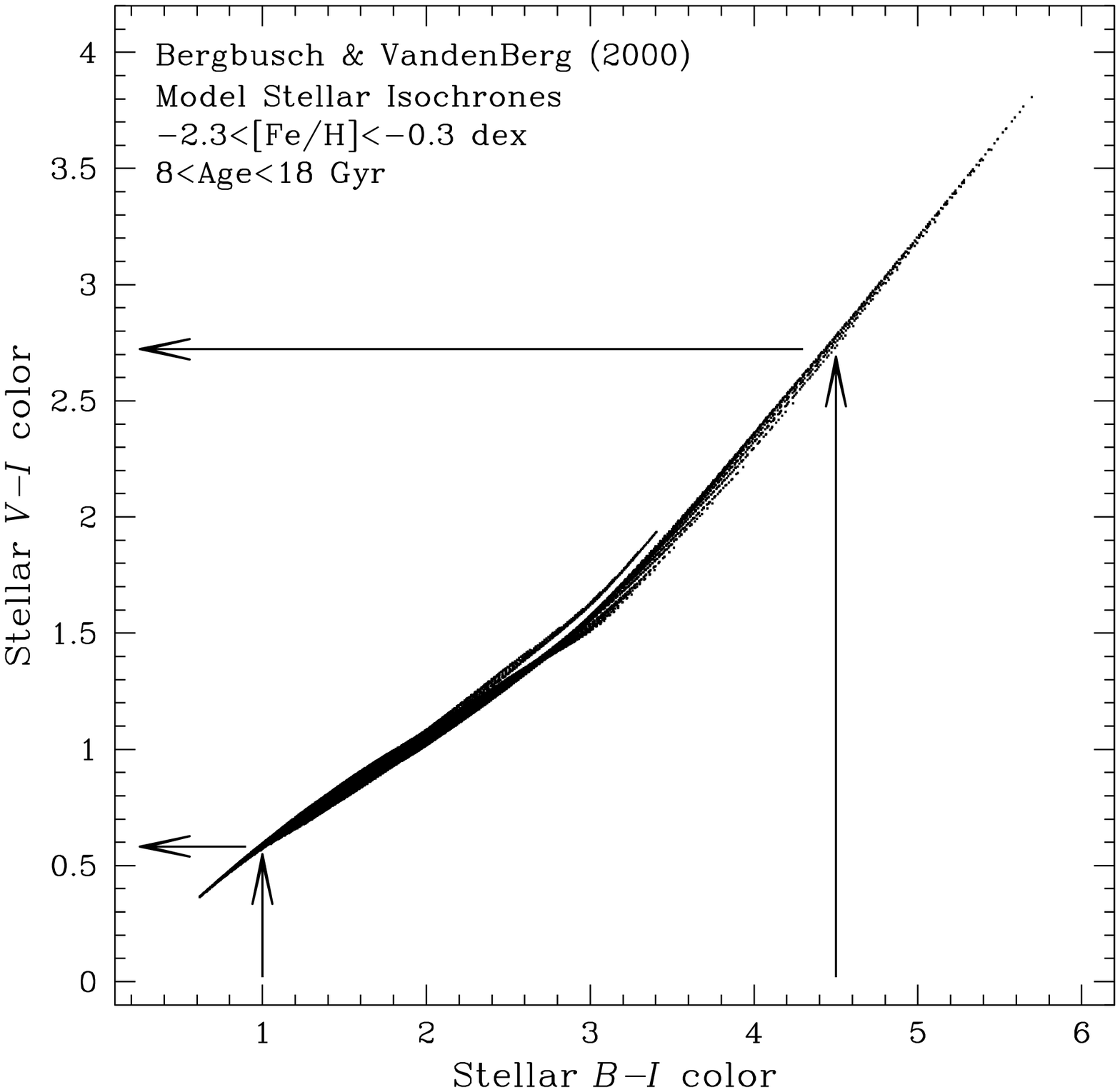}}
\figcaption[Reitzel.fig10.eps]{\label{bvi_iso}{Color-color diagram of $V-I$
versus $B-I$ for the Bergbusch \& VandenBerg (2001) model stellar isochrones
with metallicities in the range $\rm-2.31\leq[Fe/H]\leq-0.3$~dex and ages in
the range $\rm8.0\leq{t}\leq18.0$~Gyr. The arrows demonstrate how the
apparent $V_{\rm synth}$ magnitude of a star is estimated by interpolating
between the measured (and subsequently dereddened) $B$ and $I$ apparent
magnitudes.  The range of $B-I$ over which this interpolation is carried out
is spanned by the two~pairs of arrows.  Even though the model isochrones span
a wide range of metallicity and age, they map onto a well-defined, narrow
locus in the ($V-I,~ B-I$) plane.}}

\centerline{\epsfxsize=7in \epsfysize=9in
\epsfbox{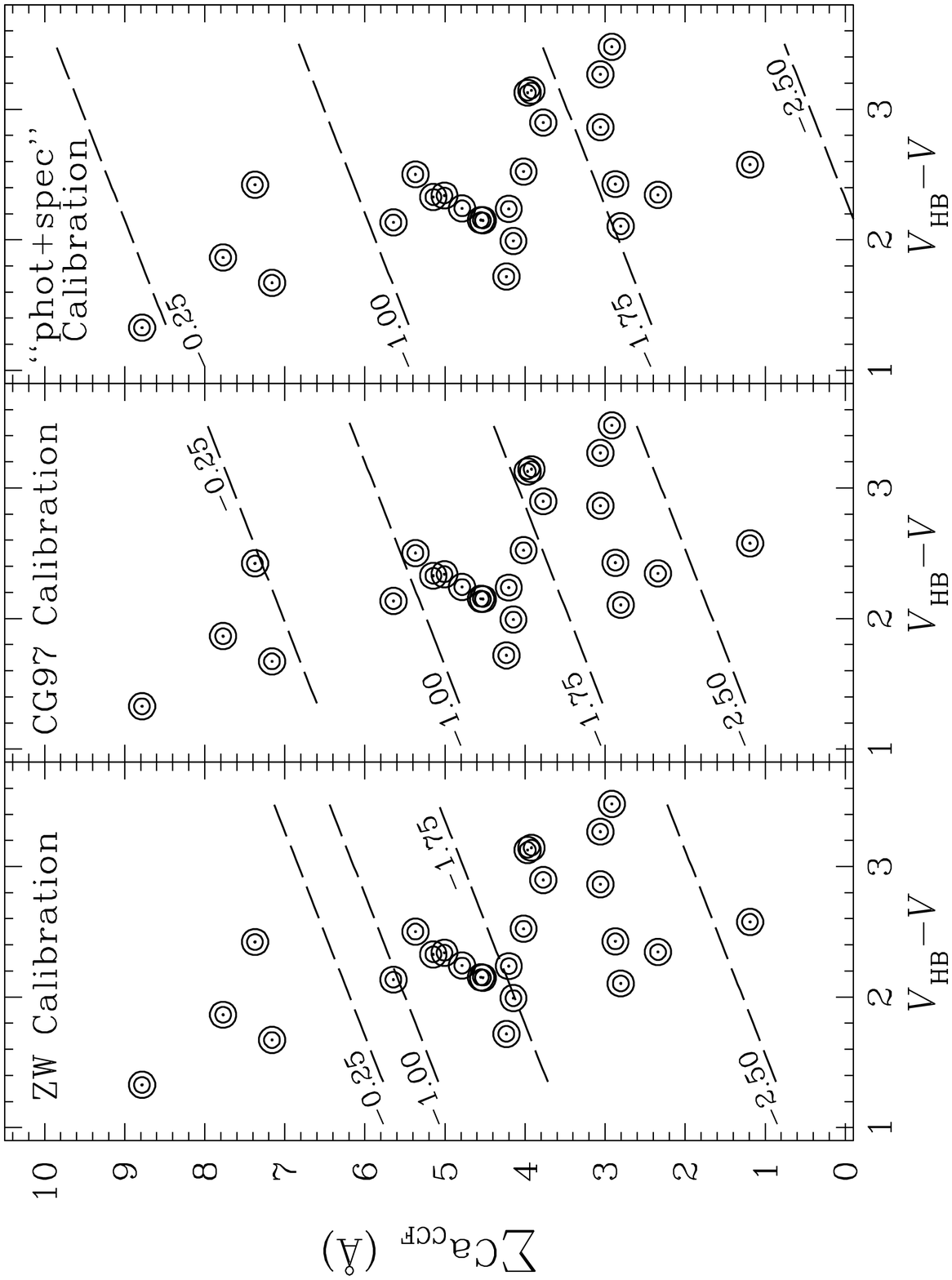}}
\figcaption[Reitzel.fig11.eps]{\label{w_vs_vhb}{The effective weighted sum
of the equivalent widths of the Ca\,{\smcap ii} lines, $\rm\Sigma{Ca}_{CCF}$,
plotted versus the ``distance'' (in mag) above the level of the horizontal
branch, $V_{\rm HB}-V$, for objects with $v\leq-220~\rm km~s^{-1}$.  The
quantity $\rm\Sigma{Ca}_{CCF}$ is derived from the strength of the
cross-correlation peak using an empirical scaling between this quantity and
the directly-measured Ca\,{\smcap ii} equivalent width
(\S\,\ref{feh_spec_ccf_sec}).  Dashed lines of constant [Fe/H] are shown for
the Zinn-West and Carretta-Gratton metallicity scales (left and middle
panels, respectively), and for an empirical calibration of the reduced
Ca\,{\smcap ii} equivalent width based on photometric metallicity estimates
dubbed $\rm[Fe/H]_{\rm phot+spec}$ (right panel).  This empirical
``phot+spec'' calibration of [Fe/H] is merely put forward to demonstrate that
the Ca\,{\smcap ii} line strength is tightly correlated with the photometric
[Fe/H] estimate for M31 giants (\S\,\ref{feh_comp_sec});
$\rm[Fe/H]_{phot+spec}$ is {\it not\/} to be treated as a `true' metallicity
scale or as an alternative to ZW and CG97 calibrations.}}

\centerline{\epsfbox{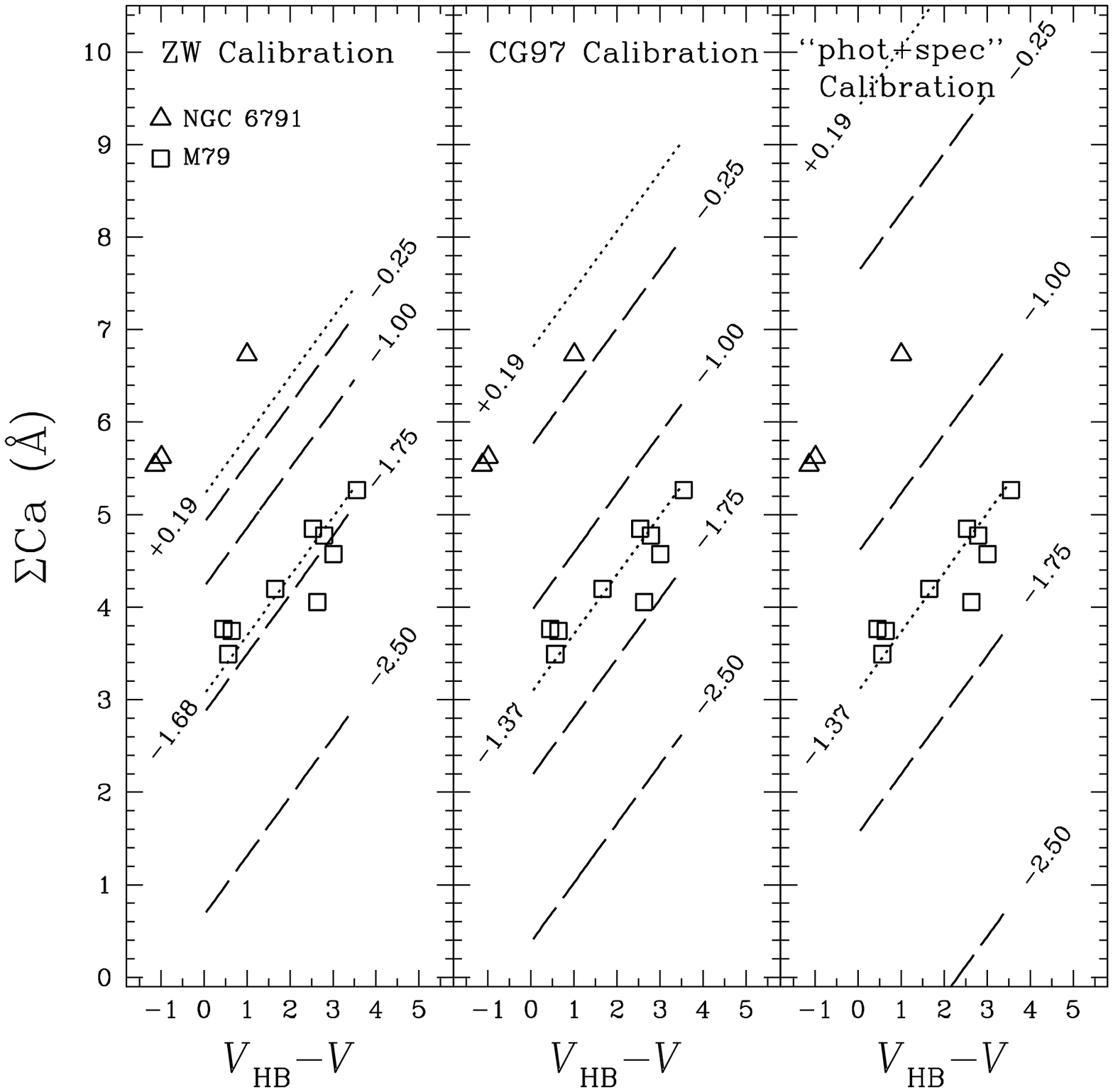}}
\figcaption[Reitzel.fig12.eps]{\label{w_vs_vhb_calib}{Same as
Fig.~\ref{w_vs_vhb} for red giants in two Galactic star clusters of known
metallicity that are used to check the calibration of the spectroscopic
[Fe/H] scale: M79 with $\rm[Fe/H]_{ZW}=-1.68$~dex and
$\rm[Fe/H]_{CG97}=-1.37$~dex (squares) and NGC~6791 with
$\rm[Fe/H]_{ZW}=+0.19$~dex (triangles).  There is no [Fe/H] measurement for
NGC~6791 on the Carretta-Gratton scale, so the Zinn-West value of $+0.19$~dex
is used in all three panels.  The CG97 value for M79 is adopted for the
empirical ``phot+spec'' calibration.  The dotted lines indicate the
metallicity of M79 and NGC~6791 for each calibration, marking the expected
location of cluster red giants in the plots.  The large discrepancy between
the expected versus observed location of NGC~6791 giants in the ``phot+spec''
panel indicates that this empirical calibration relation is inaccurate at the
high-metallicity end.}}

\centerline{\epsfxsize=7.5in \epsfysize=9in
\epsfbox{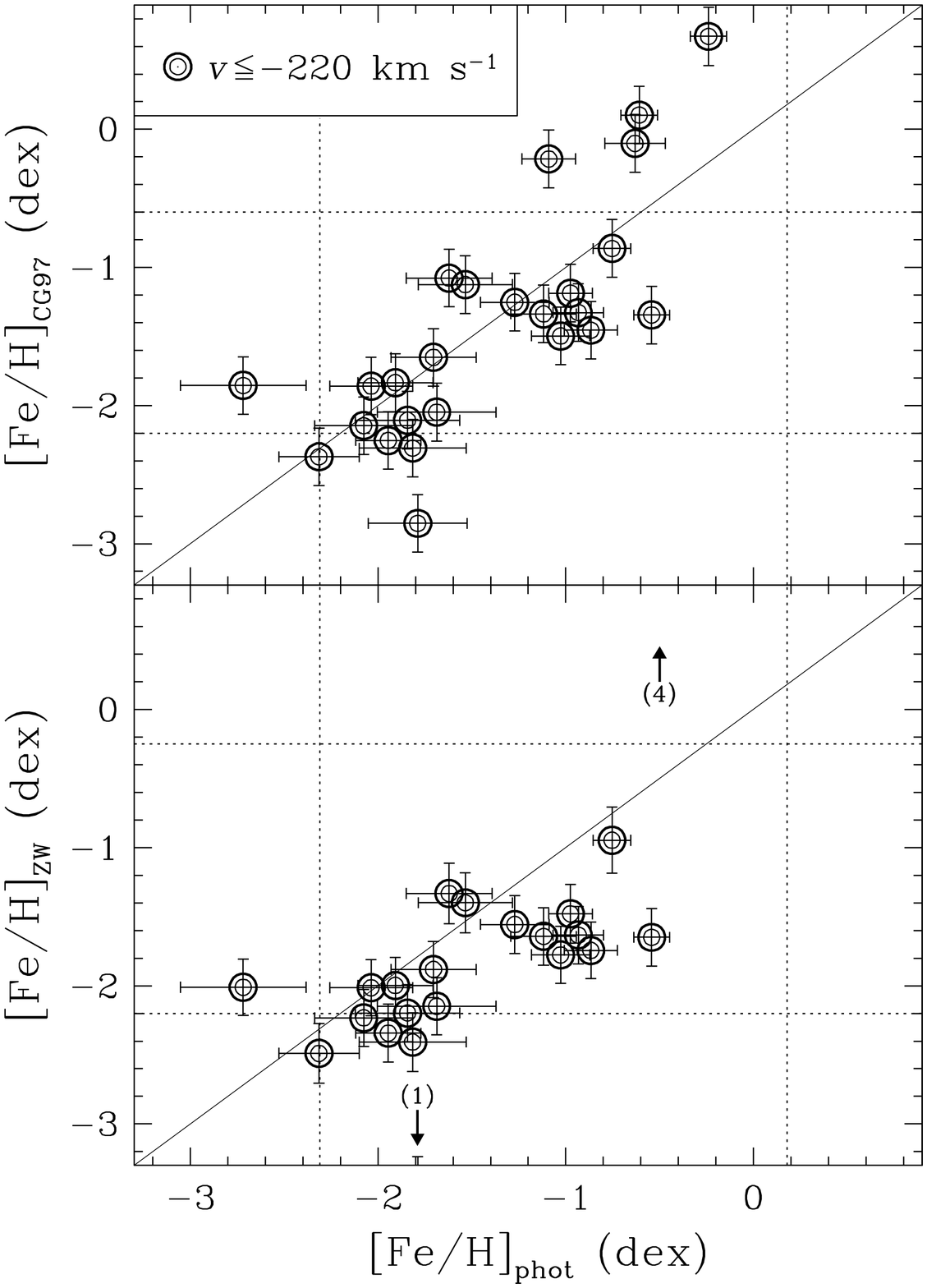}}
\figcaption[Reitzel.fig13.eps]{\label{fehcomp_mem}{A star-by-star comparison
of the photometric versus spectroscopic [Fe/H] estimates for stars with
$v\leq-220~\rm km~s^{-1}$.  The Carretta-Gratton and Zinn-West calibrations
of the spectroscopic metallicity scales are shown in the upper and lower
panels, respectively.  The photometric estimate is based on the location of
the star in the ($B-I$, $I$) CMD.  Also indicated are $1\sigma$ errors in
each of the three~[Fe/H] estimates (\S\S\,\ref{feh_phot_uncert_sec} and
\ref{feh_spec_uncert_sec}).  Four~objects (possible M31 disk stars/metal-rich
debris---\S\,\ref{poss_disk_sec}) are above the upper end and one~object is
below the lower end of the displayed $y$-axis range in the lower panel (see
arrows).  The calibration range for each method is indicated by dotted lines.
The $\rm[Fe/H]_{CG97}$ estimate correlates reasonably well with
$\rm[Fe/H]_{phot}$ over a wide range of metallicities; $\rm[Fe/H]_{ZW}$ tends
to be somewhat lower than $\rm[Fe/H]_{phot}$ within the calibrated range and
substantially higher than $\rm[Fe/H]_{phot}$ beyond the upper end of the
calibrated range where the cubic ZW relation diverges.}}

\centerline{\epsfxsize=7.5in \epsfysize=9in
\epsfbox{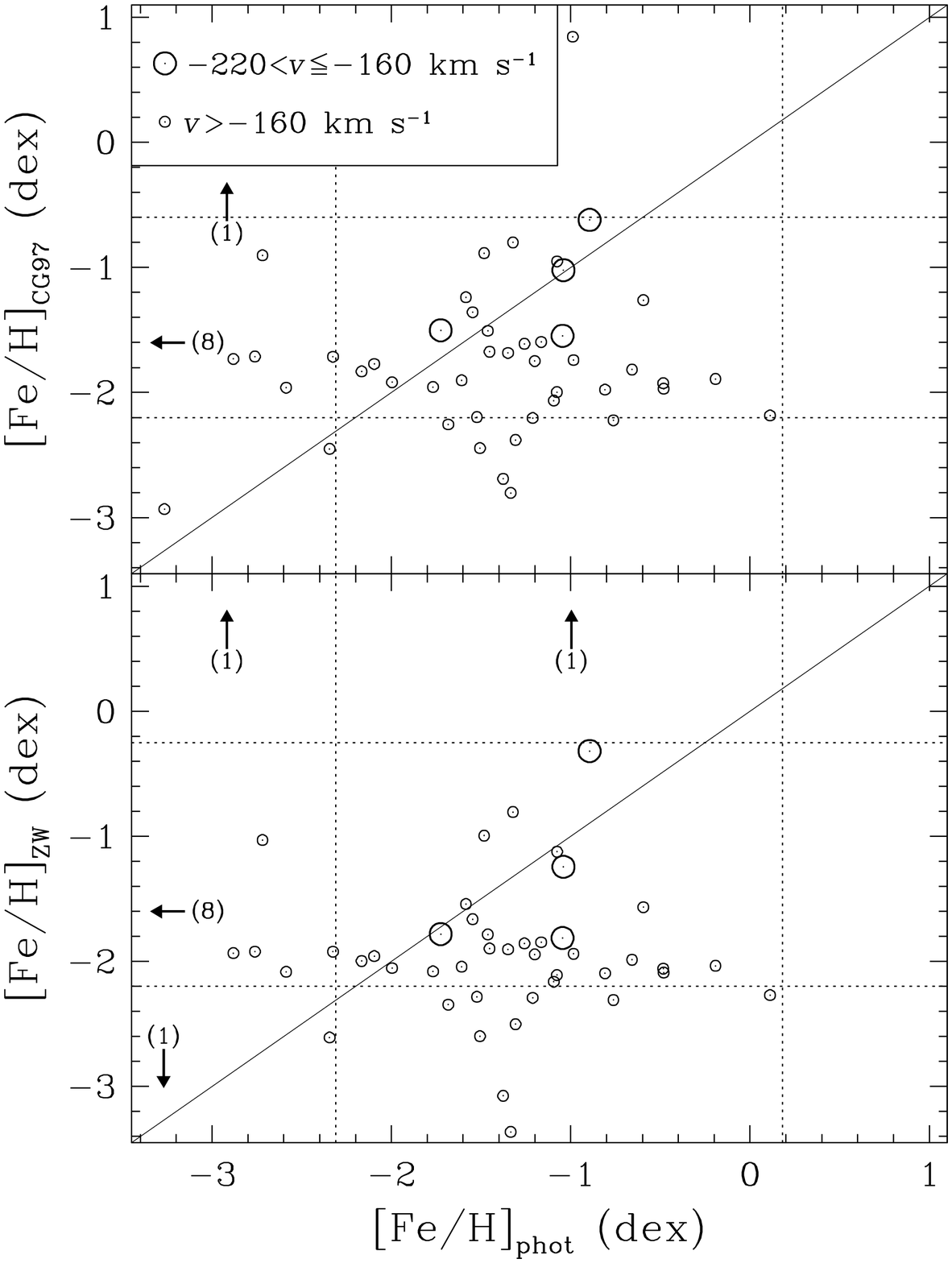}}
\figcaption[Reitzel.fig14.eps]{\label{fehcomp_nmem}{Same as
Fig.~\ref{fehcomp_mem} for objects with $v>-160$~km~s$^{-1}$ (small thin
circles) and $-220<v\leq-160$~km~s$^{-1}$ (large bold circles).  Objects
which lie beyond the displayed range of the plots are are indicated by arrows
with the number of objects in parentheses.  Note the lack of an overall
correlation in this plot in contrast to Fig.~\ref{fehcomp_mem}.  It is likely
that a handful of M31 red giants ($<10$) are hidden in this plot, and while
they might lie close to the $x=y$ line, their correlation is swamped by the
vast majority of contaminating foreground Milky Way stars which are
understandably uncorrelated.  In particular, the four~stars with
$-220<v\leq-160$~km~s$^{-1}$ that appear in this plot (red subset) are
expected to be M31 red giants, and it is reassuring that these lie relatively
close to the $x=y$ line.}}

\centerline{\epsfbox{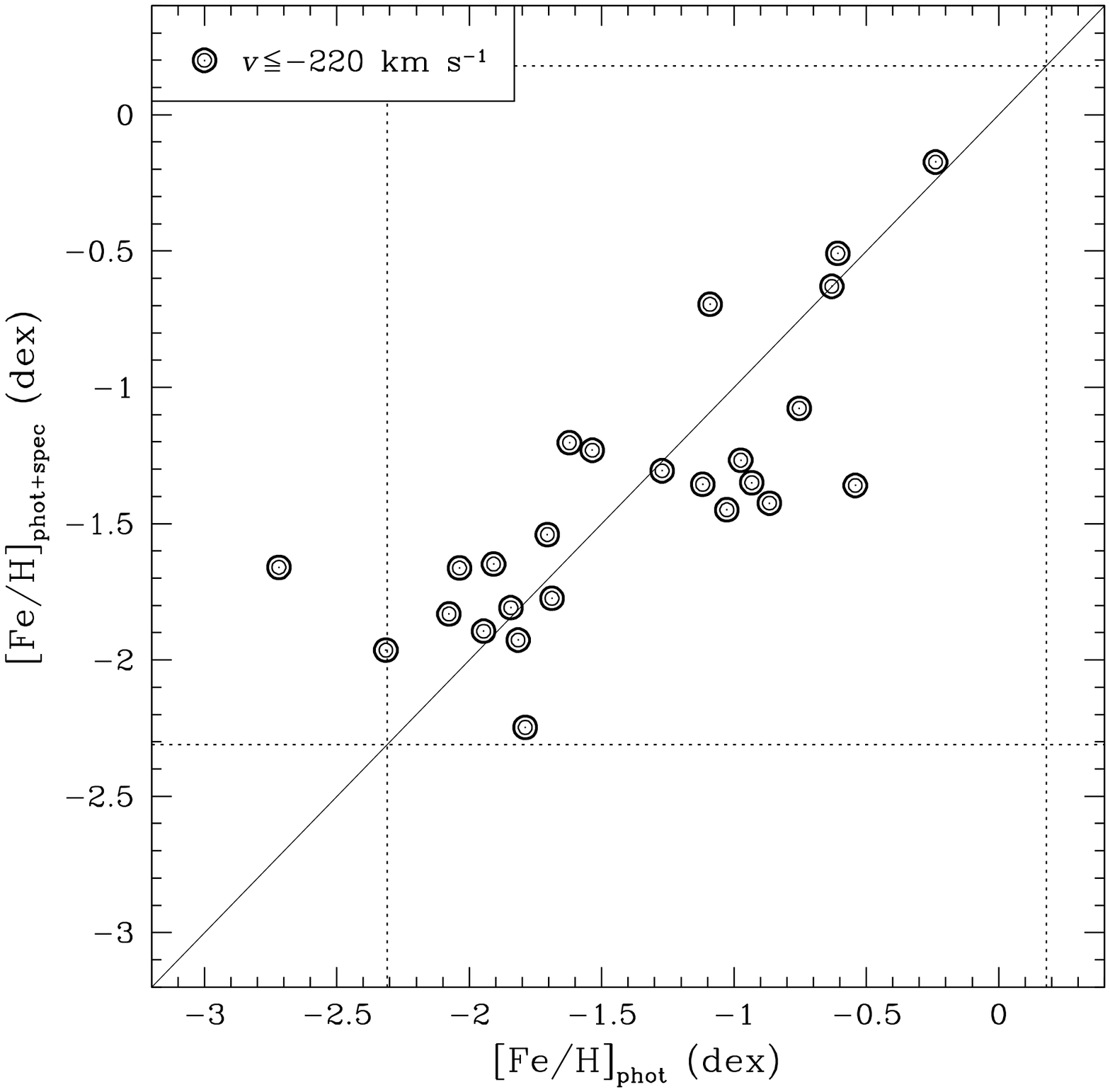}}
\figcaption[Reitzel.fig15.eps]{\label{feh_photspec}{Spectroscopic estimate of
the metallicity using the empirical ``phot+spec'' calibration relation,
$\rm[Fe/H]_{phot+spec}$ (\S\,\ref{feh_comp_sec}), plotted versus the
photometric estimate of the metallicity, $\rm[Fe/H]_{phot}$, for stars with
$v\leq-220$~km~s$^{-1}$.  Calibration ranges for the two methods are
indicated by the dotted lines.  The clear trend seen in this plot indicates
that the Ca\,{\smcap ii} line strength of a star correlates reasonably well
with its location in the ($B-I$, $I$) CMD.}}

\centerline{\epsfbox{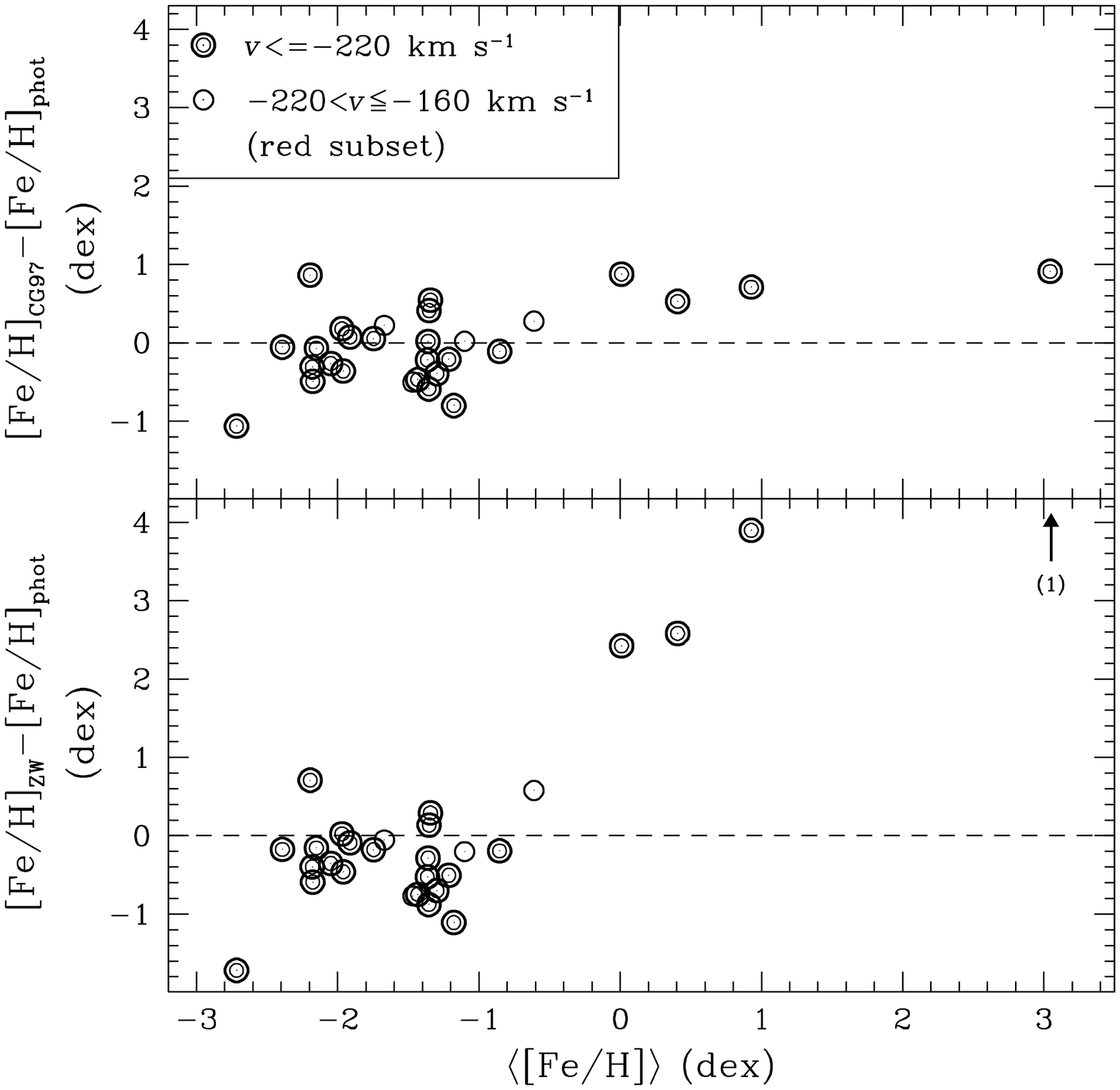}}
\figcaption[Reitzel.fig16.eps]{\label{feh_diff}{({\it Upper
panel\/})~Difference between the spectroscopic metallicity estimate on the
Carretta-Gratton scale and the photometric estimate plotted versus
$\rm\langle[Fe/H]\rangle$, the average of three~metallicity estimates, two
spectroscopic (CG97 and ZW) and one photometric, for the secure sample of
29~M31 red giants.  This mean [Fe/H] value is only used to provide a common
scale on which to rank stars in our comparison of the different metallicity
measurement methods.  There are four~stars with solar/super-solar
$\rm\langle[Fe/H]\rangle$ values that lie systematically above the zero line
(\S\,\ref{poss_disk_sec}).~~~
({\it Lower panel\/})~Same as upper panel, except the difference is with
respect to the spectroscopic metallicity estimate on the Zinn-West scale.
The extrapolation of the cubic ZW relation diverges at high metallicities
beyond the calibrated range; the arrow indicates that one star is located
beyond the upper end of the displayed $y$-axis range.}}

\centerline{\epsfxsize=7.5in \epsfysize=9in
\epsfbox{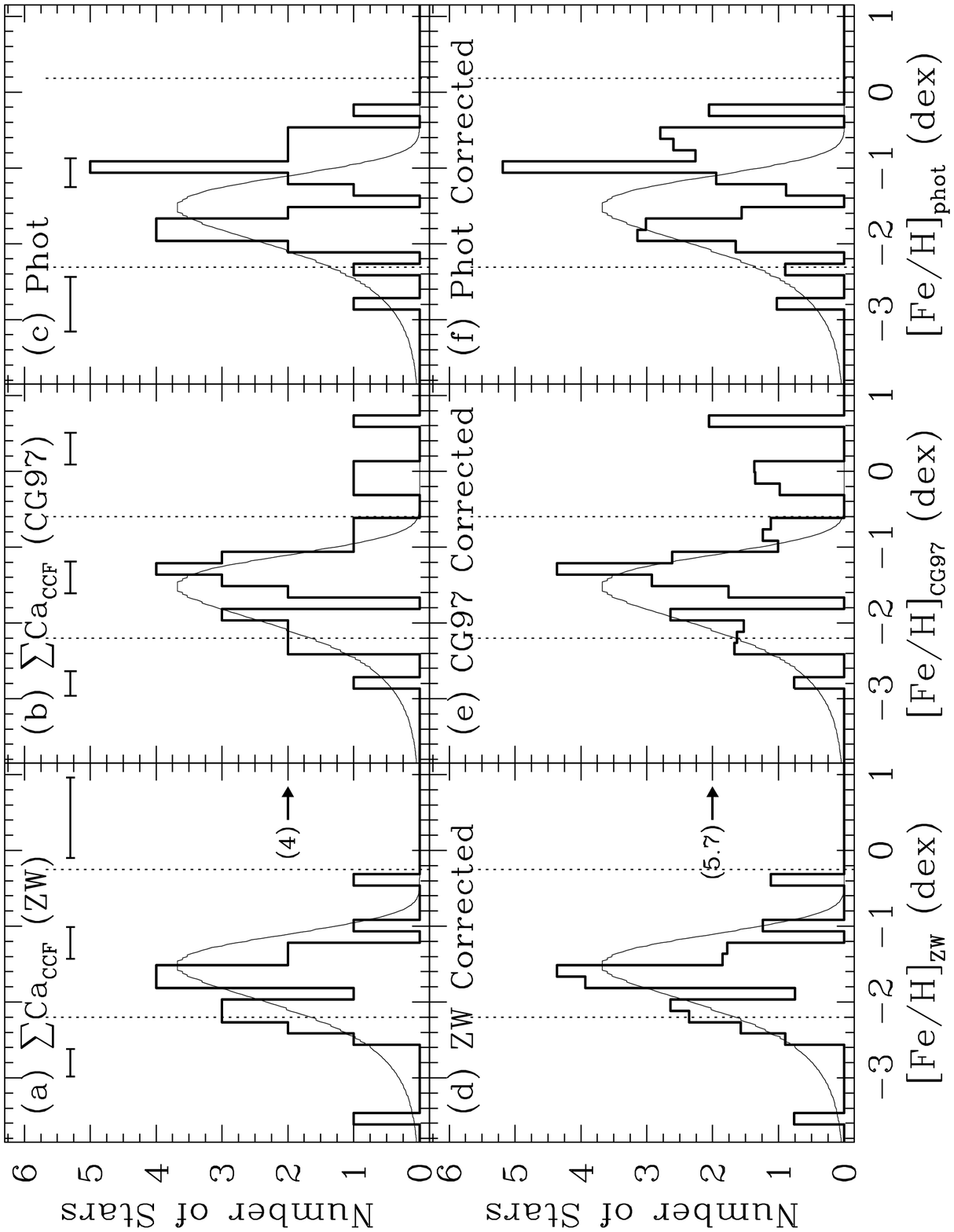}}
\figcaption[Reitzel.fig17.eps]{\label{feh_hist}{({\it a\/})--({\it
c\/})~Metallicity distribution (uncorrected) for the secure sample of 29 M31
red giants on the spectroscopic Zinn-West and Carretta-Gratton scales and
photometric scales, respectively (bold histograms).  The calibration range
for each method is indicated by the vertical dotted lines.  The typical
$1\sigma$ measurement errors in [Fe/H] for each method for objects inside and
outside the calibrated ranges are shown above each histogram.  The [Fe/H]
distribution for a steady gas loss model with $\rm[Fe/H]_0=-10.0$~dex and
$\rm\langle[Fe/H]\rangle=-1.5$~dex is shown as a thin solid curve; it is a
reasonably good match to the observations.  All three metallicity scales show
a large spread with tails extending toward the low and high ends of the
distribution.~~~
({\it d\/})--({\it f\/})~Same as ({\it a\/})--({\it c\/}), except the
metallicity distributions are corrected for the bias against high-metallicity
stars caused by the drop in $I_{\rm TRGB}$ and incompleteness at the faint
end of the secure sample.  Each star is weighted by the inverse of the
selection efficiency function (solid line in Fig.~\ref{sel_eff}) and the
histograms are normalized by average weight per star in the sample so as to
preserve the total area at the actual number of stars in the secure M31
sample, 29 (\S\,\ref{selbias}).  The arrows in the left panels indicates that
there are 4~stars beyond the upper end of the displayed $\rm[Fe/H]_{ZW}$
range, or 5.7~stars after the metallicity-bias correction.}}

\centerline{\epsfbox{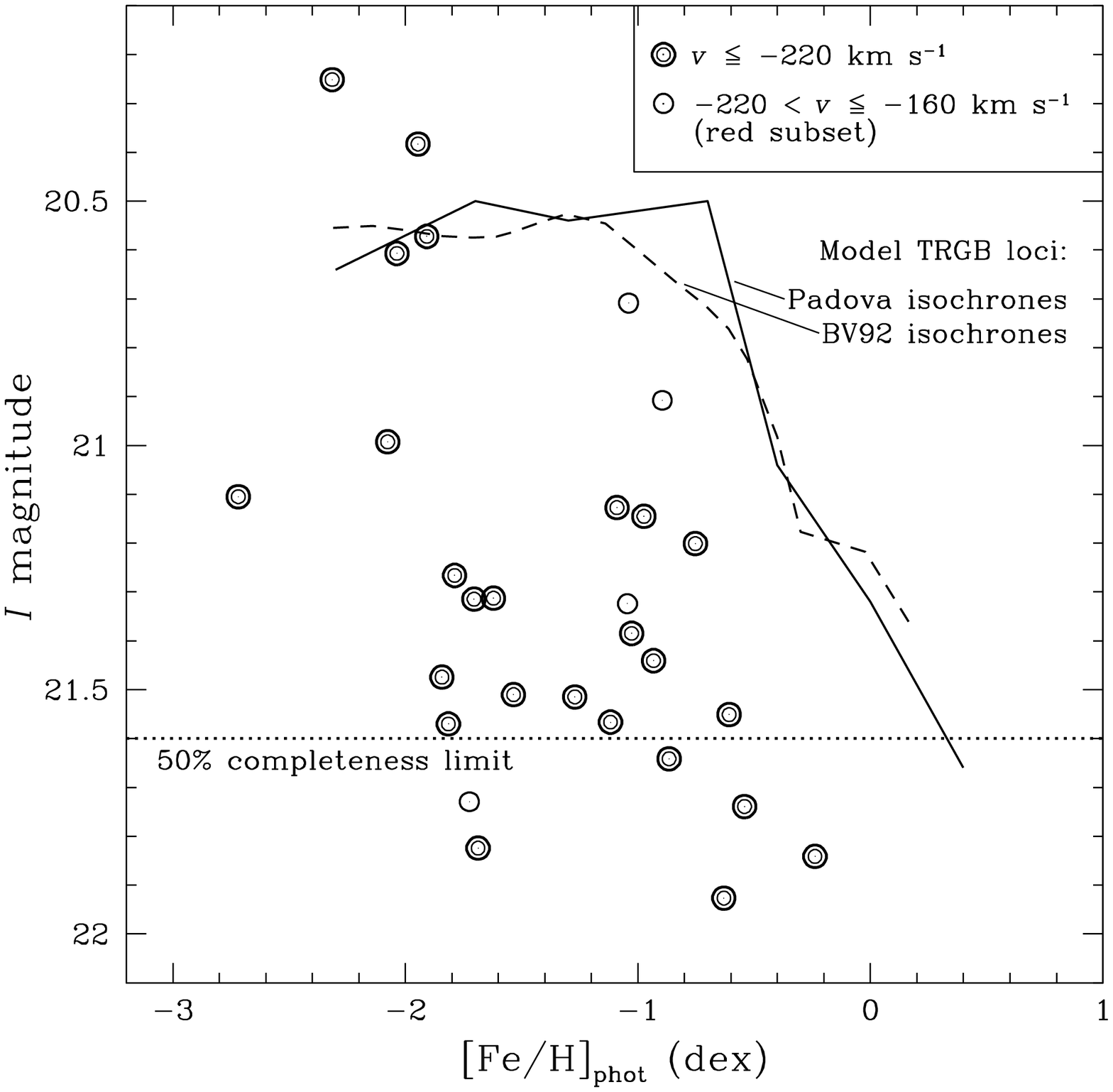}}
\figcaption[Reitzel.fig18.eps]{\label{i_vs_feh}{Apparent $I$-band magnitude
versus photometric metallicity estimate for the secure M31 giant sample.
Double circles indicate objects with $v\leq-220$~km~s$^{-1}$; open circles
represent the red subset of objects with $-220<v\leq-160$~km~s$^{-1}$.  The
expected locus of the tip of M31's red giant branch is derived from the
Padova group's (Bertelli et~al.\ 1994; Girardi et~al.\ 1996) model isochrones
(solid line) and Bergbusch \& VandenBerg (1992) model isochrones (dashed
line).  The brightness of the tip of the RGB drops significantly for
$\rm[Fe/H]\gtrsim-1$~dex, and this tends to bias any magnitude-limited
sample against high-metallicity red giants.  The dotted horizontal line marks
the estimated 50\% completeness limit for the secure sample
(\S\,\ref{selbias}).}}

\centerline{\epsfbox{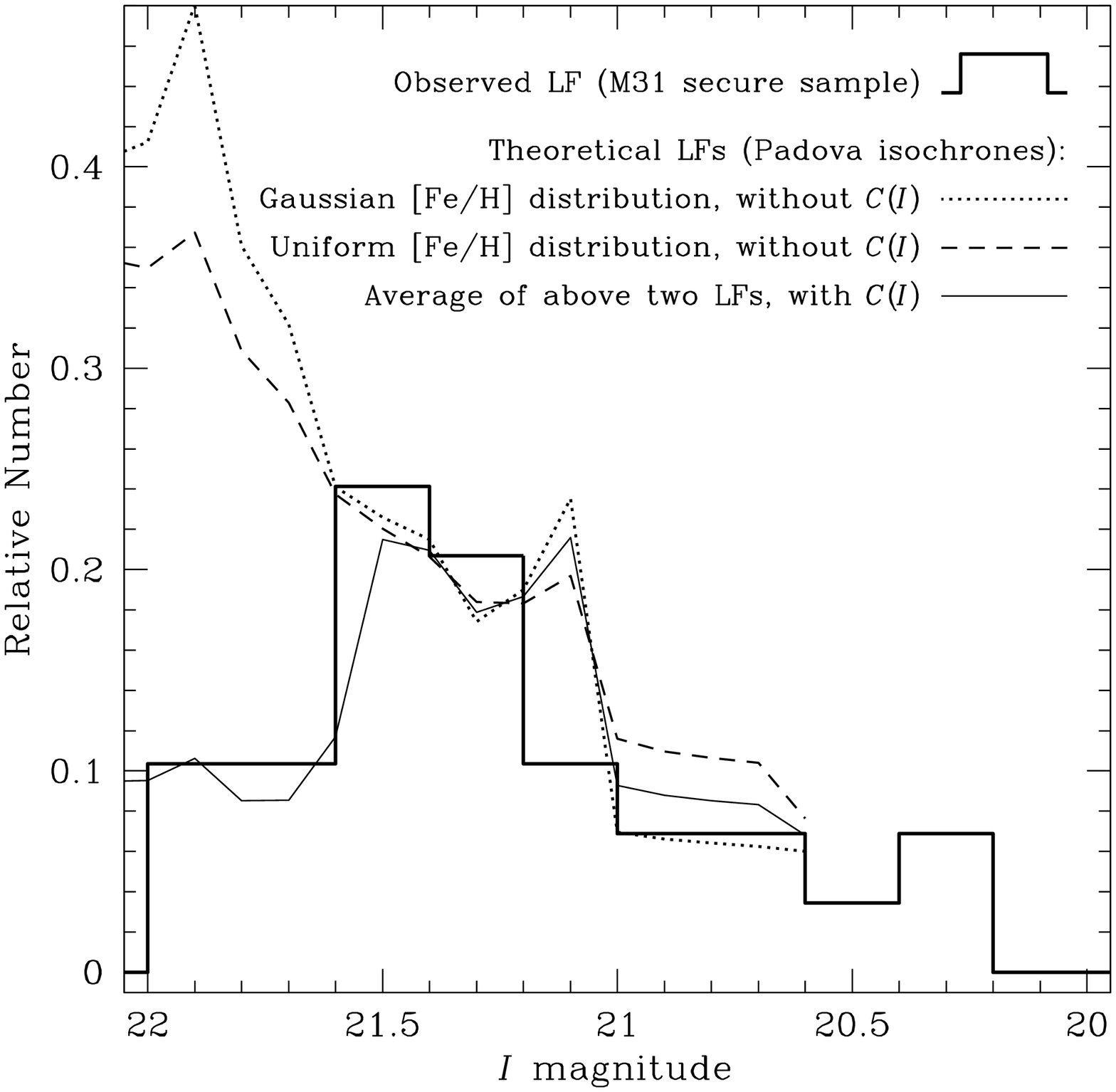}}
\figcaption[Reitzel.fig19.eps]{\label{lum_func}{Distribution of $I$-band
magnitudes for the secure M31 giant sample (bold histogram) compared to
theoretical red giant luminosity functions derived from the Padova group's
isochrones (Bertelli et~al.\ 1994; Girardi et~al.\ 1996) for two assumed
[Fe/H] distributions: Gaussian, comparable to that observed in M31's halo
(dotted line), and uniform over the range $\rm-2.3<[Fe/H]<+0.4$~dex (dashed
line).  The model LFs are normalized to match the data at the bright end.
The thin solid line is the average of these two model LFs with a faint-end
exponential cutoff applied to mimic the decrease in completeness fraction of
the secure sample with decreasing brightness; it is a good match to the
observed distribution (see \S\,\ref{selbias}).}}

\centerline{\epsfbox{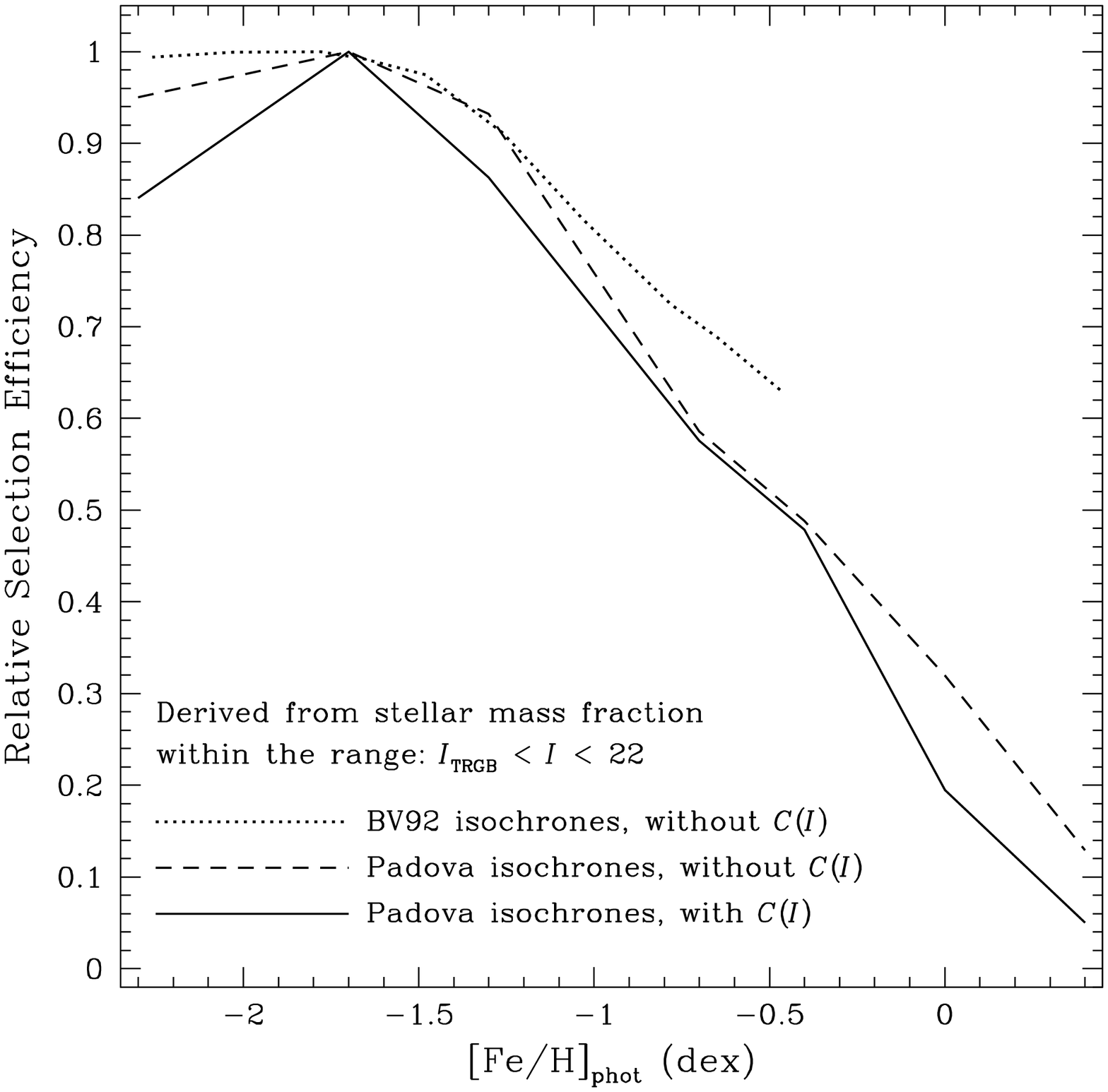}}
\figcaption[Reitzel.fig20.eps]{\label{sel_eff}{Relative selection efficiency
as a function of $\rm[Fe/H]_{phot}$ for the secure sample of M31 red giants.
The selection efficiency is derived from the mass fraction in the
apparent-magnitude range, $20<I<22$, or effectively, $I_{\rm TRGB}<I<22$, in
theoretical isochrone population functions.  The solid line is based on the
Padova group's isochrones (Bertelli et~al.\ 1994; Girardi et~al.\ 1996)
accounting for the fact that the completeness level of the secure M31 sample
falls off toward $I=22$ (\S\,\ref{selbias}); the reciprocal of this is used
to correct for the bias against high-metallicity stars.  For comparison, the
dashed line also represents a calculation based on the Padova group's
isochrones but with no correction for faint-end incompleteness [$C(I)=1$],
while the dotted line is based on the Bergbusch \& VandenBerg (1992)
isochrones and also ignores the completeness correction.  The selection
efficiency is normalized to unity at low metallicities where it is roughly
constant; it drops linearly with increasing metallicity for
$\rm[Fe/H]\gtrsim-1.5$~dex to a value of about~0.5 at $\rm[Fe/H]=-0.5$~dex.
The drop in selection efficiency toward high [Fe/H] is slightly steeper if
the completeness function is included in the calculation (solid vs.\ dashed
lines), and significantly steeper for the Padova isochrones than the
Bergbusch \& VandenBerg isochrones (dashed vs.\ dotted lines).  Details of
the [Fe/H] selection efficiency calculation are in \S\,\ref{selbias}.}}

\centerline{\epsfxsize=7.5in \epsfysize=9in
\epsfbox{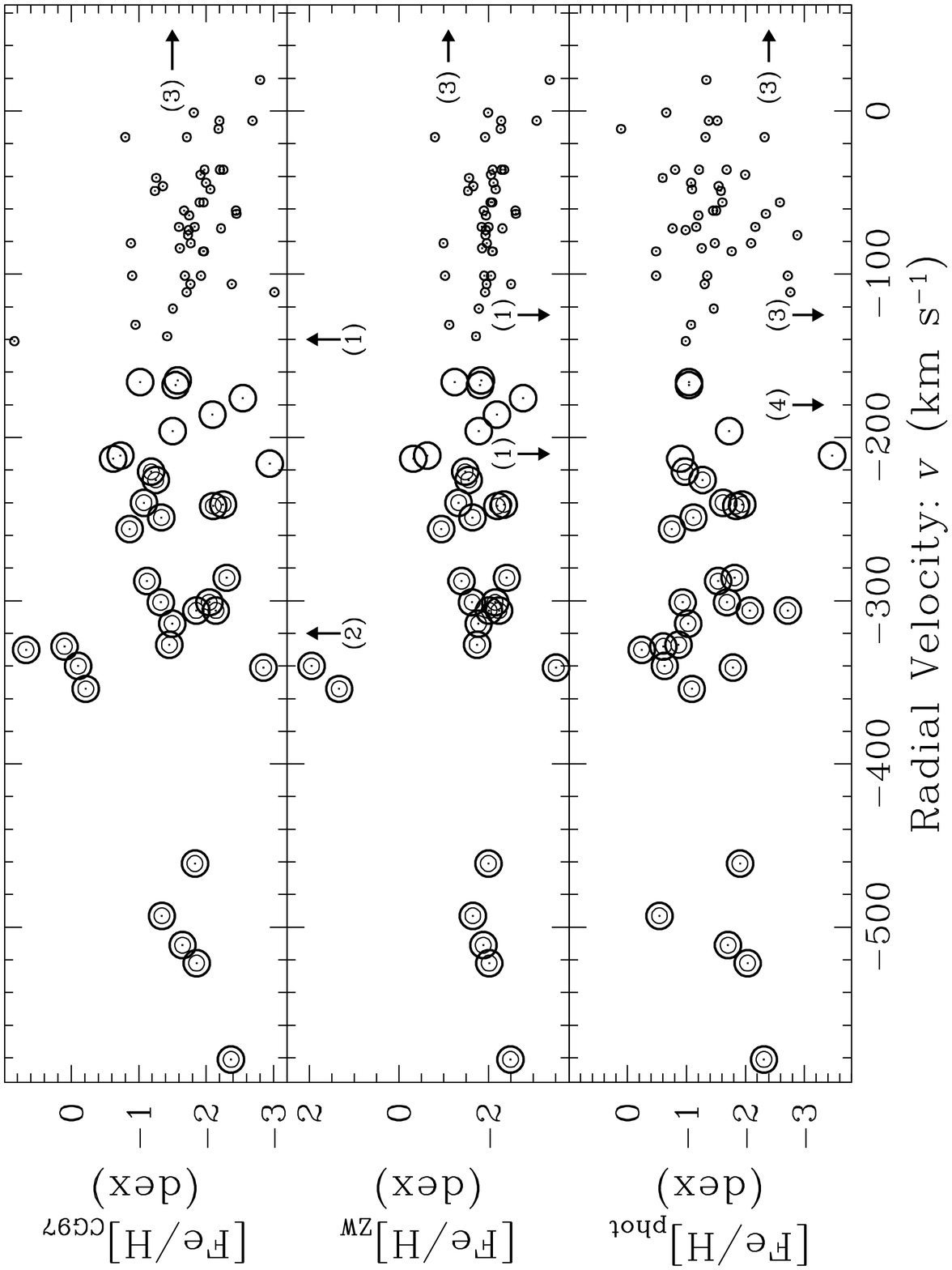}}
\figcaption[Reitzel.fig21.eps]{\label{v_vs_feh}{Metallicity on the
spectroscopic Carretta-Gratton (top) and Zinn-West (middle) scales and
photometric scale (bottom) plotted versus radial velocity:
$v>-160$~km~s$^{-1}$ (small circles), $-220<v\leq-160$~km~s$^{-1}$ (large
circles), and $v\leq-220$~km~s$^{-1}$ (large double circles).  Objects lying
beyond the limits of the plot are indicated by an arrow and the number of such
objects in each velocity category; of the three~stars beyond the right edge,
one lies beyond the upper end of the displayed $y$-axis range in the top and
middle panels, while another lies beyond the lower end of the $y$ range in the
middle panel.  There are four~high-metallicity M31 red giants in the top panel
with $v\approx-340$~km~s$^{-1}$.  They are near the upper end of the
$\rm[Fe/H]_{phot}$ distribution and are the 3rd--6th most metal-rich on the
ZW and CG97 scales (the two~objects with stronger Ca\,{\smcap ii} lines are
probably {\it not\/} M31 halo giant stars).  The velocities of these
four~metal-rich giants are consistent with that expected for M31 disk stars
on the minor axis, $v\approx{v}_{\rm sys}^{\rm M31}=-297$~km~s$^{-1}$.}}

\centerline{\epsfbox{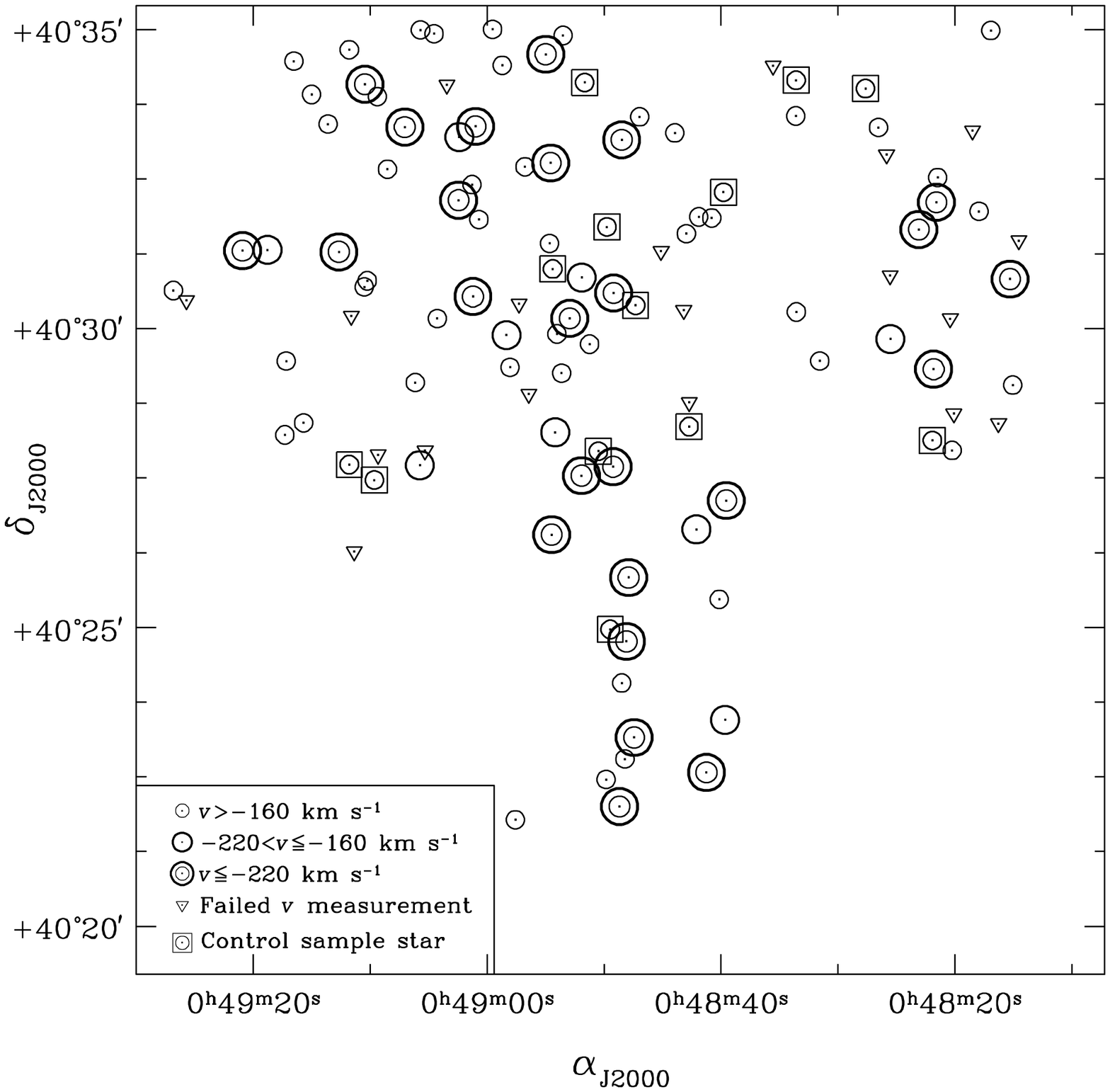}}
\figcaption[Reitzel.fig22.eps]{\label{radec_vel}{The location on the sky of
all 112~objects in the spectroscopic sample grouped by radial velocity:
$v>-160$~km~s$^{-1}$ (small circles),  $-220<v\leq-160$~km~s$^{-1}$ (large
circles), $v\leq-220$~km~s$^{-1}$ (large double circles), and objects for
which there is no reliable measurement of radial velocity (triangles).  An
extra square has been placed around each of the 13~control sample stars; all
of them have radial velocities $v>-160$~km~s$^{-1}$.  The shape of the
overall ``footprint'' of the sample is determined by the relative locations
and orientations of the 5~multi-slit masks used for the LRIS spectroscopic
observations.}}

\centerline{\epsfbox{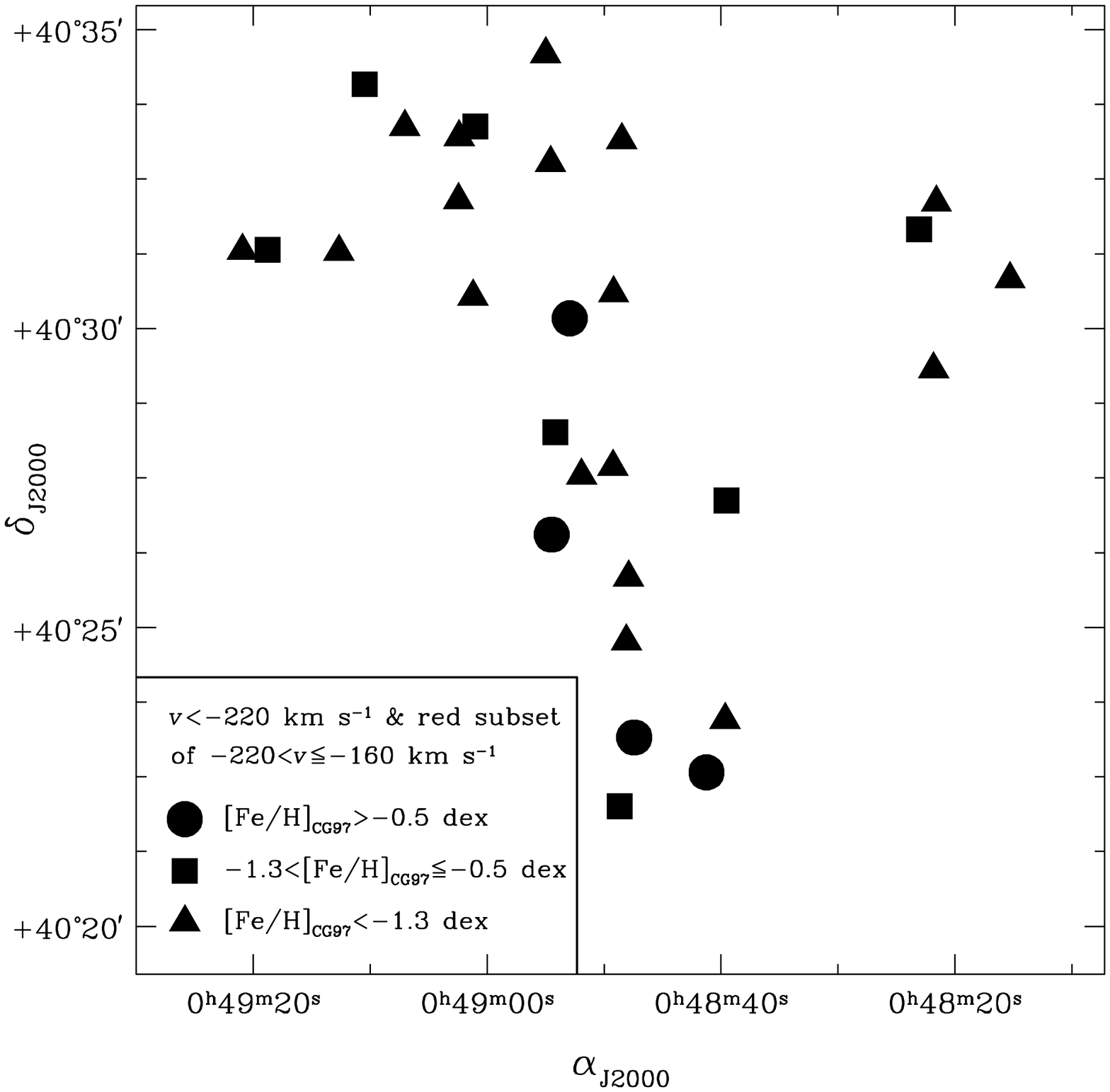}}
\figcaption[Reitzel.fig23.eps]{\label{radec_feh}{Same as Fig.~\ref{radec_vel},
except only the 29~red giants comprising the secure M31 sample are shown
grouped by metallicity on the Carretta-Gratton scale:
$\rm[Fe/H]_{CG97}>-0.5$~dex (circles), $\rm-1.3<[Fe/H]_{CG97}\leq-0.5$~dex
(squares), and $\rm[Fe/H]_{CG97}<-1.3$~dex (triangles).  The first group
(most metal-rich giants) consists of the the four~stars that possibly belong
to M31's disk or represent debris from a past accretion event (see
\S\,\ref{poss_disk_sec}).}}

\centerline{\epsfxsize=7.5in \epsfysize=9in
\epsfbox{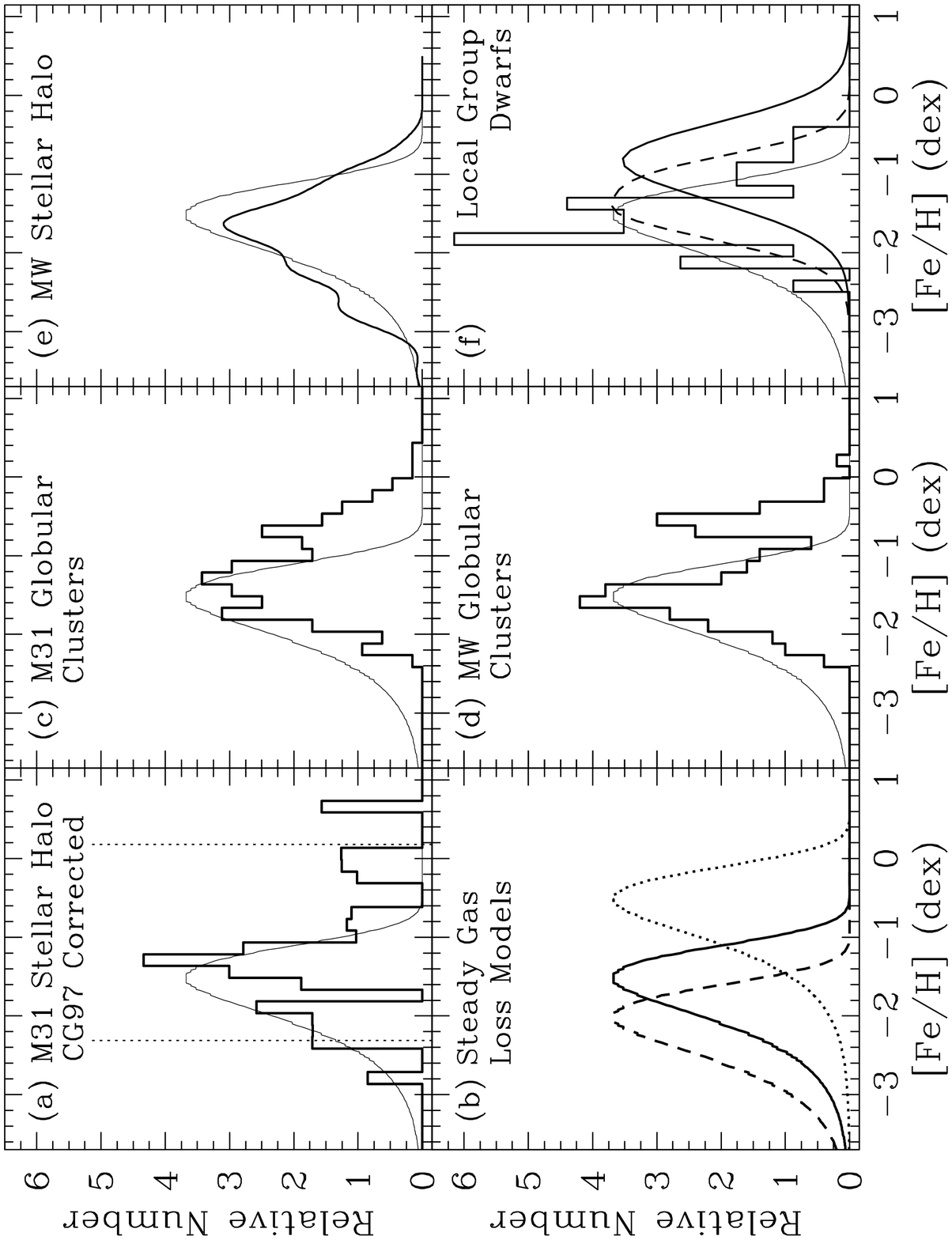}}
\figcaption[Reitzel.fig24.eps]{\label{feh_hist_comp}{Comparison of the
metallicity distribution of M31's halo to models and other M31 and Milky Way
halo samples (see \S\,\ref{halo_comp_sec} for details including data sources
for other halo tracers).~~~
({\it a\/})~Secure sample of M31 halo red giant stars on the Carretta-Gratton
scale, corrected for metallicity-selection effects caused by the drop in
$I_{\rm TRGB}$ and faint-end incompleteness [bold histogram; same as
Fig.~\ref{feh_hist}({\it e\/})].  The dotted vertical lines mark the range
over which the CG97 relation is calibrated.  The small subset of giants with
solar or higher metallicity may be members of M31's disk rather than its halo
(\S\,\ref{poss_disk_sec}).~~~
({\it b\/})~Chemical enrichment models with steady gas loss for
$\rm[Fe/H]_0=-10$~dex and $\rm\langle[Fe/H]\rangle=-2.0$ (dashed line),
$-1.5$ (solid line), and $-0.5$~dex (dotted line).~~~
({\it c\/})~The M31 globular cluster system (bold histogram).~~~
({\it d\/})~The Milky Way globular cluster system (bold histogram).~~~
({\it e\/})~Field stars in the Milky Way halo (bold solid line).~~~
({\it f\/})~Ensemble of dwarf satellite galaxies in the Local Group:
luminosity-weighted sum (bold solid line), direct sum giving equal weight to
each galaxy (bold histogram), and luminosity-weighted sum of all except the
two~luminous metal-rich satellites, the Large Magellanic Cloud and M32 (bold
dashed line).~~~
All curves are normalized to the actual number of stars in the secure M31
sample, 29.  The $\rm\langle[Fe/H]\rangle=-1.5$~dex steady-gas-loss model
[bold solid line in panel~({\it b\/})] is shown as a thin solid line in each
of the other panels to facilitate comparison among the data sets.}}

\vfill\eject
\end{document}


\begin{deluxetable}{lccccc}
\tablewidth{0pc}
\tablenum{4}
\tablecaption{Metallicity measurements}
\tablehead{
\colhead{~~~~~~~~~~Sample:} & \colhead{Main}    & \colhead{Excl.\ Red}  & \colhead{Excl.\ Poss.}  & \colhead{Excl.\ Red}    & \colhead{Poss.\ Disk} \\
\colhead{} & \colhead{} & \colhead{Intermed.} &  \colhead{Disk Stars\tablenotemark{2}} & \colhead{Intermed.\ Vel.\ \&} & \colhead{Stars\tablenotemark{2}} \\
\colhead{} & \colhead{} & \colhead{Vel.\ Stars\tablenotemark{1}} &  \colhead{} & \colhead{Poss.\ Disk Stars\tablenotemark{1,2}} & \colhead{ }}
\startdata
Number of stars	& 29 & 25 & 25 & 21 & 4 \\
\hline
$\rm [Fe/H]_{ZW}\tablenotemark{3}~~\rightarrow$ & & & & & \\
~~~~~Mean & $-1.04$ & $-1.00$ & $-1.82$ & $-1.92$ & $(+3.82)$ \\
~~~~~Median & $-1.74$ & $-1.74$ & $-1.78$ & $-1.88$ & $(+2.62)$ \\
~~~~~RMS & $\,$~~2.33 & $\,$~~2.50 & $\,$~~0.60 & $\,$~~0.53 & $\,$~~3.35 \\
~~~~~Corrected Mean\tablenotemark{6} & $-0.51$ & $-0.40$ & $-1.77$ & $-1.87$ & $(+4.57)$ \\
~~~~~Corrected Median\tablenotemark{6} & $-1.64$ & $-1.65$ & $-1.78$ & $-1.78$ & $(+3.29)$ \\
~~~~~Corrected RMS\tablenotemark{6} & $\,$~~2.98 & $\,$~~3.19 & $\,$~~0.58 & $\,$~~0.51 & $\,$~~3.23 \\
\hline
$\rm [Fe/H]_{CG97}\tablenotemark{4}~~\rightarrow$ & & & & & \\
~~~~~Mean & $-1.38$ & $-1.41$ & $-1.62$ & $-1.70$ & $+0.11$ \\
~~~~~Median & $-1.45$ & $-1.45$ & $-1.50$ & $-1.65$ & $+0.00$ \\
~~~~~RMS & $\,$~~0.80 & $\,$~~0.84 & $\,$~~0.54 & $\,$~~0.52 & $\,$~~0.39 \\
~~~~~Corrected Mean\tablenotemark{6} & $-1.22$ & $-1.22$ & $-1.56$ & $-1.65$ & $+0.20$ \\
~~~~~Corrected Median\tablenotemark{6} & $-1.34$ & $-1.34$ & $-1.50$ & $-1.50$ & $+0.10$ \\
~~~~~Corrected RMS\tablenotemark{6} & $\,$~~0.88 & $\,$~~0.93 & $\,$~~0.52 & $\,$~~0.50 & $\,$~~0.38 \\
\hline
$\rm [Fe/H]_{phot}\tablenotemark{5}~~\rightarrow$ & & & & & \\
~~~~~Mean & $-1.37$ & $-1.40$ & $-1.49$ & $-1.55$ & $-0.64$ \\
~~~~~Median & $-1.27$ & $-1.58$ & $-1.62$ & $-1.69$ & $-0.62$ \\
~~~~~RMS & $\,$~~0.60 & $\,$~~0.63 & $\,$~~0.55 & $\,$~~0.56 & $\,$~~0.35 \\
~~~~~Corrected Mean\tablenotemark{6} & $-1.25$ & $-1.26$ & $-1.42$ & $-1.47$ & $-0.56$ \\
~~~~~Corrected Median\tablenotemark{6} & $-1.05$ & $-1.09$ & $-1.27$ & $-1.62$ & $-0.61$ \\
~~~~~Corrected RMS\tablenotemark{6} & $\,$~~0.64 & $\,$~~0.67 & $\,$~~0.57 & $\,$~~0.59 & $\,$~~0.30 \\
\enddata
\tablenotetext{1}{``Red Intermed.\ Vel.\ Stars'' refers to red subset of
stars in intermediate radial velocity range, $-220<v<-160\,$km~s$^{-1}$;
these are likely to be M31 halo red giants (\S\,3.2).}
\tablenotetext{2}{``Poss.\ Disk Stars'' refers to kinematically-coherent
group of four~high-metallicity stars with velocities close to M31's systemic
velocity, similar to what might be expected for M31 disk stars or for
metal-rich debris from a past accretion event (\S\,5.3).}
\tablenotetext{3}{Spectroscopic calibration of Zinn \& West (1984)
metallicity scale from Rutledge et~al.\ (1997b); see \S\,4.2.2.  Values in
parentheses represent extrapolations well beyond the calibrated range and
should be treated as highly uncertain.}
\tablenotetext{4}{Same as~(3) for Carretta \& Gratton (1997) metallicity
scale.}
\tablenotetext{5}{Same as~(3) for a photometric calibration based on
Bergbusch \& VandenBerg (2001) model isochrones and two high metallicity
Padova model isochrones (Girardi et~al.\ 2000); see \S\,4.1.1.}
\tablenotetext{6}{Corrected for bias caused by $I_{\rm TGRB}$ becoming
fainter with increasing [Fe/H] and incompleteness at the faint end
(\S\,5.2).}
\end{deluxetable}